\documentclass[useAMS,usenatbib,times]{mn2e}

\usepackage{mathpazo} 
\usepackage[scaled]{helvet} 
\usepackage{courier} 
\normalfont
\usepackage[T1]{fontenc}

\usepackage{bm}
\usepackage{graphicx} 
\usepackage{stmaryrd}
\usepackage[usenames]{color}
\usepackage{graphicx}
\usepackage{natbib}
\usepackage{amssymb}

\definecolor{grey}{rgb}{0.75,0.75,0.75}
\definecolor{Orange}{rgb}{1.0,0.5,0.15}
\definecolor{brown}{rgb}{0.7,0.25,0.0}
\definecolor{pink}{rgb}{1.0,0.5,0.5}
\definecolor{darkerred}{rgb}{0.8,0,0}
\definecolor{darkerblue}{rgb}{0,0,0.8}
\definecolor{Blue}{rgb}{0,0.08,0.65}
\definecolor{Red}{rgb}{0.65,0.08,0.05}
\definecolor{Green}{rgb}{0.15,0.45,0.25}

\def\red{\color{Red}}

\def\equiv{{\,\triangleq\,}}

\def\d{{\mathrm{d}}}  
\newcommand{\tmmathbf}[1]{{\mathbf{#1}}}

\newcommand{\op}[1]{#1}

\newcommand{\mathd}{\mathrm{d}}

\newcommand{\mathpi}{\pi}
\def\parn{\par\noindent}

\def\iint{\int\!\!\!\!\!\int}    
\def\iiint{\iint\!\!\!\!\!\int}  

\def\PDF{{\rm PDF} }

\def\mathfrak#1{{\mathrm{#1}}}
\def\R#1{{\mathrm{#1}}}
\def\Sec#1{{Section~\ref{s:#1}}}
\def\Eq#1{{Eq.~\ref{e:#1}}}
\def\Ep#1{{~(\ref{e:#1})}}
\def\Eqs#1#2{{equations.~(\ref{e:#1})-(\ref{e:#2})}}
\def\EQN#1{\label{e:#1}}        
\def\Fig#1{{Fig.~\ref{f:#1}}}
\def\Fip#1{{~\ref{f:#1}}}

\def\M#1{{\mathbf{#1}}}
\def\MG#1{{\mbox{\boldmath $ #1$}}} 
\def\MGS#1{{{\mbox{\boldmath ${}_{#1}$}}}}
\def\M#1{{\mathbf #1}}

\newcommand{\mathe}{\mathrm{e}}

\newcommand{\half}{{{1/2}}}

\def\bfr {{\bf r}}
\def\bfv {{\bf v}}
\def\bfk {{\bf k}}
\def\bfw {{\bf w}}
\def\bfI {{\bf I}}

\def\bn {{\mathbf n}}
\def\bp {{\mathbf p}}
\def\bq {{\mathbf q}}
\def\ba {{\mathbf a}}
\def\bb {{\mathbf b}}
\def\bc {{\mathbf c}}

\def\bR {{\mathbf R}}

 \def\bv{{\bf v}} \def\br{{\bf r}} 
      
  \def\bk{{\bf  k}} \def\bx{{\bf x}} 
 \def\bI{{\bf I}}  \def\b1{{\bf 1}}
 \def\bw{{\bf w}}  \def\bO{{\bo}} \def\bG{{\bf \Gamma}}
      
\def\be{{\bf e}}

\def\bk{{\bf k}}

\def\bo{\MG{\omega}}
\def\bos{\MGS{\omega}}
\def\ie{{\frenchspacing\it i.e. }}
\def\cf{{\frenchspacing\it cf. }}

\def\eg{{\frenchspacing\it e.g. }}
\def\etc{{\frenchspacing\it etc... }}

\def\R#1{{\mathrm{#1}}}         
\def\Eq#1{{Eq.~(\ref{e:#1})}}   
\def\Ep#1{{~(\ref{e:#1})}}      
\def\Eqs#1#2{{Eqs.~(\ref{e:#1})-(\ref{e:#2})}}
\def\EQN#1{\label{e:#1}}        
\def\Fig#1{{Fig.~(\ref{f:#1})}}

\def\Fip#1{{~(\ref{f:#1})}}
\def\be{\begin{equation}}
\def\ee{\end{equation}}
\def\ba{\begin{eqnarray}}
\def\ea{\end{eqnarray}}

\def\pdrv#1#2{\frac{\partial #1}{\partial #2}}  
\def\pdrvn#1#2#3{\frac{\partial^#3 #1}{\partial #2^#3}}  
\def\drv#1#2{\frac{\d #1}{\d #2}}       

\def\M#1{{\mathbf{#1}}} 
\def\d{{\R{d}}}         
\def\MG#1{\bm{#1}} 
\def\MGS#1{\bm{#1}} 
\def\tmmathbf#1{{\mathbf{#1}}} 

\def\Comment#1{{\red skipped text}}

 \def\bv{{\bf v}}
      \def\br{{\bf r}}

 \def\bk{{\bf k}}
\def\bx{{\bf x}}
\def\bI{{\bf I}}
\def\bw{{\bf w}}

\def\bc{{\bf c}}
\def\ba{{\bf a}}
\def\bK{{\bf K}}
\def\bQ{{\bf Q}}
\def\bO{{\MG{ \Omega}}}
\def\bG{\MG{ \Gamma}}

\def\bo{\MG{\omega}}
\def\rn{{\mathrm{n}}}
\def\rN{{\mathrm{N}}}
\def\ie{{\frenchspacing\it i.e. }}
\def\cf{{\frenchspacing\it cf. }}
\def\eg{{\frenchspacing\it e.g. }}

\font\handfont=yswab 
 
\newcommand{\nicefrac}[2]{\leavevmode\kern.1em
            \raise.5ex\hbox{\the\scriptfont0 #1}\kern-.1em
      /\kern-.15em\lower.25ex\hbox{\the\scriptfont0 #2}}

\begin{document}
  \title[Dynamical flows through Dark Matter Haloes]{Dynamical flows through Dark Matter Haloes:  Inner perturbative dynamics, secular evolution, and applications}

\author[ Christophe Pichon and Dominique Aubert ]{
Christophe Pichon $^{1,2,3}$\thanks{E-mail: pichon@iap.fr},
   and Dominique Aubert$^{1,2,3}$\\
$^{1}$ Institut d'Astrophysique de Paris, 98 bis boulevard
       d'Arago, 75014 Paris, France \\
$^{2}$ Observatoire astronomique de Strasbourg, 11 rue de l'Universite, 67000
   Strasbourg, France\\
$^{3}$  Horizon, CNRS,  98 bis boulevard
       d'Arago, 75014 Paris, France}

   \date{Typeset \today ; Received / Accepted}

   \maketitle


\begin{abstract}
 We  investigate statistically  the dynamical  consequences  of cosmological
fluxes of matter  and related moments on progenitors  of today's dark matter
haloes. These  haloes are described  as open collisionless systems  which do
not  undergo strong interactions  anymore. Their  dynamics is  described via
canonical perturbation theory which accounts for two types of perturbations:
the  tidal field  corresponding  to  fly-bys and  accretion  of dark  matter
through the halo's outer boundary.

The non-linear evolution of both the  entering flux and the particles of the
halo  is  followed  perturbatively.   The  dynamical  equations  are  solved
linearly, order by order, projecting on a biorthogonal basis to consistently
satisfy  the  field equation. 
Since our perturbative solution of
the Boltzmann Poisson is explicit, we obtain, as a result,
 expressions for the N-point correlation  function of the halo's response to
the  perturbative environment.   It allows  statistical predictions  for the
ensemble distribution of  the inner dynamical features of  haloes. 
We demonstrate the feasibility of the implementation via a simple 
example in the appendix.  We argue
that the  fluid description  accounts for the  dynamical drag and  the tidal
stripping  of  incoming structures.   We  discuss  the  realm of  non-linear
problems which  could be addressed statistically  by such a  theory, such as
differential  dynamical friction, tidal  stripping and  the self  gravity of
objects within the virial sphere.

The secular evolution of open galactic haloes is investigated: we derive the
kinetic  equation which governs  the quasi-linear  evolution of  dark matter
profile induced by infall  and its corresponding gravitational correlations.
This yields a Fokker  Planck-like equation for the angle-averaged underlying
distribution function. This equation  has an explicit source term accounting
for  the net  infall  through the  virial  sphere. Under  the assumption  of
ergodicity  we then  relate the  corresponding source,  drift  and diffusion
coefficients for the ensemble-average  distribution to the underlying cosmic
two-point statistics of the infall and discuss possible applications.

The internal dynamics of  sub-structures within galactic haloes (distortion,
 clumps   as  traced  by   Xray  emissivity,   weak  lensing,   dark  matter
 annihilation, tidal streams  ..), and the implication for  the disk (spiral
 structure, warp \etc) are then  briefly discussed.  We show how this theory
 could be  used to (i)  observationally constrain the statistical  nature of
 the  infall (ii)  predict  the observed  distribution  and correlations  of
 substructures in upcoming surveys, (iii)  predict the past evolution of the
 observed  distribution of  clumps,  and finally  (iv)  weight the  relative
 importance of the intrinsic (via the unperturbed distribution function) and
 external (tidal and/ or infall) influence of the environment in determining
 the  fate  of galaxies.   We  stress that  our  theory  describes the  {\sl
 perturbed } distribution function  (mean profile removed) directly in phase
 space.

  \end{abstract}
\begin{keywords}
Galaxies: haloes, kinematics and dynamics, statistics , Cosmology: Dark Matter  
\end{keywords}

\section{Introduction}
                                      
It now  appears clearly that  the dynamical (azimuthal  instabilities, warps,
accretion),  physical (heating, cooling)  {and} chemical  (metal-poor  cold gas
fluxes) evolution of galaxies are   processes which are 
{partly driven} by
the boundary conditions imposed by their cosmological environment.
It is therefore of prime importance {to formulate
the effects of such an interaction in a unified framework}.

Modern digital all-sky surveys, such as the  SDSS, 2MASS or the 2dF provide for the 
first time the opportunity to build  statistically relevant  constraints on the
dynamical states  of galaxies
{which can} be  used as observational input.
Other projects, like Gaia or Planck, will provide small-scale information on
our Galaxy  and its  environment and will  soon allow 
detailed confrontation  of the 
predictions of {models} with the  observations.
 We ought to be  able to draw conclusions
on the internal dynamics of the halo and its inner components
 and constrain their statistical properties.

Unfortunately, it is difficult to study the response of haloes to moderate amplitude
perturbations. Current N-body techniques suffer from resolution limitations  
(due  to particle number and drift in orbit
integration, see e.g. \citet{Power}, \citet{Binney04}, 
for a discussion of such effects)   that {hide to some extent}
 linear collective effects which  dominate the response of the halo
(\cite{weinberg2}, \cite{murali})\footnote{it has been argued
 that shadowing (\cite{Tremaine2}) will in practice allow for another
 orbit to correct for the drift, but this is of no 
help to resonant processes because
it requires that the {\sl same} orbit does not diffuse for a few libration
periods.}.
Simulations on galactic scales are also often carried 
without any attempt to {represent} the cosmological variety 
arising from the possible boundary conditions (the so-called cosmic variance problem). 
This {is because the} dynamical range required 
to describe both the environment and the inner structure is 
considerable, and can only be achieved for a limited number of simulations
(e.g. \citet{Knebe}, \citet{Gill}, \citet{Diemand}).
{By}  contrast, the method presented below {circumvents} this 
difficulty while relying on an {\sl explicit} treatment of the inner dynamics of the 
halo, in the perturbative regime. Specifically, our purpose  is to develop a tool to study  the dynamics of an
open stellar system and apply it to  the dynamic of a halo which is embedded
into  its cosmological  environment.  
 One can think of this project as an attempt to produce a semi-analytic explicit
re-simulation tool, in the spirit of what is done in N-body simulations with zoomed-in initial conditions. 

{The} concept of an initial power spectrum describing the statistical properties
of the gravitational perturbations has {proved very useful in
cosmological studies}
(e.g. \citet{Peebles}, \citet{BCGS}).
The underlying paradigm{,} 
that gravity {drives} cosmic evolution,
is likely to {be} a good description at the megaparsec scale.
We  show  below  that a  similar  approach  {to}  galactic haloes  is  still
{acceptable,} and marginally within  the reach of our modeling capabilities.
{The} description  of the  boundary is significantly  more complex,  but the
inner dynamics of  hot components is better behaved.   Here, we {describe} a
stable system  which undergoes small  interactions, rather than  an unstable
system in comoving coordinates undergoing catastrophic multi-scale collapse.

The purpose of this investigation is  to derive analytically the dynamical 
 response of a galactic
 halo,  induced  by  its (relatively weak)  interaction   with  its  near
environment.   Interaction  should be  understood  in  a  general sense  and
involve  {tidal} potential  interactions (like  that  corresponding  to a  satellite
orbiting around the galaxy), or an infall where an external quantity (virialized or not)
   is
advected  into the galactic halo.
   
     With  a suitable formalism, we
derive the  propagation of an  external perturbation from
the near  galactic environment  down to the  scale of the
 galactic disk through  the dark matter halo.   
{We essentially} solve the coupled collisionless Boltzmann-Poisson
equations as a Dirichlet initial value problem to determine the response 
of the halo to infall and tidal field.
{The} basis over which the response is projected {can} be  customized 
to, say,  the universal  profile of  dark matter  haloes, 
{which makes it possible to}  consistently and
efficiently {solve} 
the coupled dynamical and  field equations {,} 
so long as the entering
fluxes of dark matter amounts to  a small perturbation in mass compared to the
underlying equilibrium.

{In a} pair of companion papers, \citet{aubert1,aubert2}
described the statistical properties of the 
infalling distribution of dark matter at the virial radius, $R_{200}$ as a function 
of cosmic time between redshift $z=1$ and today. These papers focused on 
a description of the one- and two-point statistics of the infall towards well formed $L_\star$
dark matter haloes.
All measurements were carried for 15~000  haloes undergoing minor
mergers. 
   The  two-point  correlations were
measured both angularly  and temporally for the flux  densities, and over the
whole 5D phase space for the expansion coefficients of the source. 
 
Together with the measurements presented there, we show in this paper that the
formalism described below {will} allow astronomers
to address globally and coherently dynamical issues on galactic scales.
Most importantly it will allow them to tackle problems in a {\sl statistically representative }
manner.
This investigation has a broad field of  possible applications.   Galaxies are subject to
boundary conditions  that reflect {motions} on  larger scales  and their
dynamics may constrain the cosmology  through the rate of merging events for
example,  or the  mass distribution  of satellites.   
Halo transmission and amplification  also 
fosters communication between spatially
separated regions, (see \eg  \cite{murali}) 
 and  continuously {excites the}  disk structure.
{For example}, spirals  can  be  induced  by  encounters with  satellites  and/or  by  mass
injection (e.g. \citet{TT}, \citet{Howard}), while warps results from torque interactions with the surrounding
matter (\citet{Lopez}, \citet{JiangBinney}). Therefore  the proportion of spirals and  warps contains information
on the  structure's formation and  environment.  
The statistical link between the inner properties of galactic haloes, and 
their cosmic boundary can be reversed  to attempt and constrain the nature of the 
infall while investigating the one and two point statistics of the induced
perturbations. This is best done by transposing down to galactic scales
the classical  cosmic probes for the large-scale structures (lensing, SZ, \etc)
which have been used successfully to characterize the power-spectrum of 
fluctuations on larger scales.

The outline of this paper is the following: 
we describe in \Sec{spherical} the linear response of a 
spherical halo 
which undergoes cosmological infall of dark matter,  and compute the 
induced correlations in the inner halo;
\Sec{nonlinear} presents the second-order perturbative response of
the galactic halo to the infalling flux; (Appendix \ref{s:apendperturb}
gives the higher order corrections to the dynamics and addresses the 
issue of dynamical friction). 
\Sec{quasi1} derives the Fokker-Planck equation that the cosmic mean halo profile 
{obeys} in such an open environment. 
\Sec{applications}
describes briefly possible astrophysical applications. 
In particular, it is discussed how {the}  statistical analysis
of mean and {variance} properties of  galactic haloes and galaxies can be compared to
the quantitative prediction of  the  concordant $\Lambda$CDM cosmogony
on those scales.
We also show how to revert in time observed tidal features within our Galaxy, or in 
external galaxies.
The last section draws conclusions and discusses prospects for future work.

\section{The spherical Halo: linear response}
\label{s:spherical}

 \begin{figure} 
\centering
\resizebox{0.95\columnwidth}{0.95\columnwidth}{\includegraphics{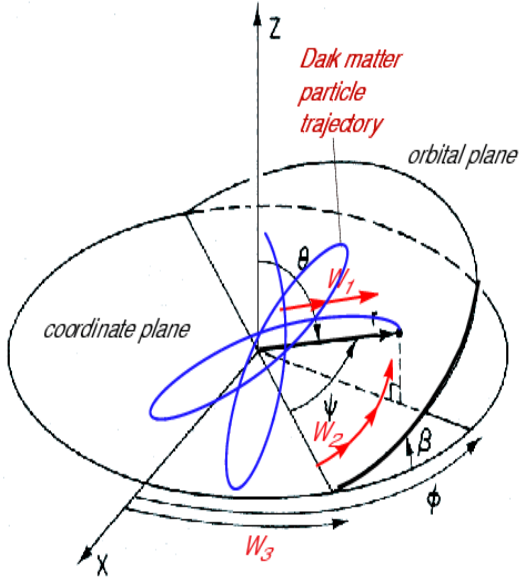}}
\caption{ The action angle, ($\bI,\bw$), - spherical coordinate, ($\br,\bv$), transformation. The dark matter 
 particle at running spherical coordinate $(r,\theta,\phi)$ describes a rosette in the 
 orbital plane orthogonal to its momentum, $\M{L}$. The line of node 
 of the orbital plane intersects the  $x-y$ plane at a constant (in spherical symmetry) angle $w_3$ with respect to the 
 $x$ axis.  The orbital plane is at an angle $\beta={\rm acos}(L_z/L)$ to the $x-y$ plane. 
  The particle polar coordinate in the orbital plane w.r.t. this line of node
 is $\psi$. The angle coordinates, $w_2$, is measured along $\psi$ but 
 varies linearly with time by construction. Finally the radial angle $w_1$ varies 
  with radius between peri-apse and apo-apse. 
   (Strongly inspired from Fig~1 of \citet{Tremaine}).}
\label{f:angles} 
\end{figure}
In the  following section, we extend  to {\sl open} spherical  stellar systems the
 formalism developed by \citet{Kalnajs2}  (for stellar disks), 
 \citet{aoki} (for gaseous disks), \citet{Frid},
 \citet{Tremaine} and e.g. \citet{Palmer}, \citet{murali}, \citet{dejonghe}, \cite{bertin} by adding a
 source  term  to  the  collisionless Boltzmann  equation.
Since the formalism is otherwise fairly standard, we will present it relatively swiftly.
In a nutshell,
the dynamical equations are solved
linearly while projecting over a biorthogonal basis 
to consistently satisfy the Poisson equation (\eg \citet{Kalnajs1} \citet{Kalnajs2} \citet{Kalnajs4}). 
The dynamical equation of an open system characterized by its distribution
function, $ \mathfrak{F}$, together with the field equation, read formally:
\begin{equation}
  \partial_t \mathfrak{F}+ \{ \mathfrak{H} + \psi^e,\mathfrak{F}\} = s^e \hspace{0.25em}, \quad {\rm and } \quad \nabla^{2} \Psi = 4 \pi {\cal G} \int \d \bv  \mathfrak{F}\,,
  \label{e:boltzpert1}
\end{equation}
where $ \mathfrak{H}$ is the Hamiltonian of the system, $\{ \,\,\,\}$ is the usual Poisson
bracket and $(  \Psi,\psi^e, s^e )$ stands for the potential, the perturbing exterior potential and
incoming source term.
 The latter, $s^{e}(\br,\bv,t)$ accounts for the entering dark
matter \textit{at the virial radius} and is discussed in detail below (\Sec{source}, see also \citet{aubert0} and 
\citet{aubert1}).
  In  a  somewhat
unconventional manner, $ \psi^e(\br,t)$  refers  here to the external potential, \ie
the  tidal  potential  created   by  the  perturbations  {\sl  outside}
{\sl the outer boundary of the halo} ({, i.e. $R_{200}$}). 

Let us expand the Hamiltonian and the distribution function, $\mathfrak{F}$, as:
 \begin{equation}
\mathfrak{F} = F + \varepsilon f \,,\quad
  {\rm and} \quad  \mathfrak{H} = H_0 + \varepsilon   \psi
  \,, \EQN{boltzsphere0}
  \end{equation}
where  we assume that everywhere in phase space $\varepsilon \equiv m / M
\ll 1$ i.e. that the  mass of the perturbation, $m$, is small compared to the mass,
$M$, of the unperturbed halo.
In \Eq{boltzsphere0},  
  $f$  represents  the   small  response  to  the  perturbations,  $F$
represents  the  equilibrium  state  and  $ \psi$  the  small  response  in
potential.  Putting \Eq{boltzsphere0} into \Eq{boltzpert1} and reordering 
in $\varepsilon$ yields the linearized Boltzmann equation~:
\begin{equation}
\frac{\d f}{\d t}+\frac{\d H_0}{\d \bI}\cdot\frac{\d f}{\d\bw}-(\frac{\d \psi}{\d \bw}+\frac{\d \psi^e}{\d \bw})\cdot\frac{\d F}{\d\bI}=s^e,
\end{equation}
where $\bI$ and $\bw$ are conjugate canonical variables
which are described in the following section.
%
\subsection{The Boltzmann equation in action-angle}
\label{s:Boltzmann}
%
 The   most  adequate  representation {of} multiply-periodic integrable systems   relies on  the
action-angle variables\footnote{Note that $(\bw,\bI)$ are canonical variables, and as such preclude nothing 
about the evolution of the system. They simplify the expression of 
the  linearized equations, order by order.
}, since
resonant processes will dominate the response of the {live} halo,
 and are best expressed
in those variables. We will use vector notation for simplicity.
The details of the computation of these variables is discussed in  Appendix~\ref{s:implementation}
{following work by} \citet{murali}  (see also \Fig{angles}). 
This achieves 
separation of variable between the  phase space canonical variables (angle 
and actions) on the one hand, and time on the other {hand}.  
{We denote as usual the set of action variables by ${\bf I}$ and
angle variables by $\bfw$ (see Appendix~\ref{s:implementation}).
 The rates of change of angles is $\bo \equiv \d  \bfw/\d t$.
Along the {multi-periodic } orbits,
 any field, $Z$, can be
Fourier-expanded with respect to the angles as:}
{
\begin{equation}
Z(\br,\bv,t)=\sum_{\bk} Z_\bfk({\bf I},t)\exp{(\imath \bk\cdot\bw)}. 
\quad {\rm
}\quad
\end{equation}
Conversely
\begin{equation}
Z_{\bfk}({\bf I},t) =\frac{1}{(2\pi)^3}\int \d {\bw} \exp(-\imath \bk
\cdot \bw ) Z(\br,\bv,t)\,,
\end{equation}
where  $\bk\equiv(k_1,k_2,k_3)$ is the   Fourier  triple   index
conjugate to the three angles $\bw\equiv(w_1,w_2,w_3)$ . } 
Given \Eqs{boltzpert1}{boltzsphere0}, the linearized Boltzmann equation  in such a representation is:
\begin{equation}
\frac{\partial f_\bfk({\bf I},t)}{\partial t}+\imath {\bf k} 
\cdot {\bo}  f_\bfk({\bf I},t)  =
\imath {\bf k} \cdot \frac{\d{}F}{\d \bI} (\psi_\bfk ({\bf I},t)  + \psi^e_\bfk ({\bf I},t)) 
+  s^{e}_\bfk({\bf I},t).
\label{e:boltzsphere}
\end{equation}
{ Here $\psi$ is the potential perturbation created by the halo's inner
    density fluctuations
and $\psi^e$ the potential perturbation created by external fly byes}. 
The gravitational field of incoming particles is accounted for by the  source   term
{$s^{e}$}.
The solution to \Eq{boltzsphere} may then be written as:
\begin{eqnarray}
f_\bfk({\bf I},t) &=&\int_{-\infty}^{t}
\exp(\imath {\bf k} 
\cdot {\bo} (\tau-t))\times \nonumber \\
&& \hskip -0.5cm
\left[
\imath {\bf k} \cdot \frac{\d F}{\d \bI } 
\left[\psi_\bfk(\bI,\tau)+\psi^{e}_\bfk(\bI,\tau)\right] +
s^{e}_\bfk({\bf I},\tau)\, \right] \d \tau.
\label{e:DFsphere}
\end{eqnarray}
{\Eq{DFsphere}} assumes that the perturbation 
{has been} switched on a long time ago in the past  so that 
all transients have damped out.\footnote{Mathematically, 
we only retain the particular solution to \Eq{boltzsphere}, while assuming 
that the homogeneous solution did not hit long-lived resonances.   
} 
\subsection{Self-consistency }
 \Eq{DFsphere}  {can be integrated}
over velocities and {summed} over $\bk$ to {get}
the density {perturbation}:
\begin{eqnarray}
\rho(\br,t) &=&\sum_{\bk}\int_{-\infty}^{t} \d \tau
\int \d \bv  \left. \exp(\imath {\bf k} 
\cdot {\bo} (\tau-t)+\imath \bk \cdot \bw)\right.
\times \nonumber \\
&& 
\left. \hskip -0.5cm
\left[
\imath {\bf k} \cdot \frac{\d F}{\d \bI } 
\left[\psi_\bfk(\bI,\tau)+\psi^{e}_\bfk(\bI,\tau)\right] +
s^{e}_\bfk({\bf I},\tau)\, \right]\right. .
\label{e:denssphere}
\end{eqnarray} 
Let us expand the potential and the density
 over a bi-orthogonal complete {set of} basis 
functions such that:
\begin{equation}
    \psi(\br,t) =\sum_{\bn} a_{\bn}(t) \psi^{[\bn]}(\br) \, ,\quad  
    \rho(\br,t) =\sum_{\bn} a_{\bn} (t)
    \rho^{[\bn]}(\br) \, ,  \EQN{defpsiexp}
    \end{equation}
    \begin{equation}
    \nabla^{2}\psi^{[\bn]}= 4 \pi G \rho^{[\bn]}
    \, , \quad \int \psi^{[\bn]*}(\br)  \rho^{[\bp]}(\br) \d \br = 
\delta^{\bn}_{\bp}\,,
   \EQN{defbibasis}
\end{equation}
(where $ \psi^{[\bn]*}(\br) $ is the complex conjugate of  $\psi^{[\bn]}(\br) $).
{We} naturally expand the external potential {on} 
the same basis (\citet{Kalnajs1}) {as}:
\begin{equation}
\quad  \psi^{e}(\br,t) =\sum_{\bn} b_{\bn}(t) 
    \psi^{[\bn]}(\br). \EQN{poisson}
\end{equation}  
{ Thus, the coefficients  $a_{\bn}$ are representative
of the density and potential perturbations in the halo
itself, at $r < R_{200}$, while
the coefficients $b_{\bn}$ represent the potential
created in the halo by density fluctuations
at $r> R_{200}$.} 
Taking advantage of  bi-orthogonality  \Eq{denssphere}
{is multiplied}  by
$\psi_{\bp}^*(\br)$ and {integrated} over $\br$, which yields:
\begin{eqnarray}
  a_{\bp}(t) \hskip -0.25cm &=& \hskip -0.25cm  \sum_{\bk} 
    \int_{-\infty}^{t}  \hskip -0.25cm \d \tau
\iint   \d \bv \d \br \exp(\imath {\bf k} 
\cdot {\bo} (\tau-t) +\imath \bk \cdot \bw)
\psi^{[\bp]*}(\br)\times  \hskip -0.35cm\nonumber \\
&& \hskip -1.5cm
 \left[\sum_{\bn}
\imath {\bf k} \cdot \frac{\d F}{\d \bI } 
\left[ a_{\bn}(\tau)+ b_{\bn}(\tau) \right]  \psi^{[\bn]}_\bk(\bI) +
{s}^{e}_\bfk({\bf I},\tau)\, \right] \,.
    \label{e:ap1}
\end{eqnarray}
We may now swap from position-velocity to action-angle variables.
{Since this}
transformation is canonical
$ \d \bv \d \br= \d \bw \d \bI$.
In \Eq{ap1} only $\psi^{[\bp]}(\br)$ depends on $\bw$, so we may carry the 
$\bw$ integration over $\psi^{[\bp]*}$, which yields $\psi^{[\bp]*}_\bk(\bI)$.
{Eq. (\ref{e:ap1}) then becomes :}
\begin{eqnarray}
     a_{\bp}(t) &=& (2\pi)^ 3 \sum_{\bk} 
    \int_{-\infty}^{t} \d \tau
\int   \d \bI  \exp(\imath {\bf k} 
\cdot {\bo} (\tau-t))\times    \label{e:ap2} \\
&& \hskip -2cm
\left[
\sum_{\bn}
\imath {\bf k} \cdot \frac{\d F}{\d \bI } 
\left[ a_{\bn}(\tau)+ b_{\bn}(\tau) \right]\psi^{[\bp]*}_\bk(\bI)  
\psi^{[\bn]}_\bk(\bI) +
{s}^{e}_\bk({\bf I},\tau)\, \psi^{[\bp]*}_\bk(\bI)  \right] \,.
 \nonumber
\end{eqnarray}
At this point, it seems natural to expand the source term {on} a basis too, but
unlike  the previous  one, this  basis  should also  describe  velocity
space. 
We admit for now that such a basis $ \phi_{n}(\br,\bv)$  exists,  and write:
\begin{equation}
s^e(\bfr,\bfv,t)=\sum_{\bn} c_{\bn}(t) \phi^{[\bn]}(\br,\bv) \, ,
\quad {\rm so }
\end{equation}
\begin{equation}
 \quad {s}^{e}_\bk({\bf I},\tau)\,= \sum_{\bn} c_{\bn}(\tau) {\sigma}^{[\bn]e}_\bk({\bf I}), \EQN{defsum}
\end{equation}
where $ {\sigma}^{[\bn]e}_\bk({\bf I})$ is the angle transform of $\phi^{[\bn]}(\br,\bv)$ (see \Eq{defsek} below).
{The coefficients $c_{\bn}$ are representative of
the mass exchange between the halo
and the external world. The }
sum in \Eq{defsum} spans velocity space as well as configuration  space, and 
therefore involve significantly more terms.
 Such an expansion is performed in \citet{aubert1} to constrain
the source function measured in cosmological simulations. 
 Calling ${\bf a}(\tau) 
=[a_{1}(\tau),\cdots, 
a_{\bn}(\tau)\cdots]$,  ${\bf b}(\tau) =[b_{1}(\tau),\cdots, b_{\bn}(\tau)\cdots]$, 
and ${\bf  c}(\tau)   =[c_{1}(\tau),\cdots,  c_{\bn}(\tau)\cdots]$, we define  two
matrices, $\bK$ and $\bQ$. The matrix $\bK$ has elements
${ K}_{\bp,\bn}$ defined by:
\begin{equation}
  { K}_{\bp,\bn}(\tau) =
  (2\pi)^ 3\! \sum_{\bk} \!\!
\int \!\! \d \bI   \exp(\imath {\bf k} 
\cdot {\bo} \tau)
\imath {\bf k} \cdot \frac{\d F}{\d \bI } \psi^{[\bp]*}_\bk(\bI)  
\psi^{[\bn]}_\bk(\bI)
,
    \label{e:defkk}
\end{equation}
{ which depend } only on the halo equilibrium state.
{The matrix $\bQ$ has elements}
\begin{equation}
  { Q}_{\bp,\bn}(\tau)  =  (2\pi)^ 3\! 
   \sum_{\bk} \!
\int  \! \d \bI  \exp(\imath {\bf k} 
\cdot {\bo} \tau)
{\sigma}^{[\bn]e}_\bk({\bf I})\,  \psi^{[\bp]*}_\bk(\bI)
, 
    \label{e:defhh}
\end{equation}
{which depend} only {on the source's expansion basis}.  Equation (\ref{e:ap2}) then becomes:
\begin{equation}
 {\bf a}(t) = \int_{-\infty}^{t}\hskip -0.25cm \d \tau   \left(
{\bf K}(\tau-t) \cdot 
    \left[{\bf 
    a}(\tau)+ {\bf b}(\tau) \right]+ {\bf Q}(\tau-t)\cdot {\bf 
c}(\tau) \right).
    \label{e:SOL}
\end{equation}
The  kernels  $\bf  K$  and  $\bf  Q$ are  functions  
 of the equilibrium  state distribution  function, $F$,
and  of the two  bases, $\phi^{\bf  [\bn]}(\br,\bv)$, and  $\psi^{\bf [\bn]}(\br)$
only. {They}  may  be computed once  and for all for  a given
equilibrium model. 
Assuming  linearity and knowing  $\bf K$  and $\bf  Q$, one  can see  that the
properties  of   the  environments   ({represented by} $\bf  b$   and  $\bf   c$)  are
\textit{propagated} to the inner dynamical properties of collisionless
systems (described by $\bf a $). 
We may perform  a ``half'' Fourier transform 
{with respect to time. In the limit where the transients
may be neglected, which implies that the system should be
stable, this transform amounts to a Laplace transform with
$p=\imath \omega+\epsilon^{+}$ . } 
{Temporal} convolutions are {then}
replaced by matrix multiplications and \Eq{SOL} becomes:
\begin{equation}
  {\bf \hat a}(\omega) = (\b1 -{\bf \hat K}(\omega) )^{-1} \cdot 
  \left[{\bf \hat K}(\omega) \cdot  {\bf \hat  b}(\omega)+ {\bf \hat  Q}(\omega) \cdot {\bf 
\hat c}(\omega) \right].
    \label{e:SOLK}
\end{equation}
{In this expression,} $\b1$ is the identity matrix,  
{and} ${\bf \hat K}$ and ${\bf \hat Q}$
{include} Heaviside functions before Fourier transform
 to account for causality (see \citet{aubert0} for details).
\Sec{pertComplex} gives an  explicit expression for $\hat \bK(\omega)$.
Note the difference between $\MG{\omega}$,
the angular frequency of the orbits, defined {above} \Eq{boltzsphere}, and $\omega$, 
the half Fourier transform {variable associated with time which
appears} in \Eq{SOLK}.
Here $\bb$ and $\bc$ could be given deterministic functions of time, or 
stochastic random fields (characterized statistically in \citet{aubert1}).
In contrast, $\ba$ describes the detailed response of the halo in phase 
space within $R_{200}$.
\begin{figure} 
\centering
\resizebox{0.95\columnwidth}{0.95\columnwidth}{\includegraphics{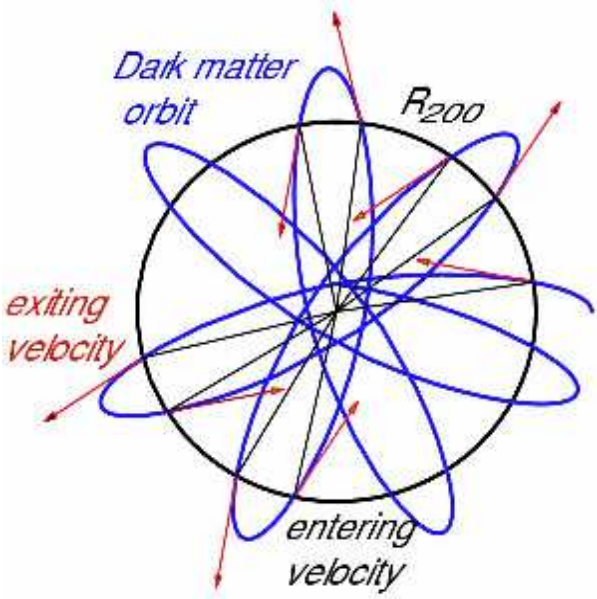}}
\caption{ A typical rosette orbit in its orbital plane; the intersection with the $R_{200}$ 
 sphere is shown, together with the corresponding velocity vectors, both entering and exiting.
 The net flux of such quantities enters \protect{ \Eq{defsedelta}} and 
 characterize the source of infall perturbing the halo. Note that by construction, in 
 the linear regime  all infalling material 
 re-exits $R_{200}$, since the perturbation evolves 
along the unperturbed orbits. This is to be contrasted to 
the situation presented in \Fig{bundle} where dynamical friction is 
qualitatively accounted for. 
   }
\label{f:enter-orbit} 
\end{figure}
%
\subsubsection{Higher moments} 
 The second moment is obtained by multiplying (\ref{e:DFsphere}) 
by $\bfv$ and by performing an integration over velocities.
 Summing over  $\bfk$ leads to:
\begin{eqnarray}
\rho \bar \bv (\br,t) &=&\sum_{\bk}\int_{-\infty}^{t} \d \tau
\int \d\bv \exp(\imath {\bf k} 
\cdot {\bo} (\tau-t)+\imath \bk \cdot \bw)
 \times \nonumber
\\  && \hskip  -1.5cm
\left[
\imath {\bf k} \cdot \frac{\d F}{\d \bI } \bv
\left[\psi_\bk(\bI,\tau)+\psi^{e}_\bk(\bI,\tau)\right] +\bfv
{s}^{e}_\bk({\bf I},\tau)\, \right] \, . \EQN{defvrsol}
\end{eqnarray}
Using the same biorthonormal expansion as {above}, 
we may express the {mass flux as a function of the coefficients} $a_\bn$ 
{and } $b_\bn$ ({associated with the potential
perturbations of external origin}). If we define 
the following  new tensors:
\begin{equation}
{K}_{[2],\bn}(\br,\tau) =
   \sum_{\bk} 
\int \d \bv  \exp(\imath {\bf k} 
\cdot {\bo} \tau +\imath \bk \cdot \bw)
\, \bv \,\, \imath {\bf k} \cdot \frac{\d F}{\d \bI }   
\psi^{[\bn]}_\bk(\bI)
 ,
 \label{e:defK2}
\end{equation}
and
a similar expression for 
$ {Q}_{[2],\bn}(\br,\tau)$,
involving {the source's expansion basis}, {the mass flux} may {be written}
as a convolution:
\begin{eqnarray}
\rho \bar \bv (\br,t) =
  \int_{-\infty}^{t} \hskip -0.4cm \d \tau  \left( {\bf K}_2(\br,\tau-t) \cdot 
    \left[{\bf 
    a}(\tau)+ {\bf b}(\tau) \right] +{\bf Q}_2(\br,\tau-t) \cdot {\bf 
c}(\tau) \right). \nonumber
\end{eqnarray}
After half Fourier transforming with respect to time, 
we {get}
\begin{eqnarray}
\rho { \hat {\bar \bv} }
 =
    {\bf \hat K}_{[2]}(\br,\omega) \cdot 
  (\b1 -{\bf \hat K} )^{-1} \cdot 
  \left[  {\bf \hat  b}+ {\bf \hat  Q}\cdot  {\bf 
\hat  c}\right]+ 
  { \bf \hat  Q}_{[2]}(\br,\omega)\cdot {\bf \hat c }\,. 
    \label{e:solrhov}
\end{eqnarray}
We will return to \Eq{solrhov} in \Sec{applications}.

\subsection{Sinks, sources and tidal field}
\label{s:source}
%
Let us now turn to an explicit description of the 
source term ($s^{e}$, hence $\bc$), and the tidal field ($\psi^{e}$, hence $\bb$) 
entering \Eq{boltzsphere}. 
We {consider here a} source at the virial radius corresponding
to cosmic infall{. Note however} that we might 
{have considered just as well} 
sinks reflecting  
the presence of a super massive black hole at the 
center of the host galaxy  or the deflection/absorption 
of orbits due to a galactic disk.
\subsubsection{ Source of infall at $R_{200}$}
\label{s:R200infall}
A possible Ansatz for the source term
consistent with  the first two velocity  moments of the  entering
matter {has been proposed by} \citet{aubert0}.
Following {them} 
$s^e(\bfr,\bfv,t)$ {can be written as}
\begin{eqnarray}
  s^e(  \bfr, {\bf v} ,t)
   \equiv  \sum_{\bf m} Y_{\M{m}}(\bO)\delta_{\rm D}(r-R_{200}) 
  \left(\sum_{\MG{\alpha}} c_{\M{n}}(t) g_{\MG{\alpha}}(\bv) \, Y_{\M{m}}(\bO) 
  \right) , \nonumber
  \label{e:sourceexpr}
 \end{eqnarray}
where  $\M{m}$   stands  for  the   two  harmonic  numbers,   $(\ell,m)$  and
$Y_{\M{m}}(\bO)\equiv   Y_{\ell}^{m}(\bO)$ is the usual spherical harmonic.
{The Dirac function} $\delta_{\rm D}(r-R_{200})$ {appears because
the source  terms are located at the virial  radius in our representation\footnote{This choice is mainly justified by the measurments
  performed in \citet{aubert0} and \citet{aubert1} and stands as a good
  compromise between a relaxed halo inside this boundary and a low
  contribution of the orbits of relaxed particles to the flux.}.
This equation corresponds to the 
parameterization {of $\phi^{[\M{n}]}$ as}:
\begin{eqnarray}
 \phi^{[\M{n}]}(\br,\bv)  &=&g_{\MGS{\alpha}}(\bv) \, Y_{\M{m}}(\bO) \delta_{\rm D}(r-R_{200})\,, \nonumber \\ &\equiv&
 g_{{\alpha}}(v) \, Y_{\M{m}}(\bO)Y_{\M{m}'}(\bG)  \delta_{\rm D}(r-R_{200})\,,
    \label{e:defexp2}
\end{eqnarray}
of Gaussian functions, $g_{\MG{\alpha}}$, covering the radial velocity component and spherical harmonics for
the angle distribution, $\bG$, of the velocity vector and 
orientation, $\bO=(\theta,\phi)$ of the infall (see \citet{aubert1} for details). Here we have
$\MG{n}\equiv[\MG{m},\MG{\alpha}]\equiv[\MG{m},\MG{m}',\alpha]\equiv[\ell,m,\ell',m',\alpha]$.
From \Eq{defsum}
\begin{eqnarray}
{\sigma}^{[\M{n}]e}_\bk({\bf  I}) &\equiv& \frac{1}{(2\pi)^3}\int  \d^{3} {\bw}
\exp(-\imath \bk \cdot \bw ) \phi^{[\M{n}]}(\br,\bv).\EQN{defsek}
\end{eqnarray}
With Eq.(\ref{e:defexp2}), \Eq{defsek} becomes
\begin{eqnarray}
{\sigma}^{[\M{n}]e}_\bk({\bf   I})&=&   \frac{1}{(2\pi)^3}\int  \d^3   {\bw}
\exp\left(-\imath   \bk  \cdot   \bw   \right)  Y_{\M{m}}[   \bO(\bfI,\bfw)] \times \nonumber \\
&&
g_{\MGS{\alpha}}(\bv[\bfI,\bfw])\delta_{\rm D}(r(\bfI,\bfw)-R_{200}) \, .\EQN{defsedelta}
\end{eqnarray}
We can make use   of  the  $\delta_{\rm D}$   function  occurring  in
 \Eq{defsedelta}   since    $w_r\equiv   {\tilde   w}_r(r,\bI)$ (given by \Eq{defw123}).   Therefore
 \Eq{defsedelta} reads:
\begin{eqnarray}
{\sigma}^{[\M{n}]e}_\bk({\bf I})\hskip -0.3cm &=& \hskip -0.3cm\int  \frac{\d^{2} {\bw}}{(2\pi)^3}\int \d w_r
 \exp\left(-\imath   \bk  \cdot   \bw  \right)   Y_{\M{m}}[  \bO(\bfI,\bfw)]
 \times \nonumber \\
&&  g_{\MGS{\alpha}}(\bv[\bfI,\bfw])\frac{1}{|\partial {\tilde w}_r/\partial r|^{-1}}
 \delta_{\rm D}(w_r-{\tilde   w}_r[R_{200},\bI])   \,   ,   \EQN{Newinteg}   \\   \hskip -0.3cm &=& \hskip -0.3cm\int
 \frac{\d^{2}   {\bw}}{(2\pi)^3}\exp\left(-\imath  \bk  \cdot   \bw  \right)
 Y_{\M{m}}[             \bO(\bfI,\bfw,{\tilde            w}_r[R_{200},\bI])] \times \nonumber \\
&& \hskip -1.5cm
 g_{\MGS{\alpha}}(\bv[\bfI,\bfw,{\tilde    w}_r(R_{200},\bI)])\frac{\omega_r(\bI)}{
 |{\dot    r}(R_{200},\bI)|}    \exp\left(-\imath    k_r    \cdot    {\tilde
 w}_r[R_{200},\bI]\right)\,. \nonumber
\end{eqnarray}
In \Eq{Newinteg} we  sum over all intersections of the  orbit $\bI$ with the
$R_{200}$  sphere, at the  radial phase  corresponding to  that intersection
with a weight corresponding to $\omega_r/|{\dot r}|$  (see \Fig{enter-orbit}).
Note that \Eq{Newinteg}  involves $\d{}^{2}\bw \equiv \d w_{2} \d w_{3}$.

\subsubsection{ Tidal excitation from beyond $R_{200}$}
\label{s:R200tidal}
The tidal potential is given as a boundary condition on the 
virial sphere and deprojected in volume.
Let us  call $ b'_{\ell m}(t)$  the harmonic coefficients of  the expansion of
the  external potential on  the virial  sphere. We
expand    the   potential   over    the   biorthogonal    basis,   $(u_n^{\ell
m},d_n^{\ell,m})$ (see Appendix~\ref{s:implementation}), so that
\begin{eqnarray}
\psi^e(r,\bO,t) &=&\sum_{n,\ell,m}     b'_{\ell    m}(t)\,    Y_{\ell}^m(\bO)\,
 \left(\frac{r}{R_{200}}\right)^\ell\,,  \nonumber \\  &=&\sum_{\M{n}}     b_{\M{n}     }(t)
 \psi^{[\M{n}]}(\br) \,, \label{e:defexp}
\end{eqnarray}
where $ \psi^{[\M{n}]}(\br)  \equiv Y_{\ell}^m(\bO) u_{j}^{\ell m}(r)$.  The
first equality in \Eq{defexp} corresponds to the inner solution of the three-dimensional potential  whose boundary  condition is given  by $Y^\ell_m(\bO)
b'_{\ell m}$ on  the sphere of radius $R_{200}$  (defined below).  Since the
basis is biorthogonal, it follows that
\begin{equation}
b_{\M{n}       }(t)      =       \left(\int      d_n^{\ell       m}(r)      \,
\left(\frac{r}{R_{200}}\right)^\ell \d r \right) b'_{\ell m}(t)\,.
\end{equation}
It  is  therefore straightforward  to  recover  the  coefficient of  the  3D
external potential from that of the potential on the sphere.

\subsection{Induced correlations in the halo}
\label{s:correlLin}
%
{Our} purpose is to characterize {\sl statistically} the response of the 
dark matter halo to tidal perturbation and infall. 
This is best done by computing the  N-point statistics of the perturbed density field. 
Let us  start with the two-point correlation.
{From Eq.(\ref{e:SOLK}) the } variance-covariance matrix of the 
response is {given by}
\begin{eqnarray}
   \left\langle {\bf \hat a}\cdot {\bf \hat a}^{*\top}  \right\rangle &&
   = \big\langle 
\left[{\bf \hat 
    K} \cdot  {\bf \hat  b}+ {\bf \hat  Q} \cdot {\bf 
\hat c}\right]\cdot  ({\b1} 
-{\bf \hat 
    K} )^{-1 }\cdot 
    \nonumber \\&&
  ({\b1} -{\bf \hat K} )^{-1 *\top} \cdot 
  \left[{\bf \hat K} \cdot  {\bf \hat  b}+ {\bf \hat  Q} \cdot {\bf 
\hat c}\right]^{\top *}  
\big\rangle.
    \label{e:correlA}
\end{eqnarray}

{This expression of the}
$ \bn \times \bn $ matrix,  $\langle\, {\tmmathbf{  \hat  a}}\cdot  {\tmmathbf{ \hat  a}}^{*\top}
\rangle $  involves autocorrelation {terms
like the {components }}  of  $\langle  \, {\tmmathbf{ \hat b}}\cdot
{\tmmathbf{  \hat b}}^{*\top}\rangle  $ (the tidal field)  and $\langle \,  {\tmmathbf{ \hat  c}}
\cdot
{\tmmathbf{ \hat c}}^{*\top} \rangle $ (the source of infall), 
but also  cross-correlation terms such
as the components of  $\langle\, {\tmmathbf{ \hat b}}\cdot {\tmmathbf{ \hat c}}^{*\top} \rangle $.
For a spherical harmonic basis, the induced density perturbation reads
 (see \Eq{harmexp} in Appendix~\ref{s:implementation})
\begin{equation}
  \rho ( r, \MG{ \Omega}, t ) =\sum_\bn a_\bn \rho_\bn(\br)= \sum_{n \ell 
m} a^{n}_{ \ell m} ( t ) Y^m_{\ell} (
  \MG{ \Omega} ) d^n_{\ell m} ( r )\,,
\end{equation}
{The functions} $ d^n_{\ell m} ( r )$ 
{depend on the chosen basis. An example } is given by \Eq{harmexp2}.
{Again,} $\bn$ stands here for $n,\ell,m$, 
respectively the radial and the two angular `quantum numbers'.
As a consequence the two-point correlation function for the perturbed density reads
\begin{eqnarray}
 \hskip -0.3cm   \left\langle \rho ( r, \MG{\Omega} + \Delta \MG{\Omega}, t + 
\Delta t )
  \rho ( r', \MG{  \Omega}, t ) \right\rangle = \hskip -0.25cm \sum_{n \ell  m n' 
\ell' m'}
  Y^m_{\ell} (  \MG{\Omega} )\times  \nonumber \\ 
 Y^{m'*}_{\ell'}  ( \MG{\Omega} 
  + 
\Delta \MG{
  \Omega} ) d^n_{\ell m} ( r ) d^{n'}_{\ell' m'} ( r' ) \langle 
a^{n}_{\ell m}
  ( t ) a^{n' *}_{ \ell' m'} ( t + \Delta t ) \rangle.
\end{eqnarray}
{The}  statistical averages,  $\langle 
a^{n}_{\ell m}( t ) a^{n' *}_{ \ell' m'} ( t + \Delta t ) \rangle$  are given by
the temporal inverse Fourier transform of  \Eq{correlA}. 
If the perturbation is stationary and statistically rotationally 
invariant
\(
  \langle a^{n}_{ \ell m} ( t ) a^{n'}_{ \ell' m'} ( t + \Delta t ) \rangle \equiv
  C^{n n'}_{\ell} ( \Delta t) \delta_m^{m'} \delta_{\ell}^{\ell'}.
\)
{The} correlation function {then} obeys 
\begin{eqnarray}
&& \hskip -0.3cm  \left\langle \rho ( r, \MG{\Omega} + \Delta \MG{\Omega}, t + 
\Delta t ) \rho ( r',
  \MG{\Omega}, t ) \right\rangle =  \nonumber \\  && 
  \sum_{n n' \ell m} P_{\ell} ( 
\cos ( \gamma ))
  d^n_{\ell m} ( r ) d^{n'}_{\ell m} ( r' ) C^{ n 
n'}_{\ell} ( \Delta t )\,, \EQN{corpos}
\end{eqnarray}
where $\gamma$ stands for the angle between $\bO$ and $\bO'$.
Evaluating \Eq{corpos} for $\gamma=0$, $\Delta t=0$, $r=r'$ gives a measure of 
the cosmic variance of the amplitude of the response of the halo 
as a function of radius $r$. The full-width half maximum (FWHM) of 
$ \left\langle \rho ( r, \MG{\Omega} + \Delta \MG{\Omega}, t  ) \rho ( r,
  \MG{\Omega}, t ) \right\rangle$
  is a measure as a function of time, $t$, and radius, $r$,
 of the angular extent of the ensemble average mean  polarization.
 Conversely, the FWHM of $ \left\langle \rho ( r, \MG{\Omega}, t  
 ) \rho ( r+\Delta r,
  \MG{\Omega}, t ) \right\rangle$ is a measure of its radial extent
  in  the direction $\bO$.
Note that \Eq{DFsphere} together with \Ep{SOL} yield 
a description of the 
response  both in position and velocity. For instance,
\Eq{solrhov} allow us to predict the induced correlations amongst streams.
Applications of \Eqs{correlA}{corpos}  (and their non-linear generalization in
\Sec{NLcorrel})
will be discussed in greater details
in \Sec{applications}. The actual implementation of \Eqs{correlA} is carried in a
simplified framework in \Sec{statprogex}.
%
\subsubsection{Link with propagators}
\label{s:propagM}
%
Let us  emphasize  that the splitting of the gravitational
field into  two components,  one {originating} 
outside of  $R_{200}$, and one {from} the  inside, via
point particles obeying  the distribution $s^{e}(\M{r},\M{v},t)$ is somewhat  {\it ad hoc}
 from the  point of  view  of the  linear dynamics. 
 It is convenient from the point of view of the measurements, and 
 crucial for the non-linear evolution (described below), 
 or the ensemble average, as shown above \footnote{
 One  should
account for the fact that $\psi_e$ should be  switched on long before any particles enter
$R_{200}$ since no particle is created at the boundary. }.
{It}  allows us to specify the statistical characteristics of the infall
without having to refer to the properties of the object on  which
{this} infall occurs. 

We discuss in appendix~\ref{s:propag} the formulation of the 
response of a self-gravitating sphere {in terms of} a propagator (\ie 
the Green function of the collisionless Boltzmann-Poisson 
equation).  \onecolumn
This formulation is mathematically equivalent to the approach 
described above, but there we
 relied  on Gauss's theorem to reproject all the
  information beyond $R_{200}$  back onto the virial sphere.
  This information involves two contributions: one
   relative to particles beyond $R_{200}$, which contribute to the tidal field,
    the other {relative}
     to particles entering $R_{200}$ which contribute to $s^{e}(\bO,\bv)$)

  The main asset of this formulation is to localize the boundary, 
 which is possible since the interaction is purely
  gravitational, at the expense of having {\sl two} sources of different nature. 
In particular, this implies that the environment may be 
characterized once and for all, independently of the detailed 
nature of the inner halo.

\vskip 0.4cm
In this section, we assumed that the polarisation of the halo 
was {\sl linear}.  This hopefully provided some insight for some
aspects of the dynamics, but
 effectively ignores non-linear phenomena such as 
dynamical friction or tidal stripping. 
Let us now expand perturbation theory to higher orders.

%
\section{Non-linear perturbative response }
\label{s:nonlinear}
%
In the following, we describe, {using}  perturbation theory, the non
linear response of the halo to {material entering}  
at the virial sphere. It is assumed that
the perturbation is first-order in the hierarchy, and that the halo is
dynamically stable. This should warrant the validity of the expansion. We use
the angle-action variables of the {\sl unperturbed} system as canonical variables and
investigate the non-linear evolution of the infall and the tidal excitation.\parn
In essence, the key is to expand the potential onto the biorthogonal
potential density basis which allows us to decouple position-velocity and time
({i.e. perform a separation of variable}), and solve in turns each order of the perturbation
expansion.
%
\subsection{Perturbative expansion}
\label{s:perturb}
%
Recall that the dynamical equation of an open system characterized by its distribution 
function, $ \mathfrak{F}$  is given by \Eq{boltzpert1}.
 Let us expand again {$F$ as}
 \begin{equation}
\mathfrak{F} = F + \sum_\rn \varepsilon^\rn f^{(\rn )} \,,\quad
  {\rm and} \quad  \mathfrak{H} = H_0 + \sum_\rn \varepsilon^\rn \psi^{( \rn
)}\,,\end{equation}
 where the unperturbed equilibrium is  characterized by the
distribution function, $F (\bI)$. Note that $(\rn)$, the order of the expansion 
should not be confused with 
$\bn \triangleq (n,\ell,m)$. 
Finally, it is assumed that the {external} perturbation 
{enters as a} first-order only, \ie
$s^e \propto \varepsilon $ and $\psi^e \propto \varepsilon $.
In short, the rewriting of \Eq{boltzpert1} to order $\varepsilon^\rn$ yields:
\begin{eqnarray}
 \frac{\partial f^{( \rn )}_{\bk}}{\partial t}+ 
   \imath\bk\cdot
  \MG{\omega}  f^{( \rn )}_{\bk}
  =\Bigg(\frac{\partial F}{\partial
  \bI} \cdot\imath\bk[ \hspace{0.25em} \psi^{( \rn
  )}_{\bk} + \delta_\rn^1 \psi_{\bk}^e ] - 
   \sum^{\rn - 1}_{k =
  1} \{ \psi^{( k )}, f^{( \rn - k )} \}_{\bk}  + \delta_\rn^1
  s_{\tmmathbf{\bk}}^e \Bigg) . \EQN{eqnn}
\end{eqnarray}
In the following, we  solve \Eq{eqnn} recursively, order by order,  to recover
the perturbative response of the halo to {the} tidal interaction and infall.
We expand both the potentials and the source term over a
biorthogonal basis, so that, with $^{( \rn )}$ referring 
to the order in the hierarchy and
$^{[ \bp ]}$ to the label in the basis
\begin{equation}
 \psi_{\bk}^{( \rn )} (\bI, t ) = \sum_\bp a_\bp^{( \rn )} ( t ) \psi_{\bk}^{ [ \bp ]}
   (\bI) \hspace{0.25em}, \quad
    \psi_{\bk}^e (\bI, t
   ) = \sum_\bp b_\bp ( t ) \psi_{\bk}^{ [ \bp ]} (\bI) \hspace{0.25em},
   \quad \hspace{2em} s_{\bk}^e (\bI, t ) = \sum_\bp c_\bp ( t ) \sigma_{\bk}^{
   [ \bp ]} (\bI) \hspace{0.25em} .  \EQN{expandnBasis}
    \end{equation}
Recall also that the superscript, $^{[ \bp ]}$, in \Eq{expandnBasis} spans discretely a 3D or 5D space
    depending on the type of function basis.}
The first-order solution for $a_\bp$ was given \Eq{ap2}. Let us turn
to the higher order equations. 
\subsubsection{Second-order perturbation theory} 
\label{s:2nd}
%
The second-order equation {for $a^{( 2 )}_\bp$}  reads
\begin{eqnarray}
  a^{( 2 )}_\bp ( t ) &=& (2\pi)^3 \sum_\bn \int_{-\infty}^{t} \d \tau \hspace{0.25em} a^{( 2 )}_\bn ( \tau
  ) \left( \sum_{\bk} \int \d\bI \psi^{[ \bn ]}_{\bk}
  (\bI) \psi^{[ \bp ]*}_{\bk} (\bI) \hspace{0.25em}
  \frac{\partial F}{\partial \bI} \cdot\imath\bk \exp (
 \imath\bk \cdot \MG{\omega} [ \tau - t ] ) \right) + \nonumber \\
  && (2\pi)^3 \int_{-\infty}^{t} \d \tau \sum_{\bk} \int \d\bI \exp (\imath\bk
  \cdot \MG{\omega} [ \tau - t ] ) \left\{ f^{( 1 )} (
  \tau,\tmmathbf{w},\bI), \psi^{( 1 )} (
  \tau,\tmmathbf{w},\bI) \right\}_{\bk} \psi^{[ \bp
  ]*}_{\bk} (\bI) \hspace{0.25em}   ,\EQN{defA2}
\end{eqnarray}
where $\{ f^{( 1 )}, \psi^{( 1 )} \}$ is the Poisson bracket of the
perturbation to first-order. 
Now for a set $( f, \psi )$ we have
\begin{equation}
  \left\{ f, \psi \right\}_{\bk} = \int \d\tmmathbf{w} \exp (
  -\imath\bk \cdot \tmmathbf{w}) \left\{ \sum_{\bk_1}
  f_{\bk_1} (\bI) \exp ( \imath \bk_1 \cdot
  \tmmathbf{w}), \sum_{\bk_2} \psi_{\bk_2} (\bI) \exp ( \imath
  \bk_2 \cdot \tmmathbf{w}) \right\} \hspace{0.25em} .
\end{equation}
Therefore
\begin{equation}
  \left\{ f, \psi \right\}_{\bk} = \sum_{\bk_1 +
  \bk_2 =\bk} \left( \psi_{\bk_2}
  \frac{\partial f_{\bk_1}}{\!\! \!\partial \bI} \cdot
\imath   \bk_2 \hspace{0.25em} - f_{\bk_1} \frac{\partial
  \psi_{\bk_2}}{\!\!\! \partial \bI} \cdot \imath  \bk_1
  \hspace{0.25em} \right) \equiv \sum_{\bk_1 + \bk_2
  =\bk} \left\llbracket f_{\bk_1}, \psi_{\bk_2}
  \right\rrbracket_{} \hspace{0.25em} , 
\end{equation}
where the sum is over $\bk_{1}$ with $\bk_{2}=\bk-\bk_{1}$.
Given \Eq{DFsphere},
\Eq{defA2} may be rearranged as 
\begin{eqnarray}
  a^{( 2 )}_\bp ( t ) &=&(2\pi)^{3} \sum_{\M{q}_1} \int_{-\infty}^t \d \tau_1 \hspace{0.25em} a^{( 2
  )}_{\M{q}_1} ( \tau_1 ) \left( \sum_\bk \int \d\bI \psi^{[ \M{q}_1 ]}_\bk
  (\bI) \psi^{[ \bp ] *}_\bk (\bI) \hspace{0.25em} \frac{\partial
  F}{\partial \bI} \cdot\imath\bk \exp (\imath\bk \cdot
  \MG{\omega} [ \tau_1 - t ] ) \right)
 + 
 \nonumber   \\ &&
 (2\pi)^{3}  \int^t \d \tau_1 \int^{\tau_1} \d \tau_2 \left( \sum_\bk \int \d\bI
  \exp (\imath\bk \cdot \MG{\omega} [ \tau_1 - t ] ) \sum_{\M{q}_1, \M{q}_2}
  [ a^{( 1 )}_{\M{q}_1} ( \tau_1 ) + b_{\M{q}_1} ( \tau_1 ) ] \times  
  \nonumber  \right.\\ && \left.
\sum_{\bk_1 + \bk_2 = \bk } \left\llbracket \exp
  (\imath \bk_1 \cdot \MG{\omega} [ \tau_2 - \tau_1 ] ) \left[
  \frac{\partial F}{\partial \bI} \cdot \imath \bk_1
  \hspace{0.25em} [ a^{( 1 )}_{\M{q}_2} ( \tau_2 ) + \hspace{0.25em} b_{\M{q}_2} (
  \tau_2 ) ] \psi^{[ \M{q}_2 ]}_{\bk_1} (\bI) + c_{\M{q}_2} [ \tau_2
  ] \sigma^{e, [ \M{q}_2 ]}_{\bk_1} (\bI) \right],
  \psi_{_{\bk_2}}^{[ \M{q}_1 ]} \right\rrbracket \psi^{[ \bp
  ]*}_{\bk} \right) \,.\label{e:defa22}
\end{eqnarray}
{Note that the r.h.s. of Eq.(\ref{e:defa22}) is linear
in $\ba^{( 2 )}$ while it is quadratic in  $\ba^{(1)}$, $\bb^{(1)}$, $\bc^{(1)}$,
involving products such as $\ba \, \ba$, $\ba \,\bc$, $\bb\,\bc$ and so on.
More generally, the perturbation theory at order $(n)$ is linear in $a^{(n)}$.} 
Note {also}
that \Eq{defa22} involves a double ordered time integral over $\tau_1$ and $\tau_2$
of the source coefficient, $c_{\M{q}_1}(\tau_1)$ and $c_{\M{q}_2}(\tau_2)$,
which accounts for the  fact that, non-linearly, the 
relative phase of the accretion {events} matter 
(\Eq{defa22Complex} gives the analog to \Eq{defa22} in the complex frequency plane).
Eq~(\ref{e:defa22}) includes in particular {a term like}
\begin{eqnarray}  
   \exp ( \imath
   ( \bk_1 + \bk_2 ) \cdot \MG{\omega} [ \tau_1 - t ] )
   \exp ( \imath \bk_1 \cdot \MG{\omega} [ \tau_2 - \tau_1 ] ) \psi^{[\bp
   ]*}_{\bk_1 + \bk_2} 
   \sum_{\M{q}_1, \M{q}_2} \left( \frac{\partial \sigma^{e, [ \M{q}_2
   ]}_{\bk_1}}{\partial \bI} \psi^{[ \M{q}_1 ]}_{\bk_2}
   - \frac{\partial \psi^{[ \M{q}_1 ]}_{\bk_2}}{\partial \bI}
   \sigma^{e, [ \M{q}_2 ]}_{\bk_1} \right) \left( a^{( 1 )}_{\M{q}_1} (
   \tau_1 ) + b_{q_{_1}} ( \tau_1 ) \right) c_{q_{_2}} ( \tau_2 ) 
\end{eqnarray}
which involves the rate of change of the source term with respect to action
variation (via $\partial \sigma^{e, [ \M{q}_2 ]}_{\bk_1} / \partial
\bI$) modulated twice over time as $\exp ( \imath( \bk_1 +
\bk_2 ) \cdot \MG{\omega} [ \tau_1 - t ] ) \exp ( \imath
\bk_1 \cdot \MG{\omega} [ \tau_2 - \tau_1 ] )$.

{The second order solution can be synthetically written
by introducing  tensors $\bK_2$ and $\bQ_2$ similar to those 
defined in \Eq{defkk} and \Eq{defhh} to express the 
first order solution as \Eq{SOL}. These latter tensors will
now be referred to as $\bK_1$ and $\bQ_1$. Specifically,
the components of these tensors are defined as:}
\begin{equation}
  ( \bK_1 )_{\bp, \M{q}_1} [ \tau_1 - t ]\equiv (\bK )_{\bp, \M{q}_1} [ \tau_1 - t ]  = (2\pi)^3\sum_\bk  \int \d\bI \exp (
 \imath\bk  \cdot \MG{\omega} [ \tau_1 - t ] ) \psi^{[ \bp ]*}_\bk 
  \hspace{0.25em} \psi^{[ \M{q}_1 ]}_\bk  \hspace{0.25em} \frac{\partial
  F}{\partial \bI} \cdot\imath\bk  \hspace{0.25em}, \label{e:defK1}
\end{equation}
\begin{equation}
  ( \bK_2 )_{\bp, \M{q}_1, \M{q}_2} [ \tau_1 - t, \tau_2 - \tau_1 ] = (2\pi)^3\sum_\bk  \int
  \d\bI \exp (\imath\bk  \cdot \MG{\omega} [ \tau_1 - t ] )
  \sum_{\bk_1 + \bk_2 =\bk} \left\llbracket \exp (
  \imath \bk_1\cdot \MG{\omega} [ \tau_2 - \tau_1 ] ) \frac{\partial
  F}{\partial \bI} \cdot \imath \bk_1 \hspace{0.25em} \psi^{[
  \M{q}_{_2} ]}_{\bk_1}, \psi_{\bk_2}^{[ \M{q}_1 ]}
  \right\rrbracket \psi^{[ \bp ]*}_{\bk} \hspace{0.25em}, \label{e:defK2}
\end{equation}
while $\bQ_i$ involves replacing $\psi^{[\M{q} ]}_{\bk} \hspace{0.25em}
\partial F / \partial \bI \cdot \bk$ by
$\sigma_{\tmmathbf{m
}}^{e, [\M{q} ]}$.  
For instance, 
\[
  ( \bQ_2 )_{\bp, \M{q}_1, \M{q}_2} [ \tau_1 - t, \tau_2 - \tau_1 ] =(2\pi)^3 \sum_\bk  \int
  \d\bI \exp (\imath\bk  \cdot \MG{\omega} [ \tau_1 - t ] )
  \sum_{\bk_1 + \bk_2 =\bk} \left\llbracket \exp (
 \imath \bk_1 \cdot \MG{\omega} [ \tau_2 - \tau_1 ] ) \sigma^{e,[
  \M{q}_{_2} ]}_{\bk_1}, \psi_{\bk_2}^{[ \M{q}_1 ]}
  \right\rrbracket \psi^{[ \bp ]*}_{\bk} \hspace{0.25em}, \label{e:defQ2}
\]
This implies in particular that $\bQ_1 \equiv \bQ$ given by \Eq{defhh}.
Note that each component of $\bK_2$ has the
same complexity as $\bK_1$, i.e. the perturbation theory is linear order by
order; on the other hand it involves \textsl{all} the  couplings in
configuration space, hence the double sum in
$\bk$.
With these definitions, Eqs. (\ref{e:ap2}) and (\ref{e:defa22}) read formally
\begin{equation}
  \ba^{( 1 )} = \bK_1 \cdot [ \ba^{( 1 )}_1 + \bb ] + \bQ_1 \cdot \bc
  \hspace{0.25em},   \EQN{defdiag1}
\end{equation}
\begin{equation}
  \ba^{( 2 )} = \bK_1 \cdot \ba^{( 2 )} + \bK_2 \cdot [ \ba^{( 1 )} + \bb ] \otimes
  [ \ba^{( 1 )} + \bb ] + \bQ_2 \cdot [ \ba^{( 1 )} + \bb ] \otimes \bc \,.\EQN{sol2}
  \hspace{0.25em} \EQN{defdiag2}
\end{equation} 
{where the dot operator  is not merely a tensor
contraction, but also involves a time convolution. For example,
$\M{Z}$ being a given field:}
\begin{equation}
  ( \bK_1 \cdot \M{Z} )_\bp ( t ) \equiv \sum_\M{q} \int_{- \infty}^t \d \tau {(K_{1})}_{ \bp, \bq}
  ( \tau - t ) Z_\M{q} ( \tau ) \hspace{0.25em}, \EQN{defcontract1}
\end{equation}
{and similarly the} higher order contraction rule 
over the fields $\M{Z}^{1} \otimes \cdots  \otimes \M{Z}^{\rn}$ {is defined as:}
\begin{equation}
  ( \bK_\rn \cdot \M{Z}^1 \otimes \cdots \otimes \M{Z}^\rn )_\M{p} ( t ) \equiv \sum_{\M{q}_1,
  \cdots \M{q}_\rn} \int^t \d \tau_1 \cdots \int^{\tau_{\rn - 1}} \d \tau_\rn ( K_\rn
  )_{\bp, \M{q}_1, \cdots \M{q}_n} ( \tau_1 - t, \cdots, \tau_\rn - \tau_{\rn - 1} )
  Z^1_{\M{q}_1} ( \tau_1 ) \cdots Z^\rn_{\M{q}_\rn} ( \tau_\rn ) \hspace{0.25em}.
  \EQN{defcontractn}
\end{equation}
Note that the order of the argument does matter. (\ie
\Eq{defcontractn}  defines a  non-commutative algebra).
Note also that the sum of the order in each term corresponds 
to the order of the perturbation. 
For instance, in \Eq{defdiag2}, the second term involves
the product of two first-order terms, while the first term is a single 
second-order term. 
Note finally that the contraction for the $\bQ_{\rm n}$ involve a summation
over 5 indices, $\ell,m,\alpha,\ell',m'$,  (whereas contraction over  $\bK_{\rm n}$ involves only 3 indices: $n,\ell,m$).
We illustrate and discuss in \Fig{diag1} through synthetic diagrams 
the corresponding expansion. (See also  \Fig{diag2}  in Appendix~\ref{s:apendperturb}
for an expansion to higher order).
\twocolumn
In appendix~\ref{s:apendperturb}, we show in \Eqs{soln}{defKn} how to 
rewrite \Eq{sol2} to order $\rn$.
\parn
As for all expansion schemes, the issue of the truncation arises. 
Depending on the physical process investigated, the truncation order may vary.
{ For instance,  
it may be legitimate  to truncate the perturbation to second-order since the
second-order is the first-order for which dynamical
friction is taken into account.}

\subsection{Non-linear two-point correlation functions}
\label{s:NLcorrel}
Let us now re-address the computation of the two-point correlation function 
(\cf \Sec{correlLin}) of the 
response of the halo to tidal excitation and infall while accounting for the 
non-linearities described in \Sec{2nd}.
First,  let us
reshuffle the hierarchy in a format which is best suited for the 
statistical average of  the non-linear response.
\subsubsection{Reordering in $\bb$ and $\bc$ }

Let us define 
 ${\cal F }( \omega, t ) \equiv \exp ( \imath \omega t )$ the Fourier operator, so that ${\cal F }\cdot \M{Z}$
 and ${\cal F }^{\top} \cdot \M{Z} $ are respectively the half-Fourier and inverse half-Fourier transform
 of their argument $\M{Z}$.
Calling
\begin{equation}
  \bR_1 \equiv{\cal F }^{\top} \cdot ( \b1 - \hat {\bK}_1 )^{- 1} \cdot {\cal F } \hspace{0.25em},
\EQN{defR1}
\end{equation}
\Eq{sol2} (and its generalization \Eq{soln}) reads like a recursion:
\begin{equation}
  \ba^{( \rn )} \equiv \bR_1 \cdot \M{\cal K} [ \ba^{( \rn - 1 )} \cdots, \ba^{( 1 )}, \bb, \bc ]
  \hspace{0.25em}, \hspace{2em} {\rm for} \quad \rn \geqslant 2 \hspace{0.25em},\label{e:recur}
\end{equation}
where $\M{\cal K} $ stands formally for some combination of 
$\bK_{\rm n}$ and $\bQ_{\rm n}$.
Note that $\bK_1$ accounts for the self-gravity of the halo. If the halo is very
hot, this self-gravity may be neglected altogether and $\bR_1 \rightarrow \b1$.
If not, we may define $\bK'_i \equiv \bR_1 \cdot \bK_i$, $\bQ'_i \equiv \bR_1 \cdot
\bQ_i$, and rewrite the recursive relations \Eq{recur} with $\bK_1 \equiv
0.$
For instance:
\begin{equation}
  \ba^{( 1 )} = \bR_1 \cdot \left( \bK_1 \cdot \bb + \bQ_1 \cdot \bc \right) =
  \bK'_1 \cdot \bb + \bQ'_1 \cdot \bc 
  \hspace{0.25em},
  \EQN{reorder1}
\end{equation}
which we can rearrange as:
\begin{equation}
\ba^{( 1 )} \equiv A_b \cdot \bb + A_c \cdot \bc
\end{equation}
where $A_b \equiv \bK'_1$ and $A_c \equiv \bQ'_1$. 
Let us also introduce $\bK''_1
= \bK'_1 + \b1$.
{ Similarly, the contribution of $b$'s and $c$'s 
to the second order term for $a$ can be expressed as:}
\begin{equation}
  \ba^{( 2 )} \equiv A_{bb} \cdot \bb \otimes \bb + A_{cc} \cdot \bc \otimes \bc +
  A_{cb} \cdot \bc \otimes \bb + A_{bc} \cdot \bb \otimes \bc \,. \EQN{reorder2}
\end{equation}
where
\begin{eqnarray}
 A_{bb} \!\!\! &=&  \!\!\!  \bK'_2 \circ \bK''_1\,,
\quad 
 A_{cc} =  \bK'_2 \circ
   \bQ'_1 + \bQ'_2 \circ [ \bQ'_1, \bI ] \,, 
  \\
     A_{cb} \!\!\!  &=&  \!\!\! \bK'_2 \circ [
   \bQ'_1, \bK''_1 ]\,,  \,\,\,\,\, A_{bc} = K'_2 \circ [ \bK''_1, \bQ'_1 ] + \bQ'_2
   \circ [ \bK''_1, \bI ]   \nonumber
   \,. \EQN{defAbb}
   \end{eqnarray}
   Here the bracket, $[\,\,,\,\,]$ accounts for the differential composition, 
   so that,
\begin{eqnarray}
 A_{cb}\cdot \bb \otimes \bb &=& \bK'_2\cdot ( \bQ'_1\cdot \bb  ) \otimes 
  (  \bK''_1 \cdot \bb) \nonumber \\
  &=& \bR_1\cdot \bK_2\cdot ( \bR_1\cdot \bQ_1\cdot \bb  ) \otimes 
  ( [\b1 + \bR_1] \cdot \bK_1 \cdot \bb) . \nonumber 
  \end{eqnarray}
   In appendix~\ref{s:reorderannexe}, we also show 
how to 
{write an equation similar to \Eq{reorder2} for the third order
contribution and more generally for an arbitrary order (see Eq.\ref{e:defapn}).}

\subsubsection{Non-linear correlators}
We may now complete the calculation of, say, 
the two-point correlation function of the density,
$ C^{\rho}_2$ :
\begin{eqnarray}
  C^{\rho}_2 \equiv \langle \rho ( x_1 ) \rho ( x_2 ) \rangle = 
   \sum_\rn \sum^\rn_{p = 1} \varepsilon^{\rn} \langle \rho^{( p )} (
  x_1 ) \rho^{( \rn - p )} ( x_2 ) \rangle  \,,
\end{eqnarray}
where $x_i = ( \br_i, \tau_i )$, i=1,2. 
Following \Eq{expandnBasis}, let us also expand  the response in density, $\rho$, over  the basis function
$\{ \rho^{[\M{q} ]} ( \br ) \}_\M{q}$,
so that 
\begin{eqnarray}
  C^{\rho}_2 = \sum_\rn \varepsilon^{\rn} \sum^\rn_{p = 1} \sum_{\M{q}_1, \M{q}_2} \rho^{[
  \M{q}_1 ]} ( \br_1 ) \rho^{[ \M{q}_2 ]} ( \br_2 ) \langle a_{\M{q}_1}^{( p )} ( \tau_1 )
  a_{\M{q}_2}^{( \rn - p )} ( \tau_2 ) \rangle  \,. \nonumber
\end{eqnarray}
Now, given \Eq{reorder1} and \Ep{reorder2}, we may rearrange this equation
as:
\begin{eqnarray}
  C^{\rho}_2\!\!\! &=&\!\!\!  \varepsilon^2 \sum_{\M{q}_1, \M{q}_2} \rho^{[ \M{q}_1 ]} ( \br_1 ) \rho^{[ \M{q}_2 ]} ( \br_2 )
  \left[  C_{2}^{\{2\}}  +  \varepsilon C_{2}^{\{3\}} 
   + \ldots  \right]\,. \EQN{expC2}
\end{eqnarray}
where {$C_{2}^{\{2\}}$ is a simple reshuffling of  \Eq{correlA}, i.e:}
\begin{eqnarray}
C_{2}^{\{2\}}  \hskip -0.3cm &=& \hskip -0.3cm A_b \!\!\times\!\! A_b \cdot \langle \bb \otimes \bb
  \rangle + A_c \!\!\times\!\! A_c \cdot \langle \bc \otimes \bc \rangle + A_b \!\!\times\!\! A_c
  \cdot \langle \bb \otimes \bc \rangle + \nonumber \hskip -0.5cm  \\ && 
  A_c \!\!\times\!\! A_b \cdot \langle \bc \otimes \bb
  \rangle \,, \label{e:defC2e2}
\end{eqnarray}
and:
\begin{eqnarray}
  C_{2}^{\{3\}} &=& ( A_{bb} \!\!\times\!\! A_b + A_b \!\!\times\!\!
  A_{\op{bb}} ) \cdot \langle \bb \otimes \bb \otimes \bb \rangle + \nonumber \\ && \hskip -0.cm
   ( A_{cc} \!\!\times\!\!
  A_c + A_c \!\!\times\!\! A_{\op{cc}} ) \cdot \langle \bc \otimes \bc \otimes \bc \rangle +\nonumber \\ && \hskip -0.cm
( A_{bc} \!\!\times\!\! A_c + A_b \!\!\times\!\! A_{\op{bc}} ) \cdot \langle \bb \otimes \bc
  \otimes \bc \rangle +  \nonumber \\ &&  \hskip -0.cm
  ( A_{bb} \!\!\times\!\! A_c + A_b \!\!\times\!\! A_{\op{bc}} ) \cdot
  \langle \bb \otimes \bb \otimes \bc \rangle + \nonumber \\ && \hskip -0.cm
  ( A_{cb} \!\!\times\!\! A_b + A_c \!\!\times\!\! A_{bb} ) \cdot \langle
  \bc \otimes \bb \otimes \bb \rangle \,. \EQN{defC2e3}
\end{eqnarray}
{The} $\times$  operator is  non-commutative and guaranties  that the 
order is preserved in the dot contraction.
Recall that $A_b \equiv \bK'_1$ {\rm and } $ A_c \equiv \bQ'_1\,$, 
while $ A_{\op{bb}}, A_{\op{cc}} , A_{\op{cb}} $ and $ A_{\op{bc}} $ 
 are given by \Eq{defAbb} (or in terms of the underlying distribution function, 
 $F_{0}(\bI)$, and the basis function, $\psi^{[\bn]}(\br)$
 via Eqs (\ref{e:defK1}),  (\ref{e:defK2}) and \Ep{defR1}
 through the definitions of $\bK_1, \bQ_{1}, \bK_{2}$ and $\bQ_2$).
It follows from \Eq{defC2e3} that the non-linear two-point correlation will involve at least the
three-point correlation of the incoming flux {and of the external potential}. We will see in \Sec{applications}
that this is a generic consequence of mode coupling.
Now the three point correlation of the incoming flux, $\bc$, and the tidal field, $\bb$ may be reexpressed 
in terms of the mean and the two-point correlations of those fields while relying on 
 Wick's theorem, since we showed in \citet{aubert1} that these fields were approximately
 Gaussian. 
Appendix~\ref{s:npointcorrel}  presents formally the generalization of \Eqs{defC2e2}{defC2e3} 
for the N-point correlation function to arbitrary order.

Equations such as \Eq{defdiag2} or its reordered version \Eq{reorder2} 
might look deceivingly simple. One should nevertheless keep in mind
that the perturbation theory involves an exponentially growing number of 
terms.
This is probably best realized by looking at diagrams  such as \Fig{diag11} 
(presented in the appendix)
while keeping in mind that {\sl each} straight line represents a triple sum over $\bk=(k_{1},k_{2},k_{3})$
and a time integral (see also Appendix~\ref{s:implementation}). 
{
The prospect of achieving 
resummation  (in the spirit of what was achieved by \eg \citet{francis} 
for the gravitational instability of the large-scale structures) given the relative complexity of the double source expansion 
is slim.  Yet it might be possible to  construct  scaling rules 
(see  \citet{Fry})
since
 gravity is also here the driving force.
 Let us stress once again
  that the perturbative expansion accounts explicitly,
   within its convergence radius, for all aspects 
 of the  non-linear physics taking place within the $R_{200}$ sphere.
 }

\subsection{Implication for dynamical friction and tidal stripping}
\label{s:dynfric}

One of the  possible assets of this perturbative
formulation is that the incoming flux may describe
a virialized object which has a finite extent, and as such will undergo internal phase mixing reflecting the fact that different points in the object will  describe 
different orbits, at different frequencies (see \Fig{bundle}). 
In the perturbative regime, dynamical friction will  also account for 
both the overall drag of the object, but also its tidal stripping (\ie the fact
that the less bound component of the object will undergo a differential more efficient friction).  Specifically, the deflection of perturbed trajectories will correctly
describe the balance (or lack thereof) between the self-gravity of the entering
flow and its  tendency to be torn by  the differential gravitational field of the halo (which
imposes the unperturbed different orbital trajectories).
{As such, the flow paradigm implemented in this paper and in
\citet{aubert0}, \citet{aubert1} should allow for the 
appropriate level of flexibility in defining what a structure is and how 
time-dependent the concept is, within  the self-gravitating halo (see also 
\Sec{satcount}).   
}

Let us briefly
 discuss how to identify substructures within the halo.

\subsubsection{Substructure counts and distribution} 
\label{s:satcount}

\begin{figure} 
\centering
\resizebox{0.95\columnwidth}{0.8\columnwidth}{\includegraphics{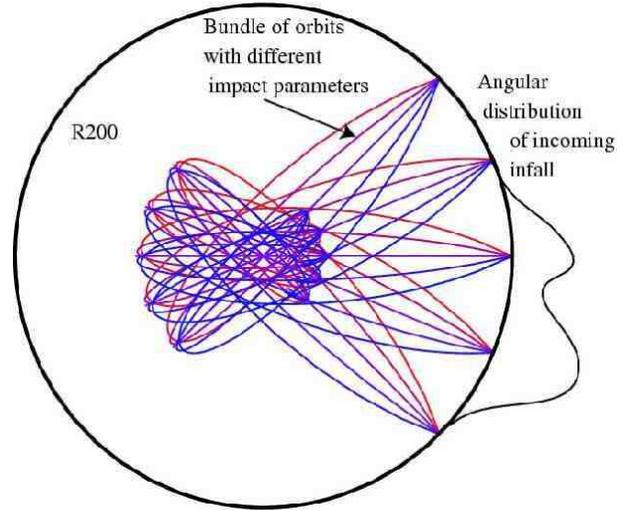}}
\caption{  displays qualitatively a bundle of orbits (in their orbital plane) which undergo dynamical friction 
and phase mixing within the $R_{200}$ radius. As expected, dark matter describing
 orbits which initially 
are at the same position, but with ``slightly'' different initial impact parameters will
end up in quite different regions at later time. 
On the right the curve represents a possible angular distribution of  a given 
entering object (for which the kinematic and angular spread has been greatly exaggerated).
The caustics corresponding to the successive
 rebound of the orbits is clearly visible here (\citet{Fillmore}).
 Note that  the amplitude of the friction
force was {\it ad hoc}, and the self-gravity within the bundle was {\sl not} taken into account.
}
\label{f:bundle} 
\end{figure}

The identification of substructures within a given halo is a very promising
 but difficult topic (see e.g. \citet{SUBFIND},\citet{MPLAMP} \citet{aubert0})  . Once the boundary flow has been propagated inwards, we 
 have in principle access to the full distribution function of the perturbation as a
 function of time. When the field $f(\bv,\br,t)$ is known inside $R_{200}$, we may attempt to identify collapsed objects and apply some form of count
 in cell statistics in order to characterize their spatial distribution
 as a function of time. 
 This would allow us in particular to put aside objects which
  have been disrupted by tidal stripping or phase mixing (indeed 
  10 \% of the mass of the halo is believed to remain in the form of 
  virialized objects, while 90 \% is  disrupted by the tidal field).  
  Recall that the disruption process  is in principle well described by the perturbative expansion.

 The criterion for the detection of objects must be carried while accounting
  for both the density contrast and the corresponding velocities (see also \citet{Arad}). 
  Indeed, we do not wish to identify as objects  local overdensities which may just correspond to caustics or
   local wave reinforcement.
   Here we are
  interested in the temporal coherence of objects.

For this purpose, we may
 coarse-grain the perturbed distribution function both in position and 
velocity, with some given smoothing function, $W(\br/{\cal R}_{\rm s},\bv/{\cal V}_{\rm s})$, and then apply some
thresholding ($W \circ f > f_{\rm min}$ where $\circ$ stands for convolution) on the amplitude of the distribution function, defining 
a set of connex regions. For each of these regions, we may then  
compute the energy of the corresponding clump. If it is negative,
the clump will be labeled as bound  for the corresponding threshold ($f_{\rm min}$), 
and coarse-graining parameters (${\cal R}_{\rm s},{\cal V}_{\rm s}$).  Note that since the response only involves 
the {\sl perturbed} density, one need not subtract the mean potential,
(which is quite a difficult task in general).

Once the bound regions are identified,    we may compute the corresponding
mass and assign it to the bottom of the local potential well.
This procedure may be applied for a range of threshold values, and standard
statistical tools for discrete sources but in spherical geometry. 
We may in particular construct in this manner the mass function of satellites as a function of radius, or, say the two-point correlation function versus mass and cosmic time.  Both issues are subjects of strong discussions when
addressed through standard N-body simulations.

 This time-dependent  identification of virialized objects
 is useful because of biasing, \ie the fact that most observational
    tracers will only be sensitive to the more massive tail of the mass function of virialized objects.

Conversely, we may want to label regions which match the thresholding 
but not the requirement on binding energy, \ie identify caustics, cusps, and 
shells (\citet{Fillmore}).   We may then characterize statistically the 
mean distance between the apoapses (see \Fig{bundle}), which will in general 
depend on ${\cal R}_{\rm s}$ ,${\cal V}_{\rm s}$ and $f_{\rm min}$ but also
on the underlying equilibrium, via $F(\bI)$ and 
on the statistical properties of $\bc$ through, say the distribution of impact
parameters.  
Note in closing that the competing effects of phase mixing, tidal stripping and
dynamical friction all assume that the underlying basis function 
reaches   sufficiently high spatial frequencies to resolve these phenomena.
In practice, since the projection of the response (both linearly and non linearly)
is achieved over a basis which has a truncation frequency, $\ell_{\rm max}$,
there is a finite time scale, $T_{\rm max}\propto \ell_{\max}/\langle\omega \rangle$ above which phase mixing
would induce winding at unresolved scales (here $\langle\omega \rangle$
represents the typical frequency of the dark matter in that region). 
Since the dynamical time is shorter in the inner region of the galaxy, 
such a threshold is going to be reached there first. 
Beyond this critical time, the dynamics is inaccurately modeled  
for the corresponding clump. 
This issue will be important for the non linear coupling of 
clumps, since substructures entering the halo at later times
will be dragged by streamers which are beyond  the accuracy 
threshold\footnote{these limitations are also clearly encountered  in classical N-body simulations}.

\section{Quasi-linear evolution of halo profile}
\label{s:quasi1}

\begin{figure*} 
\centering
\resizebox{1.95\columnwidth}{0.75\columnwidth}{\includegraphics{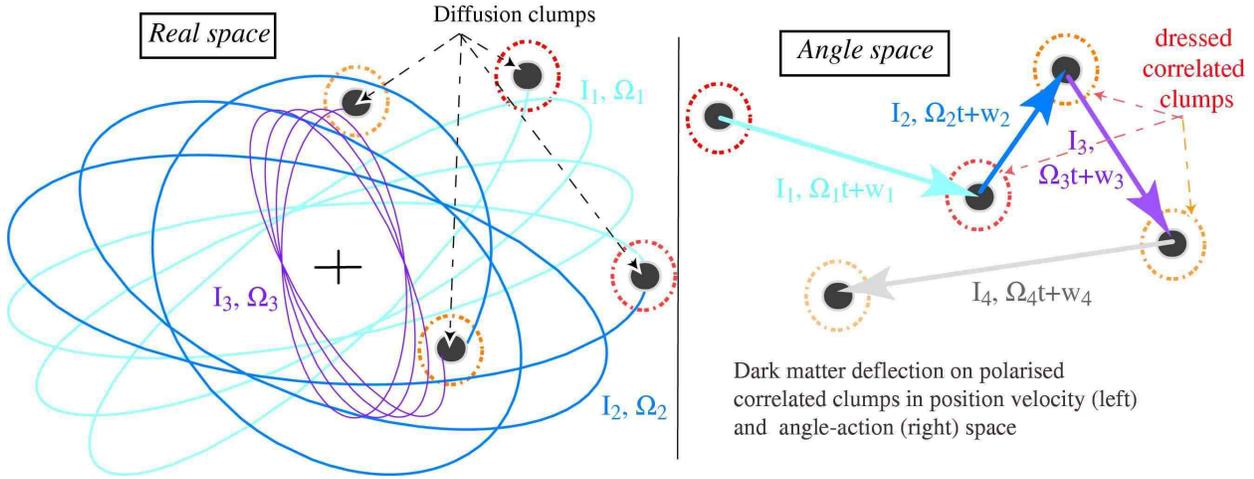}}
\caption{ 
{\sl Left panel} schematic representation of the successive deflection 
of a given orbit on correlated clumps within the halo in position space. Each clump is 
represented with its polarisation cloud. The gray scale coding in the cloud reflects
their spatial and temporal correlation within the clumps. 
During the deflections, the orbital parameters
change (though the individual change is here grossly exaggerated). 
 {\sl Right panel:} the same orbit as viewed in angle-action variables.
The dynamics in these variable is straightforward (it corresponds to straight lines 
 obeying $\bw=\bo(\bI) t+\bw_{0}$) and the diffusion process
 resembles Brownian motion obeying a Langevin equation the particle
receiving a random kick at each deflection, represented by a change in colour
which reflects the fact that the ``collision'' is instantaneous in constrast to
the time interval separating two collisions}
\label{f:deflection} 
\end{figure*}

The previous sections dealt with the halo polarization 
 while considering that perturbations were transients. 
 In practice,
  a halo undergoes    recursive excitations from its environment that will induce departures from its equilibrium state so
that  it won't remain static.
 In Appendix~\ref{s:secularAppendix}  we derive a 
   quasi linear formalism  for the collisionless open Boltzmann equation in order to take this effect in account. 
  This follows in essence the  work of  \citet{weinberg93}, \citet{weinberg01},
  or  \citet{ma}, 
  though the derivation differs. {We introduce here an explicit expression,
valid at low redshift,} 
  for the source of stochastic ``noise''. We account explicitly 
for  the correlation induced by 
  the entering material (as characterized by \citet{aubert1}) rather than 
  rely on some {\it ad hoc} assumption on its nature.
We also account consistently for the mean secular infall which adiabatically
restructures the mean profile.
\subsection{Context and derivation}

  \citet{gilbert} gives a very elegant derivation from first principles of the secular equation 
  based on a $1/N$ ($N$ being the number of particles in the system) expansion of the collisional relaxation equations presented by \citet{bogoliubov}. 
\citet{weinberg93}, \citet{weinberg01} and  \citet{ma}  rely on the same expansion scheme to 
  derive their kinetic equation for the mean halo profile.

\citet{weinberg93} focuses on the secular collective relaxation of a system
 induced by the finite number of particles within a multi-periodic  uniform medium, 
 hence transposing to collisionless stellar dynamics 
 the derivation of 
 Lenard-Balescu (\citet{lenard}, \citet{balescu})
  applied originally  to plasma physics in order to describe the secular convergence of 
  such systems towards  thermalisation.

\citet{weinberg01} derives a similar result for the spherical halo in angle and action variables,
while relying on the Kramers-Moyal (\citet{risken}) expansion, which corresponds to 
a Markovian description based on the transition probability of a change in
action induced by the interaction with a dressed particle cloud.  His
Fokker-Planck coefficients differ slightly from \Eqs{D2def}{D1def} given in the Appendix in that the spectral properties of \(\langle \hat b_\bn \hat b_{\bn'} \rangle \) are postulated in his case, while \(\bc \equiv 0\).

\citet{ma} construct a Fokker-Planck equation for the 
mean profile of a halo in 
a cosmic environment while relying on the constrained random field 
of peaks in the standard cosmological model to  derive
the drift and diffusion coefficients from first principles.
Their derivation is dynamically accurate to second-order in the
 perturbation theory (in position-velocity space)
and relate the kinetic coefficients to the properties of the underlying linear
power spectrum. In contrast
to the theory presented here, 
their kinetic equation describes the very early phase of halo formation,
whereas we focus here on the quasi-linear evolution 
(in angle action space) of fully relaxed equilibria 
at low redshift.

In Appendix~\ref{s:secularAppendix}, we account explicitly 
for the nature of the perturbation's power spectrum as defined in \citet{aubert1}
and present an explicit derivation for the 
Fokker-Planck equation obeyed {on secular time scales by the} 
distribution function in angle 
action. {It is natural to use these} variables {to describe}
a relaxed collisionless halo {since they allow}
to split the dynamics into {a} secular (phase averaged) 
and {a} fluctuating {part}.

Even though individual dark matter particles obey a collisionless {dynamics},
the phase average (``ensemble average'') 
distribution for the open system satisfies a collisional 
kinetic equation where the clumpiness of the {\sl open} 
medium breaks the mean field approximation (see also \citet{ma}).
Indeed, individually,  clumps and tidal remnants deflect the actions of the underlying
 distribution in a stochastic (but correlated) manner, 
so that in the mean ensemble sense, the coarse-grained distribution (\ie 
the distribution averaged over the angles) obeys a collisional diffusion of the Fokker-Planck type.
 In this formulation, the graininess of the system (as defined 
by the second-order closure of the BBGKY hierarchy of the N-point distribution) corresponds to the mean 
 number of clumps expected in the halo, while the detailed
 (kinetic and angular) power spectrum of the gravitational fluctuations is given by the cosmogony.

{It is usual in plasma physics to take a two-time scale approach 
to the Boltzmann equation. The short time scale describes the system's dynamics
on the dynamical (orbital) time scale, while the longer time scale 
corresponds to the secular evolution. 
The} Action-Angle variables {are best suited here. This 
time scale separation procedure leads to}
the following system of {equations}:
\begin{eqnarray} &&
    \frac{\partial f}{\partial t} + {\bo} \cdot \frac{\partial
    f}{\partial \tmmathbf{\tmmathbf{w}}} - \frac{\partial \psi}{\partial
    \tmmathbf{w}} \cdot \frac{\partial F}{\partial \tmmathbf{I}}  = 
    \frac{\partial \psi_e}{\partial \tmmathbf{w}} \cdot \frac{\partial
    F}{\partial \tmmathbf{I}} + s_e, \EQN{eqQuasi1}\\
  && \hskip -0.5cm   \frac{\partial F}{\partial T}  \!\! = \!\! \left\langle\left[
    \frac{\partial \psi}{\partial \tmmathbf{w}}+\frac{\partial
    \psi_e}{\partial \tmmathbf{w}}\right]\cdot \frac{\partial
    f}{\partial \tmmathbf{I}}\right \rangle_{T}  \!\!\! -\! \left  \langle \left[
\frac{\partial \psi}{\partial \tmmathbf{I}} \!\!+ \frac{\partial \psi^{e}}{\partial \tmmathbf{I}} 
\right]\cdot \frac{\partial
   f}{\partial \tmmathbf{w}} \right\rangle_{T}\!\!\!+\! S_e,\EQN{eqQuasi2}
\end{eqnarray}
{In \Eq{eqQuasi1} and \Ep{eqQuasi2}} 
$s_e$ and $S_e$ stand for the perturbative and secular advected source
terms, {while} $f$ stands for the fluctuating 
distribution {and}
$F$ stands for the secular 
distribution function (see  Appendix~\ref{s:secularAppendix} for details).
The bracket  around the quadratic terms  stands for a time 
average over  a secular time, $T$
which is long compared to dynamical time, $t$ (taken by
 a dark matter particle {to describe} its orbit).
 If we fix $F(\bI,T)$,
 \Eq{eqQuasi1} corresponds exactly to \Eq{boltzsphere} 
 whose solution was described in \Sec{Boltzmann}.
 This  formal solution may then be injected in the quadratic
 terms of  \Eq{eqQuasi2}. Following this route, we show in Appendix~\ref{s:secularAppendix}  how to
rearrange \Eq{eqQuasi2}  as a Fokker Planck equation:
\begin{equation}
  \frac{\partial F}{\partial T} = \langle {\rm D}_0 ( \tmmathbf{I}) \rangle -
  \langle \M{D}_1 ( \tmmathbf{I}) \rangle \cdot \frac{\partial^{}
  F}{\partial \tmmathbf{I}^{}} -\langle \M{D}_2 ( \tmmathbf{I} ) \rangle: 
  \frac{\partial^2 F}{\partial \tmmathbf{I}^2}, \label{e:eqSecularM}
  \end{equation}
where ${\rm D}_0 $, $ \M{D}_1 $ and  $\M{D}_2$  are given by \Eqs{D2def}{D0def}, while $:$ stand for the total contraction.
Note that \Eq{eqSecularM}, in contrast to \Eq{eqQuasi2} refers this time
to the driving equation for the mean halo profile since we invoqued
ergodicity to replace time averages by ensemble averages (see 
Appendix~\ref{s:secularAppendix}  for details). 
{The ${\rm D}_0$ term enters here because the halo is an open system,
which may receive or lose mass}. 
The drift term {with the factor $\M{D}_1$} 
accounts for the dynamical friction induced by 
the polarization cloud around the tidal remnants; the diffusion term
{with the factor $\M{D}_2$}
arises because of the fluctuations in the potential (both tidal 
and associated with the infalling dark matter) induced by the 
clumps. The diffusion term 
will in general {induce a spreading of the energy distribution
by accelerating some orbits to higher energies while decelerating
some other ones.}
{ The} polarization cloud will in general induce a drag on 
the clumps, {represented by the $\M{D}_1$ term}. 
Note that the former should be independent of the mass of the clump 
(since the energy is exchanged via the mean field)
while the latter will not (since more massive clump polarize the medium more). 
From the point of view of the 
entering dark matter, the net effect 
is therefore a segregation process in which the more massive
clumps fall in (as discussed by \citet{gilbert}).

According to \citet{risken},
the  corresponding Langevin (\citet{langevin}) equation reads (when the source
term, $D_{0}$ is {\sl omitted})
\begin{equation}
\pdrv{\bI}{ t}= \MG{\Delta}_{1}(\bI)  +\MG{\Delta}_{2}(\bI) \cdot \MG{\xi}(t) \,.
\EQN{langevin}
\end{equation}
Here $\MG{\Delta}_{1}(\bI)$ and $\MG{\Delta}_{2}(\bI)$ are given in terms 
of $\M{D}_{1}(\bI)$ and $\M{D}_{2}(\bI)$ 
 by
\[ \MG{\Delta}_2 = \M{D}_2^{1/2}\,,\quad {\rm and } \quad  \MG{\Delta}_1= \M{D}_1 - 
\M{D}_2^{1/2}: \nabla_\bI  \M{D}_2^{1/2}\,,
\]
where \( [ \M{D}_2]^{1/2}\) stands for the square root of the matrix \(  \M{D}_2\) which is computed via diagonalization, provided the eigenvalues are positive.
The 3D random field,  $\MG{\xi}(t)$, should have spectral properties which reflect the stochastic properties of $\bb$ and $\bc$.
 The probability distribution of the solution to the stochastic equation,
 \Eq{langevin}, obeys the Fokker Planck Equation, \Eq{eqSecularM}.
In this form, the effect of diffusion on the departure from phase mixed equilibrium is 
easily interpreted. 

\subsection{Prospects for universal halo profiles}
\label{s:Universal}

As has been suggested and illustrated by \citet{weinberg01} and \citet{ma}, 
it would be very worthwhile to use \Eq{eqSecularM} and predict the 
asymptotic dark matter profile, (and, say the cosmic evolution of 
the concentration parameter) which will be shaped in part by encounters 
and interlopers.

Note that the diffusion coefficients, $D_i$ are relatively straightforward to compute for a given halo model, $F(\bI)$ {\sl but } \Eq{eqSecularM}
 corresponds to an evolution equation for $F(\bI)$ and will in practice require re evaluating the coefficients for different values of $F$.

Let us now draw constraints on the stationary solutions of  \Eq{eqSecularM}.
Again (following \Sec{applications}), 
this may be done in one of two ways: 
take ${\rm D}_0$, $\M{D}_1$ and $\M{D}_2$ as given 
function of the actions, and
deduce  what
equation  $F$ should obey from requiring that   Eq. (\ref{e:eqSecularM}) has a stationary solution. 
(this is the route first explored by \citet{weinberg2001b});
 or, if we assume that a given model, say a universal profile,
should correspond to the asymptotic solution of  Eq. (\ref{e:eqSecularM}), we may 
find the relationship relating the corresponding asymptotic $\M{ D}_i$ coefficients.
  
For simplicity, let us illustrate this second point while neglecting  here  the  fact that  the
diffusion coefficients depend on the distribution function, and 
restricting ourselves briefly to an isotropic distribution, $F(E,T)$. Calling:
\begin{equation}
  H ( E ) = \left( \langle \M{D}_1 \rangle \cdot \bo + \langle
  \M{D}_2 \rangle: \frac{\partial \bo}{\partial \tmmathbf{I}}
   \right) / \left( \langle \M{D}_2 \rangle: \bo
  \otimes \bo \right), 
\end{equation}
 \begin{equation}
 \,\,\, {\rm and } \,\,\,\, Q ( E ) =
  \frac{\langle {\rm D}_0 \rangle}{\langle \M{D}_2 \rangle: \bo \otimes
  \bo}\,, \EQN{defhq}
\end{equation}
the stationary solution (${\partial F}/{\partial T} =0$) 
to \Eq{eqSecularM} reads formally:
\begin{eqnarray}
 &&  F ( E ) \!\!=\!\! \int_0^E\!\! \! \exp \left[ -\! \int_{e_2}^{e_3} \!\! H [ e_1 ] {de}_1
  \right] \times \nonumber \\
  && \left\{ 
   \int^{e_3} \!\! Q [ e_4 ] \exp \left[ \int_{e_2}^{e_4} \!\! H [ e_1 ]
  {de}_1 \right] {de}_4  \right\} {de}_3.
\end{eqnarray}
{This} distribution function should 
satisfy the self-consistency requirement that
\begin{equation} \rho ( r ) = 2 \sqrt{2} \int_{- \psi}^0 F ( E ) \sqrt{E + \psi} \mathd E,
   \,\,\,\, \nabla^2 \psi ( r ) = 4 \pi G \rho ( r ). \EQN{edding}
   \end{equation}
Imposing that $F ( E )$ obeys \Eqs{defhq}{edding}
yields a non-linear integral equation for the $\M{D}_i$, \ie a (admittedly indirect) constraint on the angular
correlation of the external field.  \Sec{applications} describes other means of constraining 
the power spectrum of the infalling dark matter.
  \citet{weinberg2001b} found iteratively the corresponding solution
  while making some assumptions on the spectral properties of $\bb$ in the 
  r\'egime where $\bc =0$.   In the light of his investigation, he concluded 
  that the tidal excitation drives the halo towards a less steep profile.
   It will be interesting to explore this venue with a realistic accounting 
   of the source of infall.
The setting here would be
 that the satellite problem and the cusp problem of dark matter 
haloes might be  the two sides of the same coin, so that the evolution  towards a universal profile
might be triggered by the actual infall of substructures.

Let us now return to the perturbative dynamics described in 
\Sec{spherical},~\ref{s:nonlinear}
and explore its implications for galaxies.

\section{ Applications: halo polarization, disk dynamics and inversion }
\label{s:applications}
{ \citet{aubert1,aubert2}  provided a detailed  statistical description of 
how dark matter falls onto a $L^\star$ galactic halo: how much mass
is accreted as a function of time, how is it accreted \ie in what form, 
with what velocity distribution, along which direction, and for how long?
}
Putting the theory described here 
and the tabulated measurements from that paper together, allows us 
to address globally, and coherently dynamical issues on galactic scales
 in a {\sl statistically representative } manner. With the help of the theory presented 
 in \Sec{spherical},~\ref{s:nonlinear},
we are now in a position to ask ourselves: what are the expected 
features of a  halo/galaxy induced 
by their cosmic environment. 
Specifically, we may now ``simply'' propagate
the cosmological framework  and its statistics to observables (describing the departure from 
spherical symmetry/stationarity) on galactic scales.
On these scales, the realm of astrophysical applications for the perturbative open
solution of the Poisson-Boltzmann equations 
is extremely wide. 
It is clearly beyond the scope of this paper
to attempt an exhaustive inventory. Rather, we shall here 
focus on a few specific issues, for which we show how the open 
perturbative framework improves our understanding, 
and allows for a statistical investigation.  

In particular, we shall restrict ourselves to settings 
where the detailed geometry of the infall matters, since
the theory described above does account for the configuration and 
the time lag involved in the accretion on top of $L^{\star}$ galaxies.

Recall that the purpose of the statistical propagation is threefold:
(i) constrain the properties of the infall on the basis of the {\sl observed}
distribution for the  properties of  galaxies and their environment;
(ii) {\sl predict} some of  the statistical properties of galaxies which 
are not directly observable, while relying on the   properties of the infall.
(iii) weigh the relative importance of the intrinsic properties of the disk+halo
compared to the strength of the environment.

We will distinguish three classes of problems; first we will describe 
how to transpose to galactic scales (\Sec{cosmo})
the classical probes used in cosmology to trace the large-scale structures. 
We will then explore in \Sec{galactic} the implication  for the properties of 
external galaxies, 
and in \Sec{MW} for the structures within the Milky Way halo. 
Finally, we will elaborate  in \Sec{inversion} on the prospect of inverting the 
upcoming data sets for the {\sl past} history of our own Galaxy and for field
galaxies in the local group.

\subsection{ Cosmic probes in the neighbourhood of galaxies: $R_{\rm 200}/10< R < R_{\rm 200}$ }
\label{s:cosmo}

A series of observational probes of the statistical properties of the density field
 have been devised over the years, such as weak lensing,  galaxy counts, the SZ effect, 
 X-ray  or $\gamma$-ray  emissivity maps.  In the light of large galactic surveys 
 which are available today, it becomes quite desirable to apply these probes in the 
 neighborhood of galactic haloes in order to study the dark matter distribution 
 within the $R_{200}$ radius. Some of these tracers are only sensitive to 
 the baryon density, which need not trace directly the dark matter density. 
  In this section, we will systematically assume for simplicity a simple biasing, though
  this assumption may be lifted (at the 
  expense of extra non-linearities, see \Sec{ellip}) provided the biasing law is known  (\ie 
the observables are assumed  to scale like the dark matter density,
or some power of it); we refer to \Sec{satcount} for a brief discussion of 
thresholding, which is bound to be important in practice. 

  The calculation described in the previous sections, together with the statistical measurements described in \citet{aubert0,aubert1,aubert2} {\sl should}  allow us to make statistical  predictions about observables which may be expressed in terms of the distributions of clumps within
  the galactic haloes, either via their gravitational potential,  their projected density or
  even their velocity distributions (\eg Galactic streams).

We will consider in turns observable which may be approximated as linear functions 
of the {\sl perturbed } fields, either in projected coordinates on the plane of the sky, or as 
seen by an observer at the galactic center. We will also consider observables which involve
 quadratic functions of this field (\eg the square of the electron density), or 
even more non-linear functions of the dark matter distribution  within the virial radius
(such as the locus of virialized clumps, which dissolve at a function of time).
We will in particular build the two-point statistics for these observables, since
the mean of the perturber is often zero by construction. 
Finally we will also consider  metals lines in absorptions systems, which involve the cross-correlation of 
the density and the velocity fields. 
Note that all these measurements could in principle be carried as a function
of redshift,
or a function of the mass of the halo, or while varying the anisotropy of the equilibrium for 
the halo (by varying $F(\bI)$ in \eg $\bK$ in \Eq{defK1}).  Note finally that some tracers correspond to the scales of clusters,
and we will assume here that the measurements presented in \citet{aubert1} 
could be reproduced for these objects (whereas the theory described here 
is scale independent provided the system is dynamically relaxed 
and spherical).

In this section, we will focus on a couple of probes which are supposed to scale linearly
with the dark matter density in the main text (weak lensing, SZ effect), and postpone to appendix~\ref{s:AppendAppli} a presentation of other probes
(X-ray emissivity, dark matter disintegration, metal lines in absorption spectra).

Note that all probes described below are a departure from the mean
profile of galactic haloes (just as cosmic perturbation theory describes the growth of 
structure as a departure form the mean density/expansion of the universe) and as such, assume 
that we have a good understanding of this profile.
This will undoubtedly turn out to be a serious observational constraint when
attempting to ensemble average galaxies of various size and properties. 

\subsubsection{Weak lensing in stacked haloes }


Weak lensing corresponds to the deflection of light emitted from
 background galaxies 
by the gravitational potential of structures between those galaxies and the observer.
It has recently been used 	quite successfully to constrain the 
statistical distribution of the large-scale structures. {On smaller scales, the
 effects of substructures in haloes on the lensing measurments have been demonstrated by
 e.g. \citet{Dalal}, \citet{Kocha}.}
  
In the weak lensing r\'egime (\citet{Peacock}),
the relationship between the observed convergence and 
the underlying projected dark matter profile is approximated 
to be  linear. Hence we may straightforwardly propagate 
our statistical predictions for the clumpy dark matter distribution around a dark matter 
halo (or within the neighborhood of clusters of galaxies provided some readjustment 
of the theoretical predictions described in \citet{aubert0,aubert1} on these larger scales).

The cumulative deflection angle, $ \MG{\alpha}(\MG{\theta},w)
\equiv{\delta \bx}/{r_{k}(w)} $ by which light is  deflected is given by
\begin{equation}
\MG{\alpha}(\MG{\theta},w)
=
\frac{2}{c^{2}} 
\int \d w'
\frac{r_{k}(w'-w)}{r_{k}(w)}
\nabla_{\perp}\psi(r_{k}(w')\MG{\theta},w')\,, \EQN{defLensing}
\end{equation}
where $r_{k}$ is the angular comoving distance, $\d w\equiv \d r/\sqrt{1-k r^{2}}$ ($k=0,\pm 1$) and $\bx$ 
the transverse comoving distance.
Defining the convergence, $ \kappa(\MG{\theta})$ by 
\begin{equation}
\kappa(\MG{\theta})=\frac{1}{2}\nabla_{\MGS{\theta}}\cdot \MG{\alpha}(\MG{\theta})\,,
\end{equation}
the mean ensemble average convergence of the rescaled halo, $\langle \kappa({\theta}) \rangle$,  reads:
\begin{equation}
\langle \kappa({\theta}) \rangle=
\frac{1}{c^{2}} 
\int \d w'
\frac{r_{k}(w'-w)}{r_{k}(w)}
\nabla^{2}_{\perp}\psi_{\rm NFW}(r_{k}(w'){\theta},w')
\,.
\end{equation}
Now recall that (\cf \Eq{defpsiexp} where $\psi^{[\bn]}$ is given by \Eqs{harmexp}{harmexp2} in Appendix~\ref{s:compLIN}.)
 \begin{equation}
 \delta \psi(\br)=\sum_{\bn} a_{\bn} \psi^{[\bn]}(\br)\,,
\quad { \rm hence}
\quad \delta \kappa(\MG{\theta})=\sum_{\bn} a_{\bn}  \kappa^{[\bn]}(\MG{\theta})\,,
 \end{equation} where\begin{equation}
 \kappa^{[\bn]}(\MG{\theta}) = \frac{1}{c^{2}} 
\int \d w'
\frac{r_{k}(w'-w)}{r_{k}(w)}
\nabla^{2}\psi^{[\bn]}(r_{k}(w')\MG{\theta},w')\,.
\end{equation}
It follows that the correlation function of the relative convergence obeys
\begin{equation}
\frac{\langle\delta \kappa(\MG{\theta}) \delta \kappa(\MG{\theta}') \rangle}{\langle \kappa({\theta}) \rangle^{2}}=
\frac{1}{\langle \kappa({\theta}) \rangle^{2}}\sum_{\bn,\bn'}  \langle a_{\bn} a_{\bn'} \rangle \kappa^{[\bn]}(\MG{\theta}) \kappa^{[\bn']}(\MG{\theta}')   \,. \EQN{SZ}
\end{equation}
Hence, the statistical properties of the relative convergence will depend on the 
statistical distribution of the  clumps of the halo through the  $\{a_{\bn}\}$ coefficients
which are given in \Eq{correlA} in terms of $b_{\bn}$ and $c_{\bn}$.

In practice one has to devise an observational strategy, given the expected size of the 
caustics of subclumps within haloes of galaxies or clusters, the number of 
background sources, and the expected number of foreground objects  (\ie galaxies or clusters).  

Finally, it is believed that one in a hundred large ellipticals on the sky
should undergo 
strong lensing. In the long run, the statistical properties of 
such a non-linear signal will be worth investigating within the framework described in this paper (following the non-linear steps described in say,
\Sec{ellip}). 

\subsubsection{Thermal S-Z effect  of stacked haloes} 


When the photons of the  cosmic microwave background  enter the hot
 dense gaz  within the  clusters and galactic haloes, they interact with the electrons of the gaz. 
The diffusion process transfers the energy of the photons to the electrons 
which in turn reemit this energy at a higher frequency. The corresponding 
spectral redistribution induces a local temperature decrement seen in the
 temperature map of the clusters, known as the thermal Sunyaev-Zeldovich effect
 (see e.g. \citet{Peacock}). 
The temperature decrement (at low frequency) reads as a function of the distance to the 
cluster center, $\M{R}$ :
\begin{equation}
\frac{\Delta T(\M{R})}{T_{\rm CMB}}= -2 \frac{k_{{}_{B}} \sigma_{\rm T}}{m_{e} c^{2}}
\int \d z n_{e}(z,\M{R}) T_{e}(z,\M{R}) \,,\EQN{defSZ}
\end{equation}
where  $m_{e}$, $n_{e}$ and $T_{e}$ are respectively the mass, the numerical density and the temperature of the electrons,
 while $\sigma_{\rm T}$ is the Thomson scattering section ($6.65 \times 10^{-25} {\rm cm}^{2}$), $c$ the speed of light,
  $k_{{}_{B}}$ Boltzmann's constant,  and $T_{\rm CMB}$ is the Cosmic microwave background temperature. 

Let us assume that the variation in temperature is small compared to the
variation of the electron number density\footnote{Note that we may lift this assumption at the cost of nonlinearities, 
provided we may rely on  an equation of state to 
relate it to the underlying density}. Let us also assume that the electron density is 
proportional to the dark matter density (constant biasing) as mentioned above. 
Let us define the departure from the cosmic average for the profile as:
\begin{equation}
\delta \Delta T(\M{R})=   {\Delta T(\M{R})}  -\langle  \Delta T \rangle (R)\,.
\end{equation} 
The relative fluctuation of the temperature decrement reads:
\begin{eqnarray}
\frac{1}{\langle  \Delta T \rangle^{2}(R)} \langle \delta {\Delta T(\M{R})} 
\delta {\Delta T(\M{R}')}\rangle\hskip -0.3cm &=& \hskip -0.3cm \frac{1}{\Sigma_{\rm NFW}(R)^{2}}\sum_{\bn,\bn'}
 \langle a_{\bn} a_{\bn'} \rangle \times \nonumber \\ &&  \hskip -2.5cm  \int \d z \int \d z' \rho^{[\bn]}(\M{R},z)   \rho^{[\bn']}(\M{R}',z') \,, \EQN{SZ}
\end{eqnarray}
where $\Sigma_{\rm NFW}$ is the mean { rescaled}  projected dark matter
 mass profile. Note that the double integral in \Eq{SZ} is carried over known functions
 and is just a geometric factor which will depend on $R,\Delta R$, and $\Delta\Theta$ only.
 Again, the knowledge of the statistics of the $\{a_{\bn}\}$ (which in turn only depend on 
 the equilibrium, $F_{0}$, and the statistics of $b_{\bn}$ and $c_{\bn}$ at $R_{200}$, see \Eq{SOLK}
 or \Ep{correlA}) therefore allows us to predict the 
 statistical properties of the relative fluctuations in the temperature decrement.
ALMA will soon provide detailed SZ maps of clusters for which it should be possible to apply
these techniques.

Let us note in passing that the kinetic S-Z effect  of stacked haloes
may also be investigated following the same route 
\begin{equation}
\frac{\Delta T(\M{R})}{T_{\rm CMB}}= -2 \frac{k_{{}_{B}} \sigma_{\rm T}}{m_{e} c^{2}}
\int \d z n_{e}(z,\M{R}) v_{z}(z,\M{R}).\EQN{defkinSZ}
\end{equation}

In closing, let us note that maps of SZ effects within our
own Galaxy will be available with the upcoming Planck satellite,
and will provide statistical information on the small-scale 
distribution of local clumps.
Recall finally that appendix~\ref{s:AppendAppli}  presents other
statistical probes of the outer structures found in galactic haloes
(X-ray emissivity, dark matter disintegration, metal lines in absorption spectra). Most of these probes could be used to say, probe the shape of the 
density profile in the outer parts of galaxies, or the biasing law
relating dark matter to stars or gas. 

\subsection{Galactic structure: $ R < R_{\rm 200}/10$}

\label{s:galactic}
In the previous section, we investigated  the dynamical consequences 
of the cosmic infall  in the outer region of the halo.
Let us now turn to the regions of the halo where we expect to find
 the galaxies themselves. 
  At lower redshift, the galaxies
 essentially come in two flavours, ellipticals and spirals.
 The response of ellipticals should follow closely that  of the 
 dark matter halo since both components are hot enough not to undergo 
 gravitational instabilities. In effect, describing an embedded ``spherical'' elliptical galaxy
 within a dark matter halo amounts to changing the distribution function
 to account for the presence of the elliptical and its possibly distinct 
 kinematics.  
Note that, as mentioned before the above theory could be amended to
account perturbatively for the  possible triaxiality of the elliptical (\citet{binney}). 
    
 For a disk or a very flattened spheroid, the situation becomes quite 
 different. The cooler disk is likely to be either drawn beyond its stability threshold by
 the perturbation, or will respond much more strongly to the perturber than its dark halo. 
 Hence we need to model the disk component differently.
The proto-galactic environment is likely to be extremely noisy, particularly in
outer regions, so that the halo may perturb the disk by transmitting numerous disturbances into
the inner galaxy. Moreover, the inner halo may continue to oscillate as it settles after the coalescence of advected objects.
 Halo oscillations may easily perturb the disk through the time-dependent gravitational
potential. Conversely, the structural integrity of observed disks set limits on the degree of
disequilibrium in the proto-galactic halo.

 Within the realm of features found in galactic disks, a fraction
 are known to be the result of instabilities (\eg  galactic bars),  while
  others have been 
 shown to correspond to transients (\eg galactic warps). 

With the advent of modern systematic surveys, it is possible to 
construct distributions corresponding to, say, 
 the  fraction of spirals which fall within some Hubble type,
   or the fraction of warps whose inclination is larger than some angle,
On the disk scale, we may construct the PDF of, say, the pitch angle of dynamically induced 
spirals, or the PDF of the extent of the bar, its amplitude, or less directly observed 
the PDF of pattern speeds.  Some of these processes  depend crucially on gas physics
and will not be addressed here.

\subsubsection{Pitch angle distribution for spirals}
\label{s:pitch}

For stellar disks, the stars obey formally the same equation 
as \Eqs{boltzsphere}{poisson}, but this time the modes may be unstable,
and sometimes the disk cannot be treated in isolation from the live halo in which it is 
embedded.
 On the other hand, it 
is often well approximated as an infinitely thin structure;
such a 2D system becomes integrable again with two actions, 
 $\d \M{J} \equiv \d J_r \d L_{z,_{\rm D}}$. Here $J_r $ is the radial action of the stars in the plane of the 
 disk, and $L_{z_{\rm D}}$  is the momentum of the stars in the disk.
 Following \citet{weinbergDH} and adding some source of infall at $R_{200}$,
we may describe the coupled system disk + halo in the complex plane as   
       \[
           \left(\begin{array}{c}
     {\hat \ba}_{\rm D}\\
     {\hat \ba}_{\rm H}
   \end{array}\right) = \left(\begin{array}{cc}
     {\hat \bK}_{{\rm DD}}\! & {\hat \bK}^{*}_{{\rm DH}}\!\\
     {\hat \bK}_{{\rm DH}}\! & {\hat \bK}_{{\rm HH}}\!
   \end{array}\right) \cdot \left(\begin{array}{c}
     {\hat \ba}_{\rm D}\\
     {\hat \ba}_{\rm H} \!+\! {\hat \M{b}}
   \end{array}\right) \!+\! \left(\begin{array}{c}
     0\\
    {\hat \bQ}
   \end{array}\right) \cdot\left(\begin{array}{c}
     0\\
     {\hat \bc}
   \end{array}\right)
  \,,  \]
    where 
    ${\hat \bK}_{{\rm HH}}$ is given by \Eq{defKnk},
  while 
    \[
   \left( {\hat \bK}_{{\rm DD}}\right)_{\bp,\bq}= 
 {  \sum_\bk  \int \op{\d \M{J}}
  \frac{\psi^{_{\rm D},[ \bp ]}_\bk  (\M{J})  
  \psi^{_{\rm D},[\M{q} ] *}_\bk  (\M{J})   \hspace{0.25em}}{\bk
  \cdot \bo_{_{\rm D}} - \omega}{\hspace{0.25em}}} \bk \cdot \frac{\partial
  F_{_{\rm D}}}{\partial \M{J}}, \EQN{defK1disk}
    \]
and a similar expression involving $\psi^{[ \bp ]}_\bk (\M{J}) 
  \psi^{_{\rm D},[\M{q} ] *}_\bk  (\M{J})  $ for the cross term,  ${\hat \bK}_{{\rm DH}}$.
    See \citet{pichonCannon} for details relative to  the disk.
    Here $\hat\ba_{\rm D}$ and $\hat\ba_{\rm H}$ are the coefficient 
    of the expansion for the disk and the halo respectively, 
    $ F_{_{\rm D}}$ is the distribution function of stellar stars within the disk,
    $\{\psi^{_{\rm D},[\M{q} ]}(r)\}_{\bq}$ the potential basis function over which 
    the disk response is projected,
    and $\bo_{_{\rm D}}$ the angular frequencies of the stars in the disk.
   
    Let us first assume that the unperturbed disk is stable.
Solving the coupled equation for
  $ [\hat \ba_{\rm H},\hat \ba_{\rm D}]$
  yields (after half inverse Fourier transform) the 
  temporal evolution of the spiral response as a function of time for a given tidal field,
  $\bb(t)$ and a given infall history, $\bc(t)$.
  The pitch angle, $\cal I$ of the spiral, defined by $\displaystyle \tan({\cal I})=1/\pi \int_{0}^{
  \pi} \d \theta \d \log {\cal R}/ \d \theta$
  (where ${\cal R}(\theta)$ corresponds to the crest of the spiral wave), 
  is a non-linear function of $\ba_{\rm D}$, which we may write formally as 
   ${\cal I}[\ba_{\rm D}]$. Hence we may ask ourselves
  what it's cosmic mean,  $\langle {\cal I}[\ba_{\rm D}] \rangle$ is, 
  given that $\hat \ba_{\rm D}(\omega)$ obeys
\[
 \hat \ba_{\rm D}(\omega)=    {\hat \bK}_{{\rm HD}}\cdot [\bb + \bQ \cdot \bc] \Bigg/
  {\rm Det}\left|\begin{array}{cc}
    \b1-{\hat \bK}_{{\rm DD}} & {\hat \bK}^{*}_{{\rm DH}}\\
     {\hat \bK}_{{\rm DH}} & \b1 -{\hat \bK}_{{\rm HH}}
   \end{array}\right|\,.
\]

Note that ${\cal I}[\ba_{\rm D}]$ will depend on the statistical properties of 
$\bb$, $\bc$ and also on the distribution function for the halo, $F(\bI)$, and 
the distribution function for the disk, ${ F_{_{\rm D}}}(\M{J})$. 
More generally, we may in this manner construct the full PDF of the pitch angle, 
as a function of say, cosmic time, (or relative mass in the disk or ... )
following the same route as sketched in \Sec{ellip}. 

If the disk is intrinsically unstable, we must then add to the driven response described above 
the unstable modes. The amplitude of the response will 
then depend on exactly when each 
unstable mode has been exited.  
Such a prescription is beyond the scope of this paper, but could be addressed
statistically through the description of a phase transition.

\subsubsection{Warp excitation }
As mentioned earlier  (\citet{Lopez}, \citet{JiangBinney}), warps are 
intrinsically stable modes of thin disks which respond to their
environment. 
The action of  the torque applied on  the disk of a galaxy  is different for
different angular and radial  positions of the perturbation. The
warp's   orientation  and  its   amplitude  are   functions  of   the 
external potential.

The work done by the presence of perturbations on the stellar system is 
\begin{equation}
\frac{\d{E}}{\d{t}}=-\int \d{\bfr} \nabla (\psi+\psi^{e})\cdot \rho\bfv=-\int \d{\bfr}
(\psi+\psi^{e})\nabla (\rho\bfv),
\label{e:damp}
\end{equation}
where $\psi+\psi^{e}$ is the total potential perturbation (self-response + external component).

Using \Eq{defvrsol} and  \Eq{damp}
\begin{eqnarray}
\langle \frac{\d{E}}{\d{t}} \rangle &=&-\sum_{\bn,\bn'}\int
\d{\bfr}\langle \left[a_{\bn'}(t)+b_{\bn'}(t)\right]\nabla \psi_\bn(\bfr)\cdot
\times \nonumber \\&&  \hskip -2cm \int_{-\infty}^t\!\!
\d{\tau}K_{[2],\bn}(\bfr,\tau-t)\left[a_\bn(\tau)+b_\bn(\tau)\right]+Q_{[2],\bn}(\bfr,\tau-t)c_\bn(\tau)\rangle . \nonumber  
\end{eqnarray} 
The power spectra of potential fluctuations drive  the energy
rate of change through the cross-correlation between the source and the potential.

Note in closing that the framework described in this paper 
should allow us in the future to address the possibility of warps induced 
by the accretion of gas.

\subsection{Substructures in our own Galactic halo }
\label{s:MW}
Let us now turn to the Milky Way. Our knowledge of the structure of its halo 
has increased dramatically in the course of the last decade with the advent 
of systematic imaging and spectroscopic surveys (\eg SDSS, 2dF), both in the optical 
and at longer wavelengths (\eg 2MASS), and this observational investigation will 
undoubtedly continue with efforts such as RAVE,
 or the upcoming launch of GAIA.
 This has led to the discovery of quite a few substructures within our halo, both
 in projection on the plane of the sky (tidal tails)  as star counts but also via kinematical features  (streams). The extent of the upcoming systematic
stellar surveys will allow for a systematic analysis of the  dynamical 
 properties of Galactic substructures.

\subsubsection{Extent  of tidal tails and streams in proper motion \& galactic coordinates}

The number  of stars,  $\d N$, in  the solid  angle defined by  the Galactic
longitudes and latitudes $(\ell,b)\equiv \MG{\ell}$ (within  $ \d \ell \d {(\sin b)}$), with
proper motions  $(\mu_{\ell},\mu_{b}) \equiv \MG{\mu}$ (within $\d{}  \mu_{b} \d{}
\mu_{\ell}$) at time $t$ is given by (\citet{pichonSiebert}):
\begin{eqnarray}
\d N  \equiv A_{\lambda}(
\MG{\mu}, \MG{\ell},t)\, \d \MG{\mu} \d \MG{\ell}  
  =
  \left\{
  {\int \!\!  {  {
}} 
\!\!\!\int \!\! 
\d u_r
  \mathfrak{r}^4 
  \d \mathfrak{r} f(\M{\mathfrak{r}},\M{u},t)
} \right\} \,  \d \MG{\mu} \d \MG{\ell} 
\, , \EQN{N0}
\end{eqnarray}
 The variables $\M{\mathfrak{r}},\M{u}$ are the vector position and velocity
coordinates ($u_r,u_\ell,u_b$) in phase space relative to the Local Standard
of Rest, while $\M{r}=(R,\Phi,z)$ and $\M{v}=(v_R,v_\Phi,v_z)$ are those relative
to the  Galactic centre.  In  particular, the radius  $\mathfrak{r}$ (within
$\d \mathfrak{r}$)  corresponds to the distance  along the line  of sight in
the  direction given  by the  Galactic longitudes  and  latitudes $(\ell,b)$
(within the solid angle $\d \ell \cos(b) \d b$).

These velocities are given  as a function  of the velocities measured in the
frame of the sun by
\begin{eqnarray}
    {v_{\Phi }} &=&  \frac{1}{{R}}{{ \left(  r_\odot\sin  (b )\sin  (\ell  )
      {{{u}}_{b  }}  -r_\odot\cos (b )\sin  (\ell  ) {{{u}}_r} -
    \right.  }}  \hskip -2cm\nonumber\\ && \hskip -1.7cm {{ \left. 
       r_\odot\cos
      (\ell ) {{{u}}_{\ell   }} + r\cos  (b )  \left( {{{u}}_{\ell  }} -\sin
      (\ell  ) {{{u}}_{\odot }} \right) +  \left( r_\odot +  r\cos (b ) \cos
      (\ell ) \right)  {v_{\odot }}  \right) }} \,,   \hskip -2cm\nonumber   \\
      {v_{R  }} &=&
      \frac{1}{R}\left\{  { \left( r\,\cos  (b   ) - r_\odot\,\cos (\ell   )
      \right) \,   \sin (b )\,  {{{u}}_{b   }}  - r_\odot\,\sin   (\ell  )\,
      {{{u}}_{\ell }} -} \right.  \hskip -2cm
      \nonumber \\ &&
      \left. {
     \,\cos (b )\, \left( {r}\,\cos (b ) - {r_\odot}\,\cos
      (\ell ) \right)  \, {{{u}}_r} + }\right.  \hskip -2cm \nonumber \\ &&
      \quad
      \left.  {  r_\odot\,{u_{\odot }}   -  r\,\cos  (b   ) \cos( \ell   )\,
      {{{u}}_{\odot  }} +  r\,\cos (b )\,\sin  (\ell  )\,  {{{v}}_{\odot }}}
      \right\}
      \,, \hskip -2cm \nonumber  \\ v_{z}    &=& \sin (b )\,{{{u}}_r} + \cos  (b )\, {{{u}}_{b }}
      + {w_{\odot }} \, , \hskip -2cm \EQN{vz}
\end{eqnarray}
where 
\[
\Phi = \tan^{-1}\left(
\frac{ \mathfrak{r}\, \cos (b )\,\sin (\ell ) }{R},\frac{r_\odot -\mathfrak{r}\,\cos (b )\,\cos (\ell ) }{R}\right)
 \,, \]
 \[R = \sqrt{r_\odot^2 - 2r_\odot \mathfrak{r}\,\cos (b )\,\cos (\ell )
+ \mathfrak{r}^2\cos   (b )^2}\,,
   \quad {\rm and} 
 \]
\begin{equation}
    z =  \mathfrak{r}\sin (b ) \,. 
      \EQN{defR}
\end{equation}
$R$  measures  the projected  distance  (in  the  meridional plane)  to  the
Galactic  centre, $\Phi$ the angle in the meridional plane between the 
star and the Galactic centre, while $z$  is the  height of  the star.   Here $u_{\odot},
v_{\odot},w_{\odot}$ and $r_{\odot}$ are  respectively the components of the
Sun's velocity and its distance to the Galactic centre. 

Recall that \Eq{DFsphere}  together with \Eq{SOLK} provides us with 
the full phase space distribution of the infall as a function of the actions of the 
unperturbed halo. 
Let us call  ${f}_{\bn}$, the phase space basis  defined by
\begin{equation}
{f}_{\bn}(\br,\bv,\tau) \equiv
   \sum_{\bk} 
  \exp(\imath {\bf k} 
\cdot {\bo} \tau +\imath \bk \cdot \bw)
\, \bv \,\, \imath {\bf k} \cdot \frac{\d F}{\d \bI }   
\psi^{[\bn]}_\bk(\bI)
 ,
 \label{e:defFn}
\end{equation}
so that the perturbation at time $t$ and position $(\br,\bv)$ reads
\begin{equation}
f(\br,\bv,t)=\sum_{\bn} \int^{t} \d \tau {f}_{\bn}(\br,\bv,\tau-t) a_{\bn}(\tau)\,. 
\end{equation}
We may now  seek the  characteristic signature  in observed  phase space (today \ie at $t=0$),
of a given perturbation.
\begin{equation}
A (\MG{\ell},\MG{ \mu} )= \sum_{\bn} \!\! \int \!\! \d \tau  a_{\bn}(\tau)   \!\! \int\!\!\d  \mathfrak{r} 
 \mathfrak{r}{}^{4} 
\!\!   \int \!\! \d u_{r}
 {f}_{\bn}(\br[\MG{\ell},\mathfrak{r}],\bv[ \mathfrak{r}\MG{\mu},u_{r}],\tau) \,, \EQN{defAmuell}
\end{equation}
 where $\bv[ \mathfrak{r}\MG{\mu},u_{r}]$ is given by \Eq{vz}, and $\br[\MG{\ell},\mathfrak{r}] $
is given by \Eq{defR}. 
In particular, we may compute the autocorrelation of the kinematic count defined by
\begin{equation}
  C_{\rm A}^{\MGS{\mu}} ( \Delta \MG{\ell}, \Delta \MG{\mu} ) \equiv \langle A (
  \MG{\ell} + \Delta \MG{\ell}, \MG{\mu}+ \Delta \MG{\mu} ) A (\MG{\ell},\MG{ \mu} ) \rangle\,.
\end{equation}
It involves an integral over time of the autocorrelation of the coefficients, $\langle a_{\bn}(\tau) a_{\bn'}(\tau') \rangle 
$ as
\begin{eqnarray}
C^{\MGS{\mu}}_{\rm A}\hskip -0.3cm &=&\hskip -0.3cm \sum_{\bn,\bn'}
  \iint \d \tau  \d \tau' \langle a_{\bn}(\tau) a_{\bn'}(\tau') \rangle 
\iint   \mathfrak{r}^4 
  \d \mathfrak{r}
 \mathfrak{r}'{}^{4} 
  \d \mathfrak{r'}
  \iint \d u_{r}  \d u'_{r}
\times \nonumber \\ 
&& \hskip -1.cm
 {f}_{\bn}(\br[\MG{\ell},\mathfrak{r}],\bv[ \mathfrak{r}\MG{\mu},u_{r}],\tau)  {f}_{\bn'}(\br[\MG{\ell}+\Delta \MG{\ell},\mathfrak{r}],\bv[\mathfrak{r}'\MG{\mu}+\mathfrak{r}'\Delta\MG{\mu},u'_{r}],\tau') . \nonumber
\end{eqnarray}
 Recall that  $\langle a_{\bn}(\tau) a_{\bn'}(\tau') \rangle $ can be reexpressed in 
terms of the coefficients of
 $\langle  \, {\tmmathbf{ \hat b}}\cdot
{\tmmathbf{  \hat b}}^{*\top}\rangle  $  and $\langle \,  {\tmmathbf{ \hat  c}}
\cdot
{\tmmathbf{ \hat c}}^{*\top} \rangle $ and  $\langle\, {\tmmathbf{ \hat b}}\cdot {\tmmathbf{ \hat c}}^{*\top} \rangle $
via \Eq{correlA}.
The  width of the correlation, $C^{\MGS{\mu}}_{\rm A}(\Delta \MG{\ell}, \Delta \MG{\mu})$,
both in velocity space and in position space accounts for the expected cosmic size of 
structures within the Galactic halo. 

\subsubsection{ Angular extend  of  tidal tails }

The marginal distribution  over proper motions of \Eq{N0}  yields the projection
on the sky of the perturbation:
\begin{eqnarray}
A(\MG{\ell},t) \!\!\!\! &\equiv&\!\!\!\! \iint A(\MG{\ell},
\MG{\mu} ,t) \, \d \MG{\mu}    
  = 
\!\!\!  {\iiint  {  {
}}\d u_r  \mathfrak{r}^4 
  \d \mathfrak{r} 
f(\M{\mathfrak{r}},\M{u},t) \, \d \MG{\mu}
}  
\, ,\EQN{N01}
\end{eqnarray}
which can be derived from \Eq{defAmuell} but is also found directly via 
integration over the density as
\begin{eqnarray}
A(\MG{\ell},t)=\sum_\bn a_\bn(t) 
  { \int  {  {
}}
{\tilde \rho_\bn}(\M{\mathfrak{r}},\MG{\ell}) \mathfrak{r}^2 }
  \d \mathfrak{r} \,,\quad {\rm given}\quad \nonumber \\
{\tilde \rho_\bn}(\M{\mathfrak{r}},\MG{\ell}) \equiv{ \rho_\bn}(R(\M{\mathfrak{r},\MG{\ell}}),\Phi(\M{\mathfrak{r},\MG{\ell}}),z(\mathfrak{r},\MG{\ell}))\,,
\EQN{Al1}
\end{eqnarray}
where $\rho_{\bn}$ is given by \Eq{defpsiexp} and \eg $R(\M{\mathfrak{r},\MG{\ell}})$ is given by  \Eq{defR}.
Note the generic difference between \Eq{defAmuell} and \Ep{Al1}: the former involves 
the explicit cumulative knowledge of $a_{\bn}(\tau)$ for all $\tau$ since  it involves 
a kinematical  (inertial) quantity, $\MG{\mu}$, while the latter only require the knowledge of 
the current $a_{\bn}(t)$. This difference is weaker than it seems in practice, 
since self-gravity implies that  $a_{\bn}(t)$ depends in turn on the previous $a_{\bn}(\tau)$
via \Eq{SOL}.
The corresponding angular correlation reads
\begin{eqnarray}
  C_{\rm A} ( \Delta \MG{\ell} ) &\equiv& \langle A (  \MG{\ell}+\Delta \MG{\ell}
  ) A ( \MG{\ell} ) \rangle = \nonumber \\ 
  && \hskip -2cm \sum_{\bn,\bn'} \langle a_\bn(t) \, a_{\bn'}(t) \rangle \iint 
 \d \mathfrak{r} \, \d \mathfrak{r}' \,  \mathfrak{r}'^2  \mathfrak{r}^2 
 {\tilde \rho_n}[{\mathfrak{r}},\MG{\ell}] {
  \tilde \rho_n}[{\mathfrak{r}'},\MG{\ell}+\Delta\MG{\ell}] \,.
   \EQN{defCA}
\end{eqnarray}
The FWHM of  the correlation defined by \Eq{defCA} corresponds to the``cosmic'' width of 
tidal stream projected on the sky.

\subsection{Past history of galaxies : dynamical inversion}
\label{s:inversion}
%
Let us now see how the theoretical framework presented 
in \Sec{spherical} and \Sec{nonlinear} may be applied 
to invert observed properties of  galaxies back in time 
and constrain the past infall and the tidal field on a given
 dark matter halo. In short, the idea is to notice that the perturbation 
 theory provides an explicit relationship between 
 the response and the excitation of the inner halo
 which we can tackle as an integral equation for the source.
\footnote{
Since our treatment of the
dynamics (including the self -consistent gravity polarisation) is linear order by order, we
may in principle recover the history  of the excitation. }
Let us present first the inversion  for our Milky Way (\Sec{galinv}),
and discuss briefly extra galactic stellar streams. 

\subsubsection{The Galactic inverse problem}
\label{s:galinv}


Let us rewrite formally \Eq{defAmuell} as 
\(
A (\MG{\ell},\MG{ \mu} )={ \cal A}^{\MGS{\ell},\MGS{ \mu} } \cdot \ba \,, 
\) where the dot product accounts for {\sl both} the summation over $\bn$ 
and the integration over $\tau$ (\cf \Eq{defcontract1}).
Let us assume that we have  access to kinematic star counts, \ie to a set of  measurements $\{ A_{i}\equiv
A(\MG{\ell}_{i},\MG{\mu}_{i})\}_{i\le n}$.
We want to minimize
\begin{equation}
  \chi^2   = \sum_{i} \left( A_{i} - {\cal A}_{i}^{\MGS{\ell},\MGS{ \mu} } \cdot
   \M{R}_{1}\cdot ( \mathbf{K} \cdot \mathbf{b} + \mathbf{Q}
  \cdot \mathbf{c} )  \right)^2, \label{e:defchi}
\end{equation}
subject to some penalty function. Recall that $\M{R}_{1}$ is given by \Eq{defR1}
and accounts for the self-gravity of the halo. 
Let us formally rewrite again ${\cal A}_{i}^{\MGS{\ell},\MGS{ \mu} } \cdot
   \M{R}_{1}\cdot ( \mathbf{K}_{1} \cdot \mathbf{b} + \mathbf{Q}_{1}
  \cdot \mathbf{c} ) \equiv \M{M}\cdot  {\tilde \bb} $,  with ${\tilde \bb}=[\bb,\bc]$.
  Let us also write $\M{A}=(A_{i})_{i\le n}$.
  The solution to the linear minimization, \Eq{defchi} reads
  \begin{equation}
{\tilde \bb} \equiv \M{M}^{(-1)}_\lambda\cdot \M{A}=
( \M{M}^{\top} \cdot \M{M}+ \lambda \M{P}
  )^{- 1} \cdot \M{M}^\top \cdot \M{A}\,, \EQN{sol1inv}
\end{equation}
where $\M{P}$ is some penalty which should impose smoothness for 
$\bb$ and $\bc$ both angularly and as a function of time.
For instance, For the $\bb$ field we could use (see, \eg \citet{pichonSiebert}):
\[
  P[b_{\ell m}] =\sum_\ell\left[ (\ell+1) \ell\right]^2
   \int \d \omega \omega^{2}  |\hat C_\ell|, \quad {\rm where}
  \quad 
\hat C_\ell(\omega) \equiv  \langle |\hat b_{\ell m}|^2  \rangle\,,
 \label{defpenalty} 
\]
(so that large $\omega$ and $\ell$ are less likely in the solution)
and a similar expression for the $\bc$ field which should also 
impose smoothing along velocities.
The   penalty
coefficient, $\lambda$,  should be tuned 
so  as to provide the appropriate  level of smoothing.
In practice, it might be necessary to impose further non-linear constrains on
the solution, $\tilde \bb$, such as requiring that the excitation is locally as compact and connex
as possible on the $R_{200}$ sphere. This can be done via  some form of non-linear
 band pass filter in the prior, in order to limit the effective degrees of freedom in $\tilde \bb$.

{\sl Accounting for non-linearities.}

 The non-linear solution, \Eq{reorder2} may be formally rewritten as 
 $ \ba_{2}\equiv \M{M}_{2}\cdot {\tilde\bb} \otimes {\tilde \bb}$, 
 so that the perturbative inverse reads
 \begin{equation}
 {\tilde \bb} \equiv \M{M}^{(-1)}_\lambda\cdot \M{A}- { \M{M}^{(-1)}_\lambda\cdot \M{M}_{2}\cdot
 (\M{M}^{(-1)}_\lambda\cdot \M{A})\otimes ( \M{M}^{(-1)}_\lambda\cdot \M{A}) }\,,
 \EQN{sol2inv}
 \end{equation}
(where $\M{M}_\lambda^{(-1)}$ is defined by \Eq{sol1inv}) provided the regime for the perturbative expansion applies.
If not, we may still find the best non-linear solution to the penalized likelihood problem
 of jointly minimizing $|| \M{A}- \M{M}\cdot  {\tilde \bb} - \M{M}_{2}\cdot  {\tilde \bb} \otimes
  {\tilde \bb} ||^{2}+\lambda \M{P}$, while using \Eq{sol2inv} as a starting point.

When proper motions measurements are not available (\ie we only have
access to star counts), \Eqs{defchi}{sol2inv}
still apply with some straightforward modifications, but
the conditioning of the problem
should decrease significantly, since the dynamics is less constrained.

For a data set such as GAIA, we shall have access to 
the full 6-dimensional description of phase space for some of the 
stars (via radial velocity measurements and {parallax}) or at least 5 
dimensional measurements ($\ell,b,\mu_{\ell},\mu_{b},u_{r}$).

Recall that  in practice, the
fields, $\mathbf{b}$ and $\mathbf{c}$ are respectively three-dimensional (2
angles  and time)  and 5 -dimensional (2  angles, time  and  3 velocities).
Consequently the inverse problem  is generically very ill-conditioned since
data space is  either two ($\ell, b$), four  ($\ell, b, \mu_{\ell}, \mu_b$),
five  -dimensional  ($\ell,  b,   \mu_{\ell},  \mu_b$,$\mathpi$),   or  six
{dimensional} ($\ell,  b, \mu_{\ell}, \mu_b$,$\mathpi$,$v_r$). In  fact it is
anticipated  that the  conditioning is  even poorer  because  the dynamical
evolution  involves  damped  modes,  implying an  exponential  decay ( which
corresponds  to a  major challenge  for extrapolation).  It remains  that the
weakly damped  modes should be tractable  back in time up  to some horizon,
which will depend on  the nature of the halo (via the conditioning
of $\M{M}_{\lambda}$ defined in \Eq{sol1inv}), the  volume of data, and the signal
to noise ratio in the measurements.

Let us close this discussion of the inverse problem by emphasizing 
again the  true complexity  of the implementation: \Eq{sol1inv}, and its 
non-linear counterpart, \Eq{sol2inv}, include via the dot product large sums 
over $\bn$ and integrals over $\tau$. $\M{M}$ and $\M{M}_{2}$ are functions 
of ${ \cal A}^{\MGS{\ell},\MGS{ \mu} } $ (which require a couple of integrals) and 
$A_{bb}$ \etc  which are themselves functions of $\bK_{i}$ (\Eq{defAbb})
(which involves the underlying distribution function, $F_{0}(\bI)$,
and the basis function, $\psi^{[\bn]}(\br)$
  via Eqs (\ref{e:defK1}),  (\ref{e:defK2}) and \Ep{defR1}).


Streamers (or tidal tails) in external galaxies may also be integrated backwards 
through the same procedure. It will involve the deprojection of the 
stream and of the underlying halo.



In contrast to the Galactic inverse problem, it will in principle be possible to reproduce the inversion process on a statistical set of haloes, which would allow us to compare directly to the predicted statistical properties  of the $\bb$ and the $\bc$  (though clearly the bias introduced by the penalised inversion would have to be accounted for).

This completes our rapid survey of possible applications for
the perturbative treatment of the dynamics of an open halo.

\section{Conclusion  }

In  the last  few  years,  with the  observational  convergence towards  the
concordant    cosmological   model,  a significant  fraction  of  the interest  has
shifted towards smaller  scales.  Indeed it now becomes  possible to project
down to  these scales some  of the predictions  of the model.  This  in turn
offers the prospect of transposing there what has certainly been a key asset
of  modern cosmology,  both observationally  and  theoretically: statistics.
This is a requirement both from the point of view of the (often understated)
variety of objects  falling onto an $L_{\star}$ galaxy,  but also because of
the  sheer  size of  the  configuration  space for  infall.   It  is also  a
requirement from  the point  of view of  the non-linear dynamics  within the
dark  matter halo  in order  to account  for the  relative time  ordering of
accretion events.  
was pointless to describe 
continuous infall, haloes are typically not in fully phase mixed equilibria,
and  the resulting  fluctuation spectrum  may  seed or  excite the  observed
properties of galaxies.  %

In this paper, we 
aimed at constructing a self-consistent description of dynamical issues  for dark matter  haloes embedded in a moderately active cosmic environment.  
It relied entirely on the assumption that the statistics of the 
infall is well-characterized, as described in  \citet{aubert1,aubert2}, 
and that the mass of the infalling material (or
 to a lesser extend that 
 of the fly-by) should be small compared to the mass of the halo.
 It also assumed that the halo was spherical and static or evolving 
 adiabatically (\Sec{quasi1}).
 The emphasis was  on  the theoretical framework, rather than the 
details of the actual implementation. 
 In other words, we aimed at  describing a self-consistent setting 
which allows us to propagate the cosmological environment into the core 
of galactic haloes.

In \Sec{spherical},
we derived the dynamical equations governing the linear 
evolution of the induced perturbation by direct infall or tidal excitation of 
a spherically symmetric (integrable)  stationary dark matter halo.
 The simplified geometry of the initial state allowed us to focus on the 
specificities of an open system.
Specifically, we  revisited 
the influence of the
external perturbations on the spherical halo, and extended the results 
of the literature by considering an advection term in the 
Boltzmann equation. This approach was compared to the classical 
Green solution in Appendix~\ref{s:propag}.
Note that both the intrinsic properties of the halo, via the distribution function, $F$ (\Eq{defkk}), and
the environment,  via $(s^e,\psi^e)$ (\Eq{defhh}),  of  the 
 infall and the tidal distortion were accounted for. 
Clearly the subclass of problems 
corresponding to tidal perturbations only will turn out to be easier to
implement at first.
Appendix~\ref{s:implementation} presents the details of the angle action variables on the 
sphere together with an explicit expression for the kernel, $\bK$,  and carried out 
a test case implementation of the statistical propagation of an
ensemble  of  radial excitations with a  powerspectrum scaling like $\nu^{-2}$.

\begin{figure} 
\centering
\resizebox{0.95\columnwidth}{0.95\columnwidth}{\includegraphics{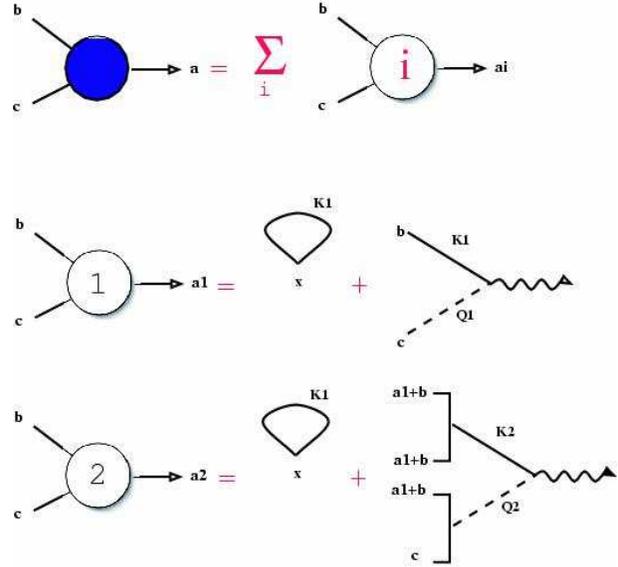}}
\caption{ diagrammatic representation
 of the expansion  to second-order given in\protect{ \Eqs{defdiag1}{defdiag2}}.
The top diagram states that one should sum over all 
orders in the coupling in order to model the non-linear
response of the halo; 
The second diagram from the top stands for \protect{  \Eq{defdiag1}} and the 
third for\protect{  \Eq{defdiag2}}. The loops correspond to the self-coupling i.e.
the self-gravity response of the halo to the perturbing flow.
The second diagram corresponds to the ``propagation'' of the 
double  excitation (see also \protect{ Appendix~\ref{s:propag}} for a discussion 
of the distribution function propagator in angle-action variables):
the input are the external potential,
$\psi^{e}$  (through 
its  $b_{\bn}$ coefficients) and the source, $s^{e}$
 (expanded over 
the $c_{\bn}$ coefficients); the output is the 
coefficient of the expansion of the inner potential. 
The coupling is achieved via the operator $K_{i}$ and 
$Q_{i}$ defined by
\protect{  \Eqs{defK2}{defQ2}} and \protect{  \Ep{defKn}},
while the contraction is achieved by \protect{ \Eq{defcontract1}},\protect{  \Ep{defcontractn} }
and is represented by the wiggly horizontal line.
  }
\label{f:diag1} 
\end{figure}

In \Sec{nonlinear}, 
we derived the non-linear response of the 
galactic halo to second-order (\Eq{defdiag2}) in the perturbation 
(and to order $\rn$ in Appendix~\ref{s:apendperturb} together with the 
corresponding N-point correlation function)
to account for tidal stripping and dynamical friction.
The dynamics was ``solved'' iteratively,
 in the spirit of the 
successful  approach initiated in cosmology by \citet{Fry} 
and considerably extended by
\citet{francis}. In particular we presented and illustrated 
a set of diagrams (\Fig{diag1},\Fip{diag2}),  each  corresponding to the contribution 
of the perturbation expansion. 
Though the actual implementation of the non-linear theory is going
to be CPU intensive, we argue that it will improve our understanding 
of the competing dynamical processes within a galactic halo.
In particular, we discussed how this {explicit theory of non-linear
  dynamics} provides the setting in which
substructure evolution (and destruction) will have to be carried, in order 
to account for \eg tidal stripping.

In \Sec{quasi1} we presented the Fokker-Planck equation
governing the quasi-linear evolution of the mean profile of the ensemble
{averaged} halo embedded in its cosmic environment. Specifically we showed
how the infall, drift and diffusion coefficients (\Eqs{D2def}{D0def}) are related to the two-point correlation of 
the tidal field, and incoming fluxes. 
Appendix~\ref{s:secularAppendix} {gives} a derivation of this equation 
from first {principles}, while in the main text, we focused on  
the bibliographic context and 
possible applications.
The key physical ingredient behind this
 secular evolution theory was  the stochastic  fluctuation 
caused by the  incoming cosmic substructures. The key technical assumption 
was that the 
two time scales corresponding to the relaxation processes and the dynamical evolution decouple.
Hence we could assume a hierarchy in time so  that the distribution function is 
constant in time when computing the polarization.

Finally, in \Sec{applications} we 
 considered in turn a few classical probes of the large-scale 
structures which had been used in the past to constrain
 the main cosmological  parameters { and } the initial power spectrum,
 which we transposed to the galactocentric context.
Note that these are built upon observables, hence 
they may be used  to constrain  the boundary power spectrum of the $a_{\bn}$.
Since  \Eq{defLensing}, \Ep{defSZ}, \Ep{defFDM}, and \Ep{defXB}  involve different
combinations of $\langle a_{\bn} a_{\bn'} \rangle$, they will constrain them at different scales with different
biases, which should ultimately allow us to better characterize the power spectrum. 
This situation is the direct analog of the cosmic situation, where the different
 tracers (weak lensing, Ly-$\alpha$ forest, CMB etc..), constrain different scales of the cosmological power spectrum (with different biases). 
Note also that  our knowledge of the statistical properties of the 
boundary (via the $b_{\bn}$ and $c_{\bn}$ coefficients) together with some 
assumptions on the equilibrium $F_{0}$ allows us to generate given realizations 
of the $a_{\bn}$ as shown in \citet{aubert1} and therefore
 virtual observables for any of these data sets, for the purpose of \eg  validating 
inverse methods. 
We investigated the consequences of the infall down to galactic scales 
and showed
how it could  be used to account for the observed distribution of disk properties
(spiral winding, warps \etc).
We demonstrated how the analytical model 
(both linear and non-linear) are
 quite useful when attempting 
to``invert'' the observations for the past accretion history a given galaxy.
This stems mainly from the fact that perturbation theory provides 
an explicit scheme for the response of the system, in contrast to the 
algorithmic procedure corresponding to N-body simulations. 

Again let us emphasize that \Eq{SOL} and its non-linear generalization \Ep{soln}
and \Ep{SOLQuasi}
yield in principle the detailed knowledge of the full {\sl perturbed} distribution (inside $R_{200}$) at later 
times. (This is to be contrasted to the situation in N-body simulations
where the response of the system is partially hidden by the
mean profile of the halo, which requires first identifying substructures (\citet{aubert0})). 
Hence we should be in a position to weigh the relative importance of the 
environment (via $s^{e}$ and $\psi^{e}$) against the inner properties of the galaxy: 
the unperturbed distribution function of the halo, $F(\bI)$, (its level of anisotropy, 
the presence of a central cusp \etc) the disk, (its mass, 
its profile, its distribution function, $F(\M{J})$ \etc).

The work presented here derives from the fact 
  that it was  realized that 
 the biorthogonal projection pioneered  by \citet{Kalnajs2} could be 
 applied order by order to the perturbative expansion of the 
 dynamical equations. Yet this in turn required the knowledge 
 of the relative phases involved in the perturbation, which 
 involves characterizing the properties of the perturber. 
 The characterization only made sense statistically in order to 
 retain the generality of the approach of  \citet{Kalnajs2}. 
 Hence the emphasis 
 on statistics.

\subsection{discussion \& Prospects}
\label{s:prospects}

Our purpose  in this paper was to address in a statistically 
representative manner 
dynamical issues on galactic scales. 
We also advocated using perturbation theory in angle-actions
in order to explicitly propagate this cosmic boundary inwards in phase space.
 As was demonstrated in the paper (and shown quantitatively in \Sec{statprogex}), this task remains
  in many respects quite challenging.
  
One of  the limitations of the above
method is the reliance on numerous expansions combined
 to  the special care required in their implementation.
One could argue that this level of sophistication might not be 
justified in the light of the weakness of some of the assumptions.
Indeed, we are limited to systems with spherical geometry 
 whereas galaxies most likely come in
a variety of shapes. This assumption could be lifted provided we compute 
the modified actions of the flattened spheroidal equilibrium using perturbation 
theory for the equilibrium 
 in the spirit of \citet{binney}, but implies a higher level of complexity; (it 
 would also require statistically specifying the orientation of the halo relative to the infall,
 as discussed in part in \citet{aubert0}).
We assumed here that the perturbation was 
  relatively light, which excludes a fraction of cosmic event
  which might dominate the distribution of some of the observables.

\Sec{Universal}, \Sec{applications}  and Appendix~\ref{s:AppendAppli} presented
 a few possible applications for the 
 framework described here and in \citet{aubert1}.
 These galactic probes would need to be further investigated,
 in particular in terms of observational and instrumental 
  constraints. The biasing specific to each tracer should be 
  accounted for. 
   The second-order perturbation theory needs to be implemented  in practice together 
  with the diffusion coefficients of \Sec{quasi1}, following \Sec{implementation}
  and extending \Sec{statprogex}. Similarly, 
  the identification and evolution of substructures within the halo
  mentioned in  \Sec{satcount} deserves more work.
  {In \citet{aubert0},
 we showed that the accretion onto L$^{\star}$ haloes was
 anisotropic; the dynamical implication of this anisotropy 
 will require some specific work in the future.
 }

We will need to demonstrate against N-body simulations 
the relevance of perturbation theory for dynamical friction;
in particular, we should explore the regime in which the second-order
truncation is appropriate, and at what cost ?
Note that truncated perturbation theory implies that
modes will ring forever. At some stage, one will therefore have to address the 
problem of energy dissipation.

Implementing a realistic treatment of the infalling gas will certainly be amongst 
the more serious challenges ahead of us. 
This is a requirement both from the point of view of the dynamics but also 
from the point of view of converting the above predictions into baryon-dependent
 observables.  The description of the 
 gas will require a proper treatment of the various 
 cooling processes, which can be quite important on galactic scales.
In particular, the thickening of galactic disks is 
most likely the result of a fine-tuning between destructive processes
such as the tidal disruption of compact substructures on the one hand,
 and the  adiabatic coplanar infall of cold gas within the disk.
In fact, the nonlinear theory presented in \Sec{nonlinear} and 
\Sec{quasi1} could  be extended to the geometry of disks
to account for the adiabatic polarization towards the plane of the disk.

Note  that we  assumed here  that  transients corresponding  to the  initial
conditions where damped out so that  the response of the system was directly
proportional to the excitation.  The  underlying picture is that of a calmer
past, which in fact is very much in contradiction with both our measurements
and common knowledge on the more violent past accretion history of galaxies.
Indeed, infalling subclumps will contribute via the external tidal potential
at  some earlier  time, and  the larger  the lookback  time,  the relatively
stronger the importance  of the perturbation (since the  intensity of infall
is  in fact  an increasing  function of  lookback time).   We  are therefore
facing   a  partially   divergent  boundary   condition.   Because   of  the
characteristics of hierarchical clustering,  the actual bootstrapping of the
analytical framework  is therefore challenging.  This could be a  problem in
particular for  non linear  dynamics, where the  coupling of  transients may
turn out to be as important as the driven response.  The importance of these
shortcomings will need to be addressed in the future.


Finally  let   us  note  that  the  theory   described  in  \Sec{spherical},
\Sec{nonlinear}  and  \Sec{quasi1} describe  perturbative  solutions to  the
collisionless Boltzmann  Poisson equation in angle-action  variables, and as
such are  not specific to the  description of dark matter  haloes.  It could
straightforwardly be  transposed to other situations  or geometries provided
the system  remains integrable.  As  mentioned in \Sec{source},  the stellar
dynamics around a massive black hole  would seem to be an obvious context in
which  this  theory  could be  applied.   For  instance,  we might  want  to
investigate the capture of streams of  stars by an infalling black hole.  In
a slightly  different context, note in  passing that the  above theory could
also  be applied  to celestial  mechanics, since  an  angle-action expansion
corresponds to an all eccentricity scheme.

Let us  close this paper  by a summary  of the pros  and cons of  the theory
presented here.
\vskip 0.2cm
\vskip 0.2cm
\parn
{\it Possible assets:}
\vskip 0.2cm
\begin{itemize}
\item fixed boundary: localized statistics;
\item fluid description : no  {\it a priori} assumption on the possibly time- dependent nature of the objects;
\item non-linear explicit treatment of the dynamics: proper account of the self-gravity of incoming objects and statistical  accounting of causality;
\item dynamically-consistent statistically-representative treatment of the cosmic environment;
\item customized description of resonant processes within the halo via angle action variables of universal profile;
\item ability to construct one- and two-point statistics for a wide range of galactic observables.
\item theoretical framework for dynamical inversion and  secular evolution
\end{itemize}

\vskip 0.2cm
\parn
{\it Possible drawbacks:}
\vskip 0.2cm
\begin{itemize}
\item weak perturbation w.r.t. spherical stationary equilibrium: not representative of \eg equal mass mergers;
\item complex time dependent 5D boundary condition;
\item {\it ad hoc} position of the boundary;
\item no obvious truncation of  two-entry perturbation theory;
\item no account of baryonic processes;
\item inconsistency in relative strength of merging events versus time;
\item non-Gaussian environment probably untractable;
\item finite temporal horizon given finite $\ell_{\rm max}$;
\item no statistical accounting of linear instabilities.
\end{itemize}

\vskip 1cm 
\parn{\handfont Acknowledgments}

{\sl  We   are  grateful  to   E.  Bertschinger,  J.~Binney,   S.   Colombi,
J.~Devriendt,    J.~Heyvaerts,   A.~Kalnajs,    J.~Magorrian,   D.~Pogosyan,
S.~Prunet,  A.~Siebert, S.~Tremaine, \&  E.~Thi\'ebaut for  useful comments
and  helpful  suggestions.  We are especially grateful to J. Heyvaerts for carefull reading
of the manuscript and
for introducing us to the  quasi-linear formalism. We thank the 
anonymous referee for constructive remarks. Support   from  the  France-Australia  PICS  is
gratefully  acknowledged.  DA thanks  the Institute  of Astronomy  and the MPA for their
hospitality and funding from a Marie Curie studentship.  }

\bibliographystyle{mn2e}
\bibliography{pichon_aubert2005}

\appendix

\onecolumn

\section{Linear propagator in action angle }
\label{s:propag}
As mentioned in \Sec{propagM}, it is useful to regard the open {collisionless} 
system 
 as a segmentation of the
source for the propagator (ie the Green function of the coupled Poisson
Boltzmann equation), where one distinguishes two contributions for the initial
distribution: the contribution at $R_{200}$ with $v_r < 0$ (what we describe
as the source term in \Eq{boltzsphere}) and the contribution beyond $R_{200}$ or at
$R_{200} $ with $v_r > 0$ (what we describe as the tidal field in the main text).
 In order to make this comparison, let us derive
generally (without any reference to  a boundary for now) the
Green function satisfying the {\sl linearized}  Boltzmann Poisson equation.
Let us call
$G(\tmmathbf{w},\tmmathbf{I},t|\tmmathbf{w'},\tmmathbf{I}',t')$ this Green function; it obeys:
\begin{equation}
\frac{\partial G}{\partial t} + \MG{\omega} \cdot \nabla_{\tmmathbf{w}} G
   + \frac{\partial F}{\partial \tmmathbf{I}} \cdot \nabla_{\tmmathbf{w}} \int
   \frac{\textrm{d} \tmmathbf{r}''}{| \tmmathbf{r}'' - \tmmathbf{r} |} d
   \tmmathbf{v}'' G ( \tmmathbf{r}'', \tmmathbf{v}'', t| \tmmathbf{w}',
   \tmmathbf{I}', t' ) = \delta_{{\rm D}} ( \tmmathbf{w} - \tmmathbf{w}' )
   \delta_{{\rm D}} ( \tmmathbf{I} - \tmmathbf{I}' ) \delta_{{\rm D}} (
   t - t' )\,,
   \end{equation}
   so that the distribution function at $(\tmmathbf{I},\tmmathbf{w},t)$ reads
\begin{equation}
f ({\tmmathbf{w},\tmmathbf{I},t)} = \int \d t' \int
   \d \M{w}' \mathrm{} \int \d \tmmathbf{I}' \mathrm{}
  {G( \tmmathbf{w},\tmmathbf{I},t|\tmmathbf{w}',\tmmathbf{I}',t')}
   f ({\tmmathbf{w'},\tmmathbf{I}' ,t')} \,. \EQN{defpropgreen}
   \end{equation}
   Let us define the linear propagator, $U_{\omega, \bk,\bw_{0}} ( \tmmathbf{I} |
\tmmathbf{I}' )$, as:
\begin{equation}
U_{\omega, \bk,\bw_{0}} ( \tmmathbf{I} | \tmmathbf{I}' ) =
   \frac{\delta_{{{\rm D}}} ( \tmmathbf{I} - \tmmathbf{I}'
   )}{\tmmathbf{k} \cdot  \MG{\omega} - \omega} - \sum_{\bn, \bn'}
   \frac{ \partial F  }{\partial \tmmathbf{I} } \cdot \tmmathbf{k} 
   \frac{\psi_{\tmmathbf{k}}^{[ \bn ]} ( \tmmathbf{I} )}{( \tmmathbf{k} \cdot
   \MG{\omega} - \omega )} \left( ( \tmmathbf{1} - \hat \tmmathbf{K} [ \omega
   ] )^{- 1} \right)_{\bn, \bn'} \sum_{\bk'} \frac{\psi_{\tmmathbf{k}'}^{[ \bn' ]} (
   \tmmathbf{I}' )}{( \tmmathbf{k}' \cdot \MG{\omega'} - \omega )}\,
   \exp(\imath \bw_{0}\cdot [\bk -\bk']) \,,
 \EQN{defpropagator}
 \end{equation}
so that the distribution function, $f ( \tmmathbf{I}, \tmmathbf{w}, t ),$ at
time $t$, with action $ \tmmathbf{I}$ and angles
$ \tmmathbf{w}$ induced by the propagation of the distribution at earlier time
$t'$, with action $\tmmathbf{I}'$, and angles $\tmmathbf{w}'$ reads
\begin{equation}
f ( \tmmathbf{I}, \tmmathbf{w}, t ) = \int {\d t}' \int \d \tmmathbf{I}'
   \int \d \tmmathbf{w}'  \sum_{\tmmathbf{k}} \int \d \omega \exp ( \imath
   \omega [ t - t' ] - \imath \tmmathbf{k} \cdot [ \tmmathbf{w} -
   \tmmathbf{w}' ] ) U_{\omega, \tmmathbf{k},\bw_{0}} ( \tmmathbf{I} | \tmmathbf{I}' )
   f ( \tmmathbf{I}',  \tmmathbf{w}', t' ) \EQN{defprop}\,.
\end{equation}
It is interesting to
 contrast \Eq{defpropagator} to the propagator found by \citet{ichimaru}
for the uniform plasma. 
In particular, the gradient of the density profile breaks the 
stationarity in $\bw-\bw_{0}$ of the propagator, \Eq{defpropagator}.
Note also that the first term on the r.h.s. of \Eq{defpropagator} corresponds to free streaming inside the halo
(\ie dark matter particles describing their unperturbed orbits), and reads in real space
\begin{eqnarray}
  G_{\rm {{free}}}{\mathrm{} }
 {(\tmmathbf{I},\tmmathbf{w},t|\tmmathbf{I}',\tmmathbf{w}',t')}
  = \sum_{\tmmathbf{k}} \int \d \omega \exp ( \imath \omega [ t - t' ] - \imath
  \tmmathbf{k} \cdot [ \tmmathbf{w} - \tmmathbf{w}'] ) \frac{\delta_{{{\rm D}}} (
  \tmmathbf{I} - \tmmathbf{I}' )}{\tmmathbf{k} \cdot 
  \MG{\omega} - \omega}  
  = \delta_{{{\rm D}}} ( \tmmathbf{I} - \tmmathbf{I}' )
  \delta_{{{\rm D}}} ( \tmmathbf{w} - \tmmathbf{w}' -
  \MG{\omega} [ t - t' ] ) \,,&  &  \nonumber
\end{eqnarray}
while the second term in \Eq{defprop}  corresponds to the self-gravitating polarization
of the halo induced by the perturbation.
 Note that 
since the field dynamical equation is solved with a right-hand side (\ie 
a source breaking the mass conservation in phase space),
 Liouville's theorem is not obeyed anymore: a new fluid is injected 
 into the halo. 
We may now assume that in \Eq{defpropgreen}, 
$f(\br',\bv',t')=f(\bw',\bI',t')$ is split in two: one contribution from
dark matter particles exiting $R_{200}$ or beyond $R_{200}$; 
another contribution describing particles on $R_{200}$ with negative 
radial velocity.  The former component may then be resumed over the 
corresponding region of  phase space with a $1/|\br -\br'| $ weight, and 
yields $\psi^{e}$. 
The latter corresponds to $s^{e}(t')$.

\section{Implementation }
\label{s:implementation}
Let us describe  in this appendix in greater details how \Sec{spherical} are 
implemented in practice, while focusing here on a simple isotropic halo (\ie $F(\bI)=F(E)$,
where $E= v^{2}/2+\Psi(r)$ is the energy, and $\Psi(r)$ the unperturbed 
potential).
We will show here how to compute the  operator, $\M{K}$, (defined by \Eq{defkk})
and elements of  $\bQ$  (defined by \Eq{defhh}),
for the corresponding basis, following e.g. \citet{Tremaine}, \citet{murali}, \citet{Seguin}. 
We will then implement in practice the average induced correlation triggered by some 
ad hoc colored radial perturbation.  
Similarly, one could compute the non-linear coefficients, $\llbracket\,\,\,\, \rrbracket$ entering 
\Eq{defa22}, but the implementation of the non linear formalism of 
\Sec{nonlinear} and \Sec{quasi1} are beyond the scope of this paper.

\subsection{Detailed angle-action linear response for isotropic spheres }


The three-dimensional nature of  galactic halo makes the implementation slightly more 
complicated than one would think at first sight. The assumption that the halo is spherical allow
us to assume that the equilibrium is integrable. Hence the action space is effectively at most two-dimensional,
but configuration space remains three-dimensional (though one angle is mute).
 In practice, this implies that integration over action space, occurring in \eg \Eq{defK1}
 is effectively two-dimensional. On the other hand, the sum over ${\bk}$ involves three indices,
 each corresponding to a degree of freedom.  



Let us define $I_1$ as the radial action, $I_2\equiv L$ as the total angular momentum and
$I_3\equiv L_z$ as the $z$-component of the angular momentum, so that 
\[
I_1=\frac{1}{ \pi }\int_{r_p}^{r_{a}}\d r
	\sqrt{2[E-\Psi(r)]-I_2^2/r^2}\, .
\]
Here $r_{a}$ and $r_{p}$ are respectively 
the apoapses and periapses of dark matter particles. 
This defines $\bI\equiv(I_{1},I_{2},I_{3})$ introduced in \Sec{Boltzmann}.
Similarly, 
Let us define  the corresponding angles, $\bw\equiv(w_{1},w_{2},w_{3})$
 (see \Fig{angles}) given by:
\begin{equation}
 w_1 = \omega_1 \int^r_{r_p(\bfI)} \frac{{\d r}}{\sqrt{2 [ E - \Psi ( r ) ] - I^2 /
   r^2}} \,, \quad
   w_2(\bfI,w_{1})=\chi -\int_{r_p(\bfI)}^{r(\bfI,w_1)}{\d r(\omega_2-I_2/r^2)\over
	\sqrt{2[E-\Psi(r)]-I_2^2/r^2}}\,,
 \quad w_3 = \phi - {\rm asin} ( \cot ( \beta ) \, \cot ( \theta
   ) ) \,, \EQN{defw123}
\end{equation}
where $\cos \beta=L_{z}/L$. 
\subsection{Computing the linear response operator}
\label{s:compLIN}
%
Following {\sl very} closely the notation of \citet{murali}, let us introduce 
a bi-orthogonal basis constructed around spherical harmonics:
\begin{equation}
\rho(\br,t)=\sum_{\ell m n} a_{\ell m n}(t) d^{\ell m}_n(r) Y_{\ell m}(\bO)\,, \quad{\rm and }\quad
\psi(\br,t)=\sum_{\ell m n} a_{\ell m n}(t) u^{\ell m}_n(r) Y_{\ell m}(\bO)\,, \EQN{harmexp}
\end{equation}
for respectively the density and the potential. 
\citet{Weinberg89} suggests the following potential-density pair
\begin{equation}
u_n^{\ell m}(r)=-\frac{4\pi G \sqrt 2}{\alpha_\bn |j_\ell(\alpha_\bn)|} R^{-1/2} j_\ell(\alpha_\bn r/R), \quad {\rm and } \quad 
d_n^{\ell m}(r)=-\frac{\alpha_\bn \sqrt 2}{|j_\ell(\alpha_\bn)|} R^{-5/2} j_\ell(\alpha_\bn r/R),\EQN{harmexp2}
\end{equation}
where $j_\ell$ stands for the spherical Bessel function and where $\alpha_\bn$ obeys the relation $\alpha_\bn j_{\ell-1/2}(\alpha_\bn)=0$.  Here $R$ is the truncation radius of the basis. 
\citet{hernos} suggest another set of (non-normalised) biorthogonal functions defined by~:
\begin{equation}
u_n^{\ell m}(r)=-\frac{-r^\ell}{(1+r)^{2\ell+1}}\sqrt{4\pi}C_n^{2\ell+3/2}(\xi), \quad {\rm and } \quad 
d_n^{\ell m}(r)=\frac{K_{n\ell}}{2\pi}\frac{r^{\ell-1}}{(1+r)^{2\ell+3}}\sqrt{4\pi}C_n^{2\ell+3/2}(\xi),\EQN{harmexp3}
\end{equation}
where $K_{n\ell}=n/2(n+4\ell+3)+(\ell+1)(2\ell+1)$, $\xi=(r-1)/(r+1)$ and $C_n^\ell(x)$ stand for ultraspherical polynomials. 

The action angle transform of the potential basis is given by :
\begin{equation}
 W_{\bfk}^{\ell n}(\bfI)\equiv \psi_{\bk}^{\bn}(\bI)={1\over 2\pi}
	\int_{-\pi}^{\pi}\d w_1\exp(-\imath k_1 w_1) u_n^{\ell k_3}(r)
	\exp[\imath k_2(\chi-w_2)],
\end{equation}
We may now rewrite \Eq{defkk} as 
\begin{equation}
K^{\bn'}_{\bn}(\tau-t)= -\delta^{\ell}_{\ell'}\delta^{m}_{m'}{(2\pi)^3\over 4\pi G}\iint \d E{L\d L\over\omega_1}
	\drv{F}{E} \sum_{\bf k} C_{\ell k_2} 
	\imath \bfk\cdot\bo \exp[\imath\bfk\cdot\bo(\tau-t)]
	W_{\bf k}^{*\ell n'}(\bfI)\biggl[W_{\bfk}^{\ell n}(\bfI)+
	{4\pi\over 3}\delta_{\ell}^{1} p_n^{\ell m} X_{\bfk}(\bfI)\biggl] ,
\label{e:kernel2}
\end{equation}
where
\begin{equation}
p_n^{1 m}=\int \d r r^2 d_n^{1 m}(r) \pdrv{\Psi}{r}, 
\end{equation}
and where
\begin{equation}
C_{\ell k_2} =
{2^{2k_2-1}\over \pi^2}
	{(\ell-k_2)!\over(\ell+k_2)!}\frac{\Gamma^2[\half(\ell+k_2+1)]}
	{\Gamma^2[\half(\ell-k_2)+1]}\,,\quad  \textrm{if  \quad $\ell+k_2 \quad{\rm even}$, else \,\,\,\ 0 }\,.
\end{equation}
Here $\Gamma$ is the standard Gamma function. 
Note that  $X_{\bfk}(\bfI)$ accounts for the fact that the response  is computed in a non
inertial referential frame. To take into account the barycentric drift of the halo, the perturbed Hamiltonian
 should include the induced inertial potential $\M{a}_b \cdot \br$, where 
 $\M{a}_b$ is the acceleration of the barycenter in the  frame of the unperturbed halo.
  Its action-angle transform is given by :
\begin{equation}
X_{\bfk}(\bfI)={1\over 2\pi}\int_{-\pi}^{\pi}\d w_1\exp(-i k_1 w_1) r
	\exp[ik_2(\psi-w_2)].
\end{equation}
As can be seen from \Eq{kernel2}, this inertial contribution is limited to the dipole component ($\ell=1$) of the response : as expected, it is equivalent to a spatially homogeneous field force.\footnote{
Technically speaking, the $\delta_{\ell  1}$ dependence arise from the fact that $\br$ is expressed as a function of $Y_{1m}(\bO)$ spherical harmonics.}


\twocolumn
\subsection{Implementation and validation}
The actual computation of the linear response of the halo to a tidal field is a two-steps procedure. First the kernel $\bK$ must be computed via \Eq{kernel2}. It involves an integration over the orbits' space and requires to Fourier transforms the the biorthogonal basis ($W$ and $X$ quantities) along orbits. It can be done by "throwing orbits" in the equilibrium potential and finding the associated sets of $(\bI,\bo)$ in the halo's model: such a procedure provides the angle dependance of the basis' functions for a given action. Knowing $W(\bI),X(\bI),\bo(\bI)$ over a given sampling of the $\bI$ space, \Eq{kernel2} can be computed. In order to achieve high computing efficiency and accurate responses, we implemented the calculation of \Eq{kernel2} in a parallel fashion, where the integrals in each-subspace of the action space are computed by a different processor.

Second, the expansion $\ba(t)$ of the halo's response is computed either by iteration or by means of a Volterra's Equation solver (e.g. \citet{NR}). We found that both methods give very similar results and differ only by their time consumption. The iterative method can be very fast if a proper initial guess is available but if it is not the case it may take a significant amount of time to achieve convergence. Conversely, the Volterra solver's time consumption is fixed for a given time resolution.

\begin{figure} 
\centering
\resizebox{0.95\columnwidth}{0.95\columnwidth}{\includegraphics{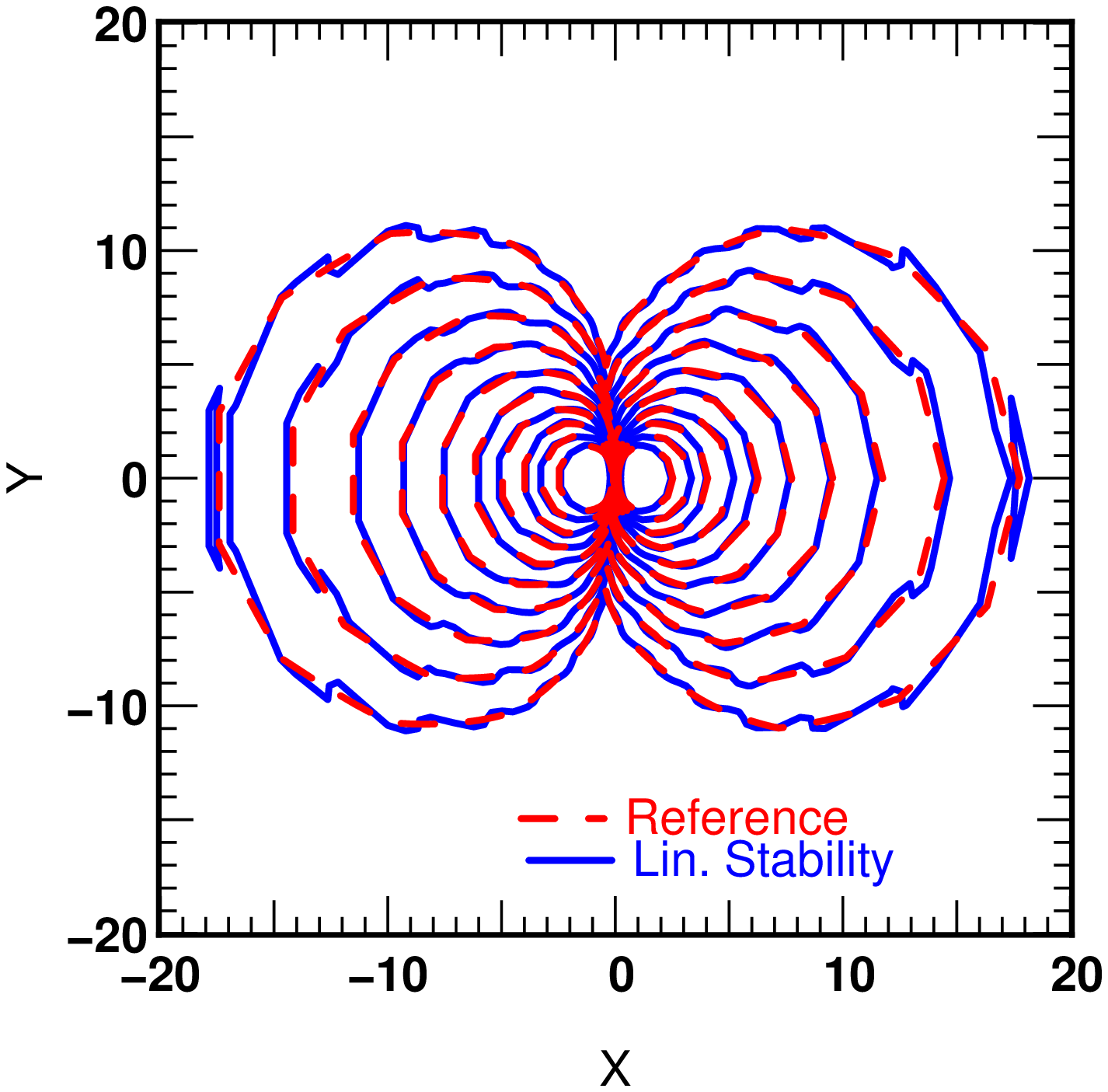}}
\caption{Isocontours in the X-Y plane of the potential's response of a Plummer Sphere embedded in a homogeneous force field (see text for details). The force field is aligned along the X axis. The dashed line stand for the prediction and the solid line stands for the linear calculation presented in this paper.}
\label{f:drift} 
\end{figure}

In order to validate our implementation, we set up two tests. The first one is  suggested by \citet{Weinberg89}. A Plummer's halo is embedded in an homogeneous force field and should experience a global drift decribed by the potential's response:
\begin{equation}
\psi(\br,t)=-\br_b(t)\cdot\nabla\Phi(\br), 
\label{e:drift1}
\end{equation}
where $\br_b$ stands for the barycenter position and $\Phi$ stands for the equilibrium potential. We chose the force field to have a $a_0\sin(\nu t)$ time dependance with $a_0=0.01$ and $\nu=0.01$. The Plummer model has a unit mass $M$ and caracteristic radius $b$. The response was computed using a $60\times60$ sampling in $(E,L)$ and 20 radial terms of the basis given by \Eq{harmexp2}. We switched off the drift compensation modelised by the $X$ term in \Eq{kernel2}.  \Fig{drift} shows the response computed at $t=10$ (in units of $\sqrt{b^3/GM}$) along with the prediction given by 
\Eq{drift1}. Clearly the two responses coincide, providing a first validation of our implementation.

\begin{figure} 
\centering
\resizebox{0.95\columnwidth}{0.95\columnwidth}{\includegraphics{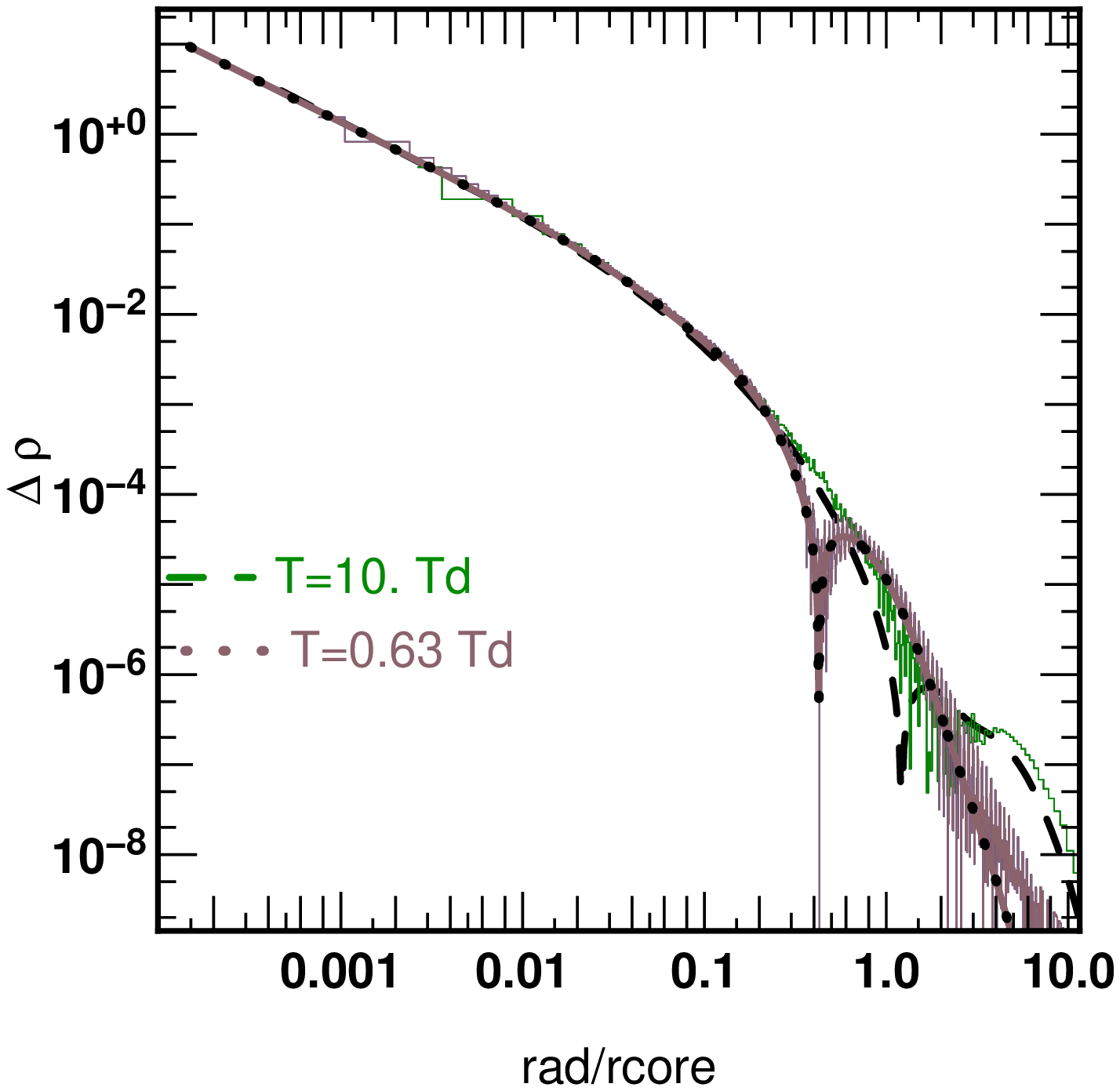}}
\caption{The radial profile of the density response of a Hernquist's halo due to a central pertuber. Lines stand for the linear response of the halo as the central pertubrer grows over a $t=0.63t_{\rm d}$ (dotted line) and $t=10t_{\rm d}$ timescale (dashed line). $t_{\rm d}$ is the dynamical time of the main halo within its core radius. Radii are given in units of the main halo's core radius. Density units are in code units. Superimposed are the calculations of the same response using a perturbative particle simulation (Magorrian, private communication). No scaling has been applied and the two methods agree quantitatively.}
\label{f:adiab} 
\end{figure}

A second test involves reproducing the contraction of a Hernquist's halo induced by a central spherical mass (which would model the presence of a galaxy for example). This central mass is assumed to follow a Hernquist's profile, whose potential is given by:
\begin{equation}
\Phi(r)=-\frac{GM}{r+a}.
\end{equation}
The halo has a unit mass M and caracteristic radius a, while the central object has a final mass of $m_p=0.001$ and $a_p=0.25$ as a constant caracteristic radius. The perturber is turned on at $t=0$ and follows a $m_p(t)=m_p(t_f)(3(t/t_f)^2-2(t/t_f)^3)$ temporal evolution, where $t_f$ is the final time step. We compare the linear response at $t=t_f$ with the simulation of the same test-case using a perturbative particle code (Magorrian, private communication). The response was computed using a $60\times60$ sampling in 13 subregions of the whole $(E,L)$ space and 21 radial terms of the basis given by \Eq{harmexp3}. Self gravity of the response is {\sl not} taken in account in both methods. \Fig{adiab} shows the comparisons between the two type of calculations, made for two different growing time $t_f$. Clearly the two methods are in good agreement. One can see that matter is dragged toward the center and the longer it takes to grow the perturber, the further from the center are the affected regions.

\subsection{Statistical propagation: a test case}
\label{s:statprogex}
In this section, we compute the two-point statistics of a halo responding to a simple type of tidal perturbation as an illustration of statistical propagation. Without any assumption on the type of perturbation, we recall that the two-point statistics of the halo's response is given by \Eq{correlA} and can  be derived directly from the perturbations' statistics. Let us simplify the correlation's computation by assuming that the halo is only tidally perturbed, so that \Eq{correlA} reduces to :
\begin{eqnarray}
   \left\langle {\bf \hat a}\cdot {\bf \hat a}^{*\top}  \right\rangle  
   = \big\langle 
\left[{\bf \hat 
    K} \cdot  {\bf \hat  b}\right]\cdot  ({\b1} 
-{\bf \hat 
    K} )^{-1 }\cdot 
  ({\b1} -{\bf \hat K} )^{-1 *\top} \cdot 
  \left[{\bf \hat K} \cdot  {\bf \hat  b}\right]^{\top *}  
\big\rangle.
    \label{e:correlsimple}
\end{eqnarray}
Furthermore, let us also (rather crudely) assume that the tidal field is  monopolar, and has a radial dependence equals to the $N$th element of the  radial basis which diagonalize the Poisson equation. Then, the tidal perturber's coefficient can be written as :
\begin{equation}
 b_{\ell m}^n(t)=b(t)\delta_{nN}\delta_{\ell 0}\delta_{m 0},
\end{equation}
where the perturbing tidal field is described by~:
\begin{equation}
\psi^e(r,\bO,t)=b(t)u_{00}^N(r).
\end{equation}
Since no radial coupling occurs, the halo's response can be simply written as~: 
\begin{equation}
\psi(r,\bO,t)=a(t)u_{00}^N(r).
\end{equation}
Consequently, the only remaining degree of freedom is the temporal variation of the tidal field.
 If we consider  an \textit{ensemble} of tidal environments and if we assume stationarity and gaussianity of the induced perturbations, it will be described by the temporal two-point correlation function of $b(t)$ coefficients, or equivalently by their temporal power spectrum $P_b(\nu)$:
\begin{equation}
P_b(\nu)=\langle{\hat b}(\nu){\hat b^*}(\nu)\rangle,
\end{equation}
where $\nu$ stands for the frequency.
If the temporal power spectrum of the response is given by $P_a(\nu)$ then \Eq{correlsimple} reduces to:
\begin{equation}
P_a(\nu)=\langle{\hat a}(\nu){\hat a^*}(\nu)\rangle=\frac{|{\hat K}^{00}_{NN}(\nu)|^2}{|1-{\hat K}^{00}_{NN}(\nu)|^2}P_b(\nu).
\label{e:propasimple}
\end{equation} 
\Eq{propasimple} simply states that the frequency structure of the 'tidal noise' is transmitted to haloes via a (scalar) transfer function given by the response kernel.

\begin{figure} 
\centering
\resizebox{0.95\columnwidth}{0.95\columnwidth}{\includegraphics{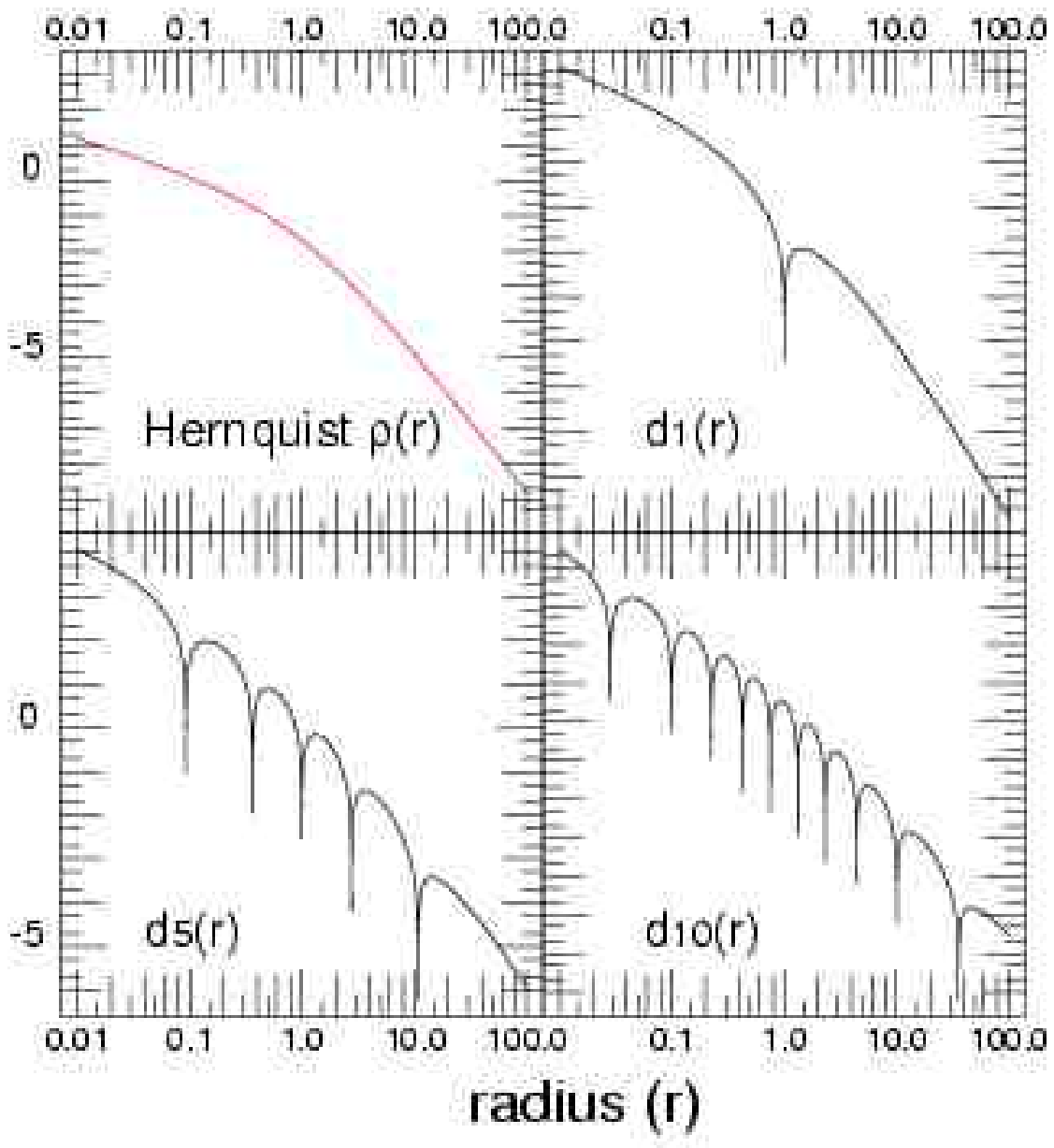}}
\caption{The halo's density profile chosen for the statistical propagation's example follows a hernquist model (top-left panel). We apply a monopolar tidal field $\psi^e(r)$ with a radial structure given by the $d^N(r)$ function of the Hernquist and Ostriker biorthogonal basis. Here are shown the corresponding density profiles $\rho^e(r)=\Delta\psi^e/4\pi G$ for $N=1,5,10$. }
\label{f:powavg} 
\end{figure}

\begin{figure} 
\centering
\resizebox{0.95\columnwidth}{0.95\columnwidth}{\includegraphics{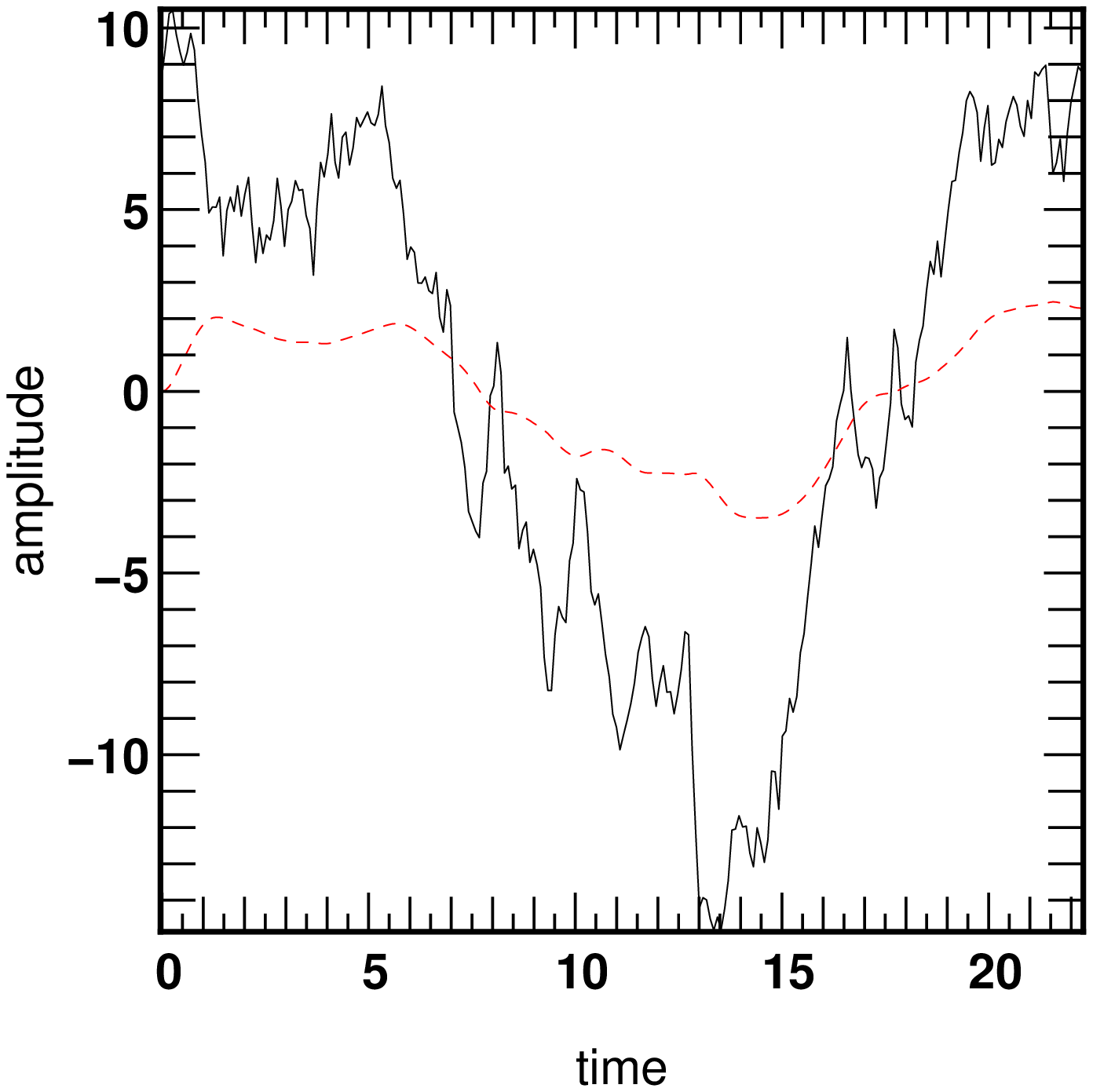}}
\caption{An example of time evolution of the tidal field's amplitude $b(t)$ (plain line). Its power spectrum $P_b(\nu)$ follows a $\nu^{-2}$ law. Assuming an N=2 radial dependance, the amplitude $a(t)$ of the induced halo's response can be computed (dashed line). The halo does not respond to high frequency features and globally its response is slightly delayed, reflecting its own inertia.}
\label{f:exresponse} 
\end{figure}

\begin{figure} 
\centering
\resizebox{0.95\columnwidth}{0.95\columnwidth}{\includegraphics{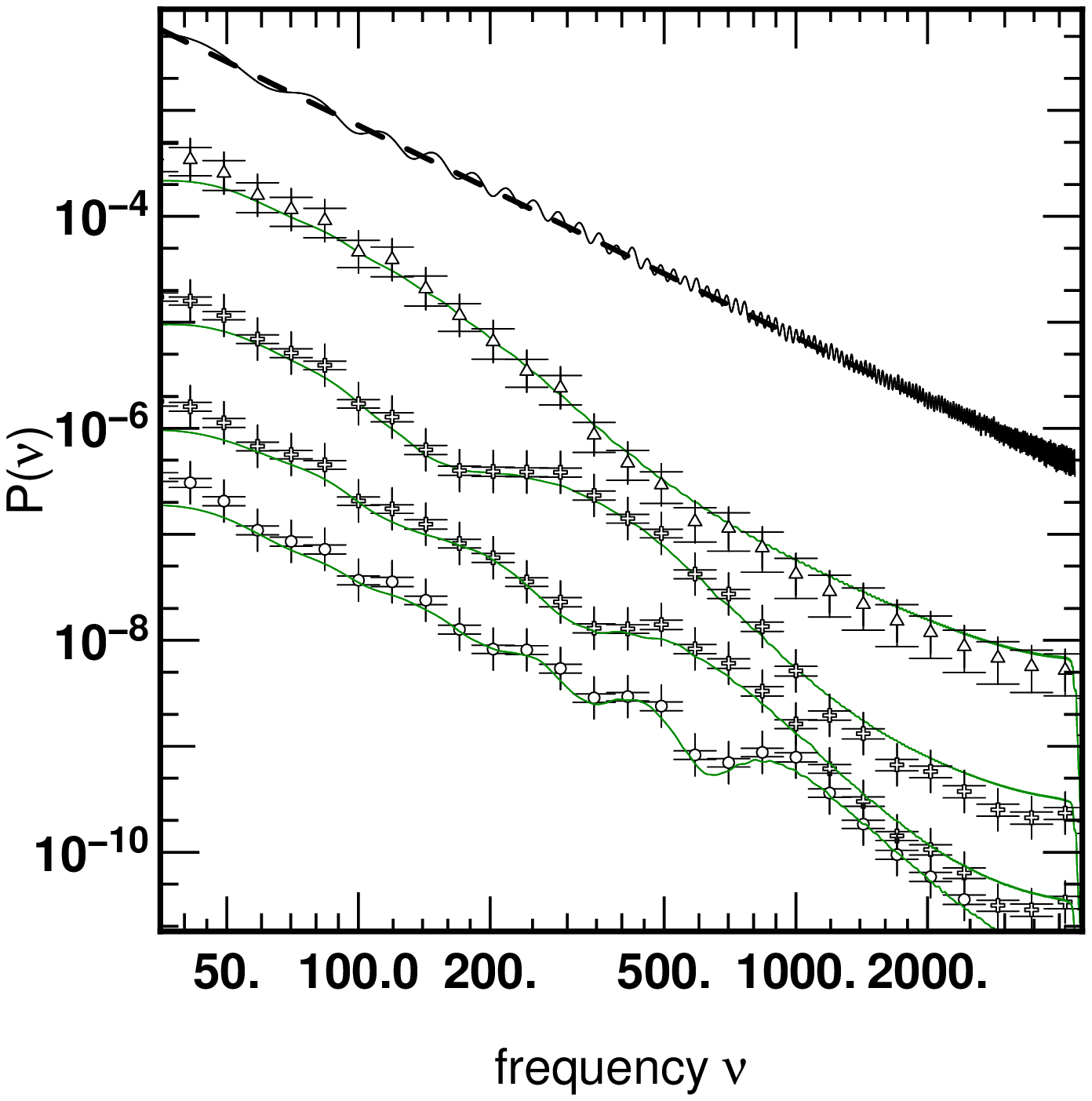}}
\caption{An example of statistical propagation. The average power spectrum of the 1000 tidal perturbations applied to the Hernquist halo (top curve) follows a $\nu^{-2}$ law (dashed thick curve). Symbols stand for power spectrum of the halo's response averaged over the 1000 realisations of $\psi^e(r,t)$  with four different radial dependences (with $N=1,3,5,10$, from top to bottom).  The superimposed curves show the direct predictions on the power spectra, following \Eq{propasimple}. For clarity these curves have been divided respectively by 1, 20, 40 and 70.  Frequencies are in code units.}
\label{f:simplebasis} 
\end{figure}

Let us further describe our test halo  again  by a Hernquist's model (\citet{hern}). The corresponding kernel is computed following the procedure described by  \Sec{compLIN}
 using the \citet{hernos} potential-density pair (see \Fig{powavg}). Further
 details can be found in \citet{murali}, \citet{Seguin}.
 The radial dependence of the tidal perturber is given by the N-th potential function $u_N(r)=u_N^{00}(r)$ of the basis described by \citet{hernos}. The associated density function is given by $d_N(r)=d_N(r)^{00}=\Delta u_N(r)/4\pi G$ and examples of such profiles are given in \Fig{simplebasis} along with the halo's profile. For simplicity, the tidal frequency distribution has been chosen to follow a power law:
\begin{equation}
P_b(\nu)\sim\nu^{-2}.
\label{e:powsimple}
\end{equation}
This power law describes the \textit{ensemble} frequency behavior and therefore a single realisation of the tidal noise may deviate from this relation as long as statistical convergence is achieved. \Fig{exresponse} shows both an example of the time dependence of such a perturber and the time dependence of the induced response. One can see that the halo acts a low pass frequency filter and do not recover all the high frequency features present in the tidal field. Also, the halo response appears as delayed in time, reflecting the effect of the halo's own inertia.

The same computation was performed for an ensemble of 1000 different tidal perturbations. \Fig{powavg} shows the power spectrum $P_b(\nu)$ averaged over all the realisations along with $P_a(\nu)$  averaged over the 1000 haloes' responses (shown as symbols with error bars). $P_b(\nu)$ departs from a power law at low frequencies ($\nu<50$ in code units) because of the finite time range over which the tidal field is applied (not shown here). At higher frequencies, the perturbers' frequency distribution follows exactly \Eq{powsimple}. Independantly, $P_a(\nu)$ is directly predicted from $P_b(\nu)$ using \Eq{propasimple}, without relying on the computations of  individual responses, and shown on the same plot as solid lines.  Clearly the predicted power spectrum of the response matches the statistically averaged one and even reproduces `bumpy' features seen at various frequencies. The filtering effect of the halo response can still be seen in the predicted spectra~:  $P_a(\nu)$ follows the $\nu^{-2}$ law at low frequencies but exhibits a steeper slope at higher frequencies. This cut-off effect is more important for large scale perturbations (low N) and reflects the fact that perturbations at high frequencies are unable to `resonnate' efficiently with the halo's large scale modes. Conversely, tidal perturbations with features on small spatial scales (large N) are more likely to induce large frequencies and preserve the frequency structure of the perturbation. Moreover, the  `bumps'  seen in the $P_a(\nu)$ curves reflect the eigen frequencies of the halo.  The pertuber's  scale-free spectrum hits resonnances which react in a stronger fashion than any other frequency. Again, these resonances occur at larger frequencies as the radial order N increases~: smaller radial scale perturbations relate to shorter caracteristic time scales.

The illustration presented in this section is admittedly simplistic but hints at the possibilities which can be foreseen for statistical propagation~: for a given set of  constrained environment, predictions on the statistics of the induced response can be made without relying on the computation of individual realisations. Predictions on spatial or spatio-temporal correlations of the halo's response can be made following the same procedure. It will possibly allow us to study the impact of the different scales of accretion  or potential, the influence of the rate of change of these pertubrations and their relative relevance on the statistical properties of matter within the halo as  discussed in \Sec{applications}.

\onecolumn

\section{Secular evolution with infall}
\label{s:secularAppendix}

Let us derive in this appendix the secular equation for the 
evolution of the ensemble average halo embedded in a  
typical cosmic environment (with infall and tidal field). 
This follows the pioneering work of  \citet{weinberg01} and \citet{ma}.
The settings in which they derive their coefficients differ: their starting point is the kinetic closure relation given by \citet{klimontovich}, \citet{ichimaru}, \citet{gilbert} who note that the Bogoliubov-Born-Green-Kirkwood-Yvon (BBGKY) hierarchy may be closed while assuming that the two-point correlation function will relax on a shorter dynamical time scale, whereas the one point distribution function evolves on a longer secular time scale\footnote{This time ordering is originally due to \citet{bogoliubov}}.
Hence if one assumes that the distribution function \(F\) entering the linearized equation, \Eq{boltzsphere}, can be considered to be constant, then the second-order equation in the BBGKY hierarchy is automatically satisfied while the r.h.s. of the first equation is proportional to the propagated (via \Eq{defpropagator}) excess correlation induced by the dressed clumps.
This kinetic theory has been successfully applied in plasma physics, leading to the so called Lenard-Balescu (\citet{lenard}, \citet{balescu}) collision term, and was also transposed by \citet{weinberg93} for a multi-periodic "stellar" system.

Note that the BBGKY hierarchy is a  \(1/N\) expansion, where   \(N\) is the number of particles in the system. Formally it would make sense here to identify   \(N\) as a measure of the clumpiness of the medium, but this definition is  qualitative only. 
We rely here on the {\sl same} time ordering hierarchy, 
but the degree of clumpiness in the system is explicitly imposed by the boundary condition. 
In this appendix, the derivation is carried from first principles, while relying
on an explicit infall and tidal field.

\subsection{Quasi linear equations in  angle action variables}

The collisionless Boltzmann equation of an open system may be written as:
\begin{equation} \frac{\partial \mathfrak{F}}{\partial t} +
\{H, \mathfrak{F} \} 
   =S^e+ s^e,  \quad {\rm with} \quad H= \frac{v^{2}}{2}+  \Psi(\bI,\bw, t,T)\,,
    \label{e:defboltzfluc} 
\end{equation}
where $\mathfrak{F}$ is defined by 
\begin{equation} \mathfrak{F}(\bI,\bw, t,T) = F(\bI,T) + f(\bI,\bw,t), 
\quad {\rm and} \quad \Psi(\bI,\bw, t,T)=  \Psi_{0}(\bI,\bw, T)+\psi(\bI,\bw, t)+\psi^{e}(\bI,\bw, t)\,, \EQN{psieseq}
\end{equation}
with $F$ describing the secular evolution of the DF and $f$ describing the
fluctuations of the DF over this secular evolution.
In \Eq{defboltzfluc}, the r.h.s. stands  for the incoming infall, both
 fluctuating  ($s^{e}[\bI,\bw,t]$) and secular ($S^{e}(\bI,T)$).
Since this system evolves secularly because of its environment,
these actions are not conserved.
 The last two  term on the r.h.s. of \Eq{psieseq} represents the fluctuating 
 component tracing the motions of clumps within the environment of the halo\footnote{We neglect here the secular drift of the external potential
 which should slowly shift the frequencies, $\MG{\omega}$ in \Eq{secular11}}.
 Note that since $F(\bI,T)$ is assumed to depend here only on the action,
it represents a coarse-grained distribution function (averaged over the 
angles) for which we make no attempt to specify where each star is along its orbit
nor how oriented the orbit is.
  Note also that  the canonical variables
$\bI$ and $\bw$ are the actions and the angles of the {\sl initial} system.
Developing the collisionless Boltzmann equation, \Eq{defboltzfluc}, 
over the secular and the fluctuating expansion leads to:
\begin{eqnarray} \frac{\partial F}{\partial t} + \frac{\partial f}{\partial t} +
   \bo \cdot  
   \frac{\partial f}{\partial \bw} -  \pdrv{\psi}{\bw}
   \cdot ( \frac{\partial F}{\partial \bI} + \frac{\partial
   f}{\partial \bI} ) - 
   \pdrv{\psi}{\bw}^e  \cdot ( \frac{\partial F}{\partial
   \bI} + \frac{\partial f}{\partial \bI} )+ \left[
\frac{\partial \psi}{\partial \tmmathbf{I}} +\frac{\partial \psi^{e}}{\partial \tmmathbf{I}} 
\right]\cdot \frac{\partial
   f}{\partial \tmmathbf{w}} = S^e+s^e . 
 \EQN{secular11}  
 \end{eqnarray}
This equation involves two time scales, $t$ and $T$. 
On the fluctuation time scale, $t$, secular
quantities can be described as static, leaving only the linearised
open collisionless Boltzmann equation (\Eq{boltzsphere}):
\begin{equation}
 \frac{\partial f}{\partial t} + \bo \cdot \frac{\partial
   f}{\partial \bw} - ( \pdrv{\psi}{\bw} + \pdrv{\psi}{\bw}^e ) \cdot \frac{\partial
   F}{\partial \bI}  
    = s^e\,,
  \EQN{eqlin}
   \end{equation}
where the amplitude of $f$ is of  first-order compared to $F$ and 
involves only  the  fluctuating part of the
external forcing, $s^{e}(\bI,\bw,t)$. On a longer time scale, $T$, the
Boltzmann equation, \Eq{secular11} can be
 $T$-averaged, considering that the average of
fluctuations are zero on such time scales. 
This leads to a second  equation:
\begin{equation}
 \frac{\partial \langle F \rangle}{\partial T}=
\left  \langle \left[
\frac{\partial \psi}{\partial \tmmathbf{w}} +\frac{\partial \psi^{e}}{\partial \tmmathbf{w}} 
\right]\cdot \frac{\partial
   f}{\partial \tmmathbf{I}} \right\rangle_{T} - \left  \langle \left[
\frac{\partial \psi}{\partial \tmmathbf{I}} +\frac{\partial \psi^{e}}{\partial \tmmathbf{I}} 
\right]\cdot \frac{\partial
   f}{\partial \tmmathbf{w}} \right\rangle_{T}+ \left\langle S^e \right\rangle_{T} . 
   \label{e:defdiffI}\end{equation}
The brackets denotes averaging over a time longer than the typical time scale
   of fluctuations:
\[
\displaystyle \langle Y \rangle_{T} \equiv 1/ \Delta T\int_{T-\Delta T/2}^{{T+\Delta T/2}}\d t  Y(t)\,.
 \]
 The time interval, $\Delta T$, should be chosen so that a given dark matter particle describing its orbit will encounter a few times
  the incoming clump at various phases along its orbit. Because the incoming clump is subject to dynamical friction, the resonance 
  will only last so long, and induce a finite but small kick, $\Delta \bI $ during $ \Delta T$. Because the infall displays some degree of temporal 
  and spatial coherence, we may not assume that the successive kicks are uncorrelated, in contrast to the situation presented by \citet{weinberg01}, 
  \citet{ma} or the classical image described in Brownian motion. 
In other words, when we write an effective microscopic Langevin counterpart to the corresponding Fokker Planck equation,
it will involve a coloured 3D random variable (see \Eq{langevin}).

    Derivative and averaging may be exchanged considering that
   $F$ and $\Psi$ evolve slowly with respect to time.
Terms involving the product of two first-order quantities survive to the time averaging because we cannot presume that 
the response in distribution function and potential within $R_{200}$ are uncorrelated.
In order to evaluate those quadratic terms, we may integrate
\Eq{eqlin}, while assuming that $F(\bI,T)$ is effectively constant w.r.t. time $t$.
The solution, \Eq{DFsphere}, may then be reinjected into the quadratic terms in 
\Eq{defdiffI}  so that they involves terms such as 
\begin{eqnarray}
  \frac{\partial f}{\partial \bI} \cdot \frac{\partial \psi}{\partial
  \M{w}} &=& - \left( \sum_{\M{k}_1,\M{k}_2}
  \mathe^{\imath ( \M{k}_1 + \M{k}_2 ) \cdot \M{w}}
  \psi_{\M{k}_2}(\bI,t) \int^t_{- \infty} \d  \tau
 \mathe^{\imath \M{k}_1
  \cdot \bos ( \tau - t )} \psi_{\M{k}_1}(\bI,\tau)  \M{k}_1 \otimes \M{k}_2 \right):  \frac{\partial^2
  F}{\partial \M{I}^2} -\nonumber \\
  && \left( \sum_{\M{k}_1,\M{k}_2} \mathe^{\imath (
  \M{k}_1 + \M{k}_2 ) \cdot \M{w}} \M{k}_1
  \psi_{\M{k}_2}(\bI,t) \M{k}_2 \cdot \frac{\partial}{\partial
  \M{I}} \int^t_{- \infty}\d  \tau \mathe^{\imath \M{k}_1 \cdot \bos
  ( \tau - t )} \psi_{\M{k}_1}(\bI,\tau)   \right) \cdot
  \frac{\partial F}{\partial \M{I}}\,. \EQN{double}
\end{eqnarray}
where we may factor  the action derivative of  $F(\bI,T)$ out of the $\tau$-time 
integral  because the secular distribution is assumed
to be constant over  a few dynamical times.
Since the l.h.s. of \Eq{defdiffI} does not depend on $\bw$, 
we may average its r.h.s. over $\d \bw$. This implies that 
in \Eq{double}, only the $\bk_{1}=-\bk_{2}$ terms remain.
We rely effectively on the averaging theorem (\citet{BT}) to convert orbit averages into angle average. The corresponding evolution equation hence depends on the actions only, as expected. Note that in doing so, we assume 
that no other resonances matter.
The secular equation, \Eq{defdiffI}, becomes finally 
 after some similar algebra for the other contributions:\footnote{note that 
when $S^{e}=s^{e}=0$
this equation is conservative by construction. }

\begin{equation}
  \frac{\partial F}{\partial t} = \langle {\rm D}_0 ( \tmmathbf{I}) \rangle -
  \langle \M{D}_1 ( \tmmathbf{I}) \rangle \cdot \frac{\partial^{}
  F}{\partial \tmmathbf{I}^{}} -\langle \M{D}_2 ( \tmmathbf{I} ) \rangle: 
  \frac{\partial^2 F}{\partial \tmmathbf{I}^2}, \label{e:eqSecular}
   \EQN{SOLQuasi}
   \end{equation}
where
\begin{equation} 
\left\langle D_0(\bI) \right\rangle = \frac{1}{( 2 \pi )^3} \int \langle S_e \rangle_{T}
 \d  \tmmathbf{w} +\left\langle \sum_{\tmmathbf{k}}  \imath
   \tmmathbf{k}\cdot \frac{\partial}{\partial
   \tmmathbf{I}} \left([ \psi_\bfk^* (\tmmathbf{I}, t ) + \psi^{e*}_\bfk(\tmmathbf{I}, t ) ] \int^t_{- \infty} \mathe^{\imath \tmmathbf{k}_1 \cdot \bos ( \tau - t )}
   s_e ( \tmmathbf{k}, \tmmathbf{I}, \tau ) \d  \tau \right) \right\rangle_{T}, 
    \label{e:D0}
    \end{equation}
while
 the drift coefficient,  $ \M{D}_{1}$, obeys
\begin{equation} \langle \M{D}_1 ( \tmmathbf{I} ) \rangle = \left\langle
   \sum_{\tmmathbf{k}_{}} \tmmathbf{k } ^{}_{}\,\,
   \bfk
   \cdot \frac{\partial}{\partial \tmmathbf{I}}\left( [\psi^*_{ \tmmathbf{k}} (
   \tmmathbf{I}, t ) + \psi^{e *}_{ \tmmathbf{k}} ( \tmmathbf{I}, t ) ]  \int^t_{- \infty}
   \mathe^{\imath \tmmathbf{k}_{} \cdot \tmmathbf{\bos} ( \tau - t )} [
   \psi_{\tmmathbf{k}_{}} ( \tmmathbf{I}, \tau ) +
   \psi^e_{\tmmathbf{k}_{}} ( \tmmathbf{I}, \tau ) ] \d  \tau \right)
   \right\rangle_{T} \,, \label{e:D1}
\end{equation}
and the  diffusion coefficient, $ \M{D}_{2}$,  is given by
\begin{equation} \langle \M{D}_2 ( \tmmathbf{I} ) \rangle = \left\langle
   \sum_{\tmmathbf{k}_{}} {\M{k}^{} \otimes \M{k}{ [} ^{}_{}
   \psi^*_{ \tmmathbf{k}} ( \tmmathbf{I}, t ) + \psi^{e *}_{ \tmmathbf{k}} (
   \tmmathbf{I}, t ) ]} \int^t_{- \infty} \mathe^{\imath \tmmathbf{k}_{} \cdot
   \tmmathbf{\bos} ( \tau - t )} [ \psi_{\tmmathbf{k}_{}} ( \tmmathbf{I},
   \tau ) + \psi^e_{\tmmathbf{k}_{}} ( \tmmathbf{I}, \tau ) ] \d  \tau
   \right\rangle_{T} . \label{e:D2}
\end{equation}

Note that the infall coefficient, $D_{0}$, includes both  the secular infall, and 
a contribution arising from the possible correlation between 
the fluctuating tidal field and the fluctuating infall. 
%
It
 may be an explicit function of time, $T$, 
 reflecting the fact that, as more mass is
accreted, the profile of  dark matter changes with time.
The coefficients
$D_{0}$, $\M{D}_{1}$ and $\M{D}_{2}$ are also an implicit function of 
time because of the time average, $\langle \,\,\,\, \rangle_{T}$ and 
via the secular distribution function, $F(\bI,T)$ 
which occurs in $\psi_{\bk}(\bI,t)$  through \Eq{SOL}.
Clearly, if the potential, and/or the source term are 
completely decorelated in time, so that $\langle \psi_{\bk}^{*}(\bI,t) \psi_{\bk}^{*}(\bI,\tau) \rangle_{T} \propto \delta_{\rm D }(t-\tau)$ and 
$\langle \psi_{\bk}^{*}(\bI,t) s^{e *}_{\bk}(\bI,\tau) \rangle_{T} \propto \delta_{\rm D}(t-\tau)$, \Eq{D2} or \Ep{D1} would vanish. Provided $\Delta T $ is long compared to the 
typical correlation time of the potential (and/or the source term),
we may take the limit $t \rightarrow \infty$ in the integrals entering 
\Eqs{D0}{D2}.
Note finally that \Eq{eqSecular} does not derive from a kinetic theory in the 
classical sense, in that it does not rely on a diffusion process in velocity space
induced by the {discrete} number of particles in the system.

\subsection{Linking the infall, drift and diffusion to the cosmic
 two-point correlations}
Up to this point we investigated the secular evolution of a {\sl given} (phase averaged) halo,
undergoing a given  inflow and tidal field accretion history.
Let us now invoke ergodicity so as to replace temporal averages by 
ensemble averages in \Eqs{D0}{D2}. In doing so, we now try and 
describe a {\sl mean } galactic halo embedded in the typical environment
presenting the most likely correlations. This involves replacing
$\langle \,\,\,\, \rangle_{T}$ with $\langle \,\,\,\, \rangle \equiv E\{\,\,\,\,\}$.
Let us  use Eq. (\ref{e:defbibasis}) to expand Eq. (\ref{e:D2}).  This yields:
 \begin{equation}
  \langle \M{D}_2 (\bfI,T ) \rangle = \sum_{\tmmathbf{k}_{}} \tmmathbf{k} ^{}_{}
  \tmmathbf{\otimes k} \sum_{\bn,\bn'_{}} \left( \int^\infty_{- \infty} \langle
  \mathfrak{a}_{\bn'}^* ( t ) \mathfrak{a}_\bn ( \tau ) \rangle \mathe^{\imath
  \tmmathbf{k}_{} \cdot \bos ( \tau - t )}  \d  \tau \right)
  \psi^{[ \bn' ]*}_{\tmmathbf{k}_{}} ( \tmmathbf{I} ) \psi^{[ \bn
  ]}_{\tmmathbf{k}_{}} ( \tmmathbf{I} )\,, \EQN{D2temp}
\end{equation}
where $\mathfrak{a}_{\bn} ( t ) \equiv a_\bn ( t ) + b_\bn ( t )$ corresponds to the
coefficient of the total (self-consistent plus external) potential.  If
the first-order perturbations are stationary, 
let us write the two-point cross-correlation of the temporal fields, $\langle \mathfrak{a}_\bn ( t ),
\mathfrak{a}_{\bn'} (\tau ) \rangle $ as $ \mathrm{C} [ \mathfrak{a}_\bn, \mathfrak{a}_{\bn'} ](t- \tau)$ so that
the integral in \Eq{D2temp} may be carried as (assuming parity for the 
correlation function):
\begin{equation}
 \int^{\infty}_0 \mathrm{C} [ \mathfrak{a}_\bn, \mathfrak{a}_{\bn'} ] ( \Delta \tau )
  \mathe^{\imath \tmmathbf{k}_{} \cdot \bos \Delta \tau} \d 
  \Delta \tau = P_{\mathfrak{a}}^{\bn,\bn'} [ \tmmathbf{k}_{} \cdot
  \bo],
\end{equation}
giving the temporal power spectrum
evaluated at the temporal frequency, $\bk \cdot \bo$. Consequently, the diffusion coefficient becomes:
\begin{equation}
\langle \M{D}_2 ( \bfI ) \rangle = \sum_{\tmmathbf{k}_{}} \tmmathbf{k} \otimes
  \tmmathbf{k} \sum_{\bn,\bn'_{}} \psi^{[ \bn' ]*}_{\tmmathbf{k}_{}} (
  \tmmathbf{I} ) \psi^{[ \bn ]}_{\tmmathbf{k}_{}} ( \tmmathbf{I} )
  P_{\mathfrak{a}}^{\bn,\bn'} [ \tmmathbf{k}_{} \cdot \bo ]. \EQN{D2def}
\end{equation}
The same procedure may be
applied to the other coefficient:
\begin{equation}
  \langle \M{D}_1 (\bfI) \rangle = \sum_{\tmmathbf{k}_{}} \tmmathbf{k} ^{}_{}
  \sum_{\bn,\bn'_{}} 
  \tmmathbf{k \cdot} \frac{\partial}{\partial \tmmathbf{I}} \left(\psi^{[ \bn' ]*}_{\tmmathbf{k}_{}} ( \tmmathbf{I} ) \psi^{[
  \bn ]}_{\tmmathbf{k}_{}} ( \tmmathbf{I} ) P_{\mathfrak{a}}^{\bn,\bn'} [
  \tmmathbf{k}_{} \cdot \bo ] \right) , \EQN{D1def}
\end{equation}
while,  for the secular correlation,  \Eq{D0}:
\begin{equation} 
\left\langle D_0(\bI) \right\rangle = \frac{1}{( 2 \pi )^3} \int \langle S_e \rangle_{T} \d  \tmmathbf{w} +\sum_{\tmmathbf{k}_{}} \tmmathbf{k} ^{}_{}
  \sum_{\bn,\bn'_{}}
  \tmmathbf{k \cdot} \frac{\partial}{\partial \tmmathbf{I}} \left(  \psi^{[ \bn' ]*}_{\tmmathbf{k}_{}} ( \tmmathbf{I} ) \sigma^{[
  \bn ],e}_{\tmmathbf{k}_{}} ( \tmmathbf{I} ) P_{\mathfrak{a} c}^{ \bn,\bn'} [
  \tmmathbf{k}_{} \cdot \bo ] \right) , 
    \label{e:D0def}
    \end{equation}
where $ P_{\mathfrak{\mathfrak{a} c}}^{ \bn,\bn'}[\omega]$ is the mixed power spectrum
given by $\langle {\hat \mathfrak{a}}^{*}_{\bn} 
{\hat c}_{\bn} \rangle=
\langle [{\hat {a}}^{*}_{\bn} +{\hat b}^{*}_{\bn} ]
{\hat c}_{\bn} \rangle$. 
Hence
\begin{equation}
P_{\mathfrak{a} c}^{\bn,\bn'} [
  \tmmathbf{k}_{} \cdot \bo ]=\left(
({\hat A}_b+1) \!\!\times\!\! (1) \cdot \langle {\hat \bb}^{*} \otimes {\hat \bc}
  \rangle + {\hat A}_c \!\!\times\!\! (1 ) \cdot \langle {\hat \bc}^{*} \otimes {\hat \bc} \rangle
  \right)[
  \tmmathbf{k}_{} \cdot \bo] \,.\EQN{defPnac}
  \end{equation}
Recall also that   (given \Eq{SOLK} and \Eq{reorder1})
\begin{equation}
P_{\mathfrak{a}}^{\bn,\bn'} [
  \tmmathbf{k}_{} \cdot \bo ]\!=\!\left(
({\hat A}_b+1) \!\!\times\!\! ({\hat A}^{*}_b+1) \cdot \langle {\hat \bb} \otimes {\hat \bb}^{*}
  \rangle + {\hat A}_c \!\!\times\!\! {\hat A}^{*}_c \cdot \langle {\hat \bc} \otimes {\hat \bc}^{*} \rangle + ({\hat A}_b+1) \!\!\times\!\! {\hat A}^{*}_c
  \cdot \langle {\hat \bb} \otimes {\hat \bc}^{*} \rangle +   
  {\hat A}_c \!\!\times\!\! ({\hat A}^{*}_b+1) \cdot \langle {\hat \bc} \otimes {\hat \bb}^{*} \rangle
  \right)[
  \tmmathbf{k}_{} \cdot \bo]. \EQN{defPna}
  \end{equation}
where $A_{b}$ and $A_{c}$ involve $\bK$
and therefore the  secular distribution function, $F$, via \Eq{defkk}.
Recall that $A_{b}$ and $A_{c}$ involve $(1-\hat \bK)^{-1}$,  which 
reflects the fact that the perturbation is dressed by the 
self-gravity of the halo.
\Eq{SOLQuasi}, together with \Eqs{D2def}{D1def} and \Eq{defPna}
provides a consistent framework in which to evolve secularly the
mean distribution of a galactic halo within its cosmic environment.
Note that it is possible via \Eq{reorder2}
to  apply non-linear corrections to the induced correlation within 
$R_{200}$.

\section{perturbation theory to higher order}
\label{s:apendperturb}

\begin{figure} 
\centering
\resizebox{0.85\columnwidth}{0.55\columnwidth}{\includegraphics{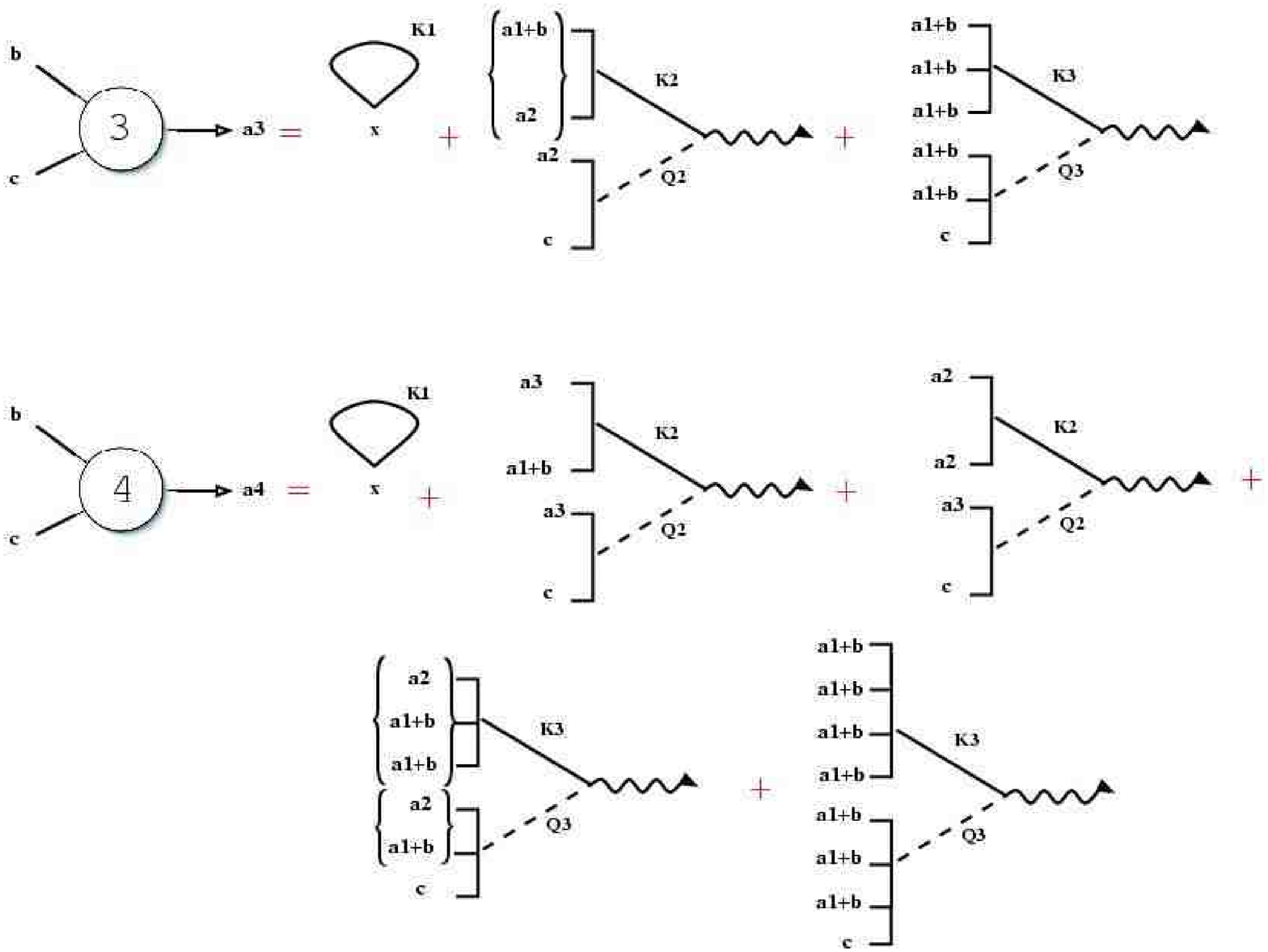}}
\caption{ diagrammatic representation of the expansion to third ({\sl top diagram},
corresponding to \Eq{defdiag3}) and fourth  ({\sl bottom diagram},
given \Eq{defdiag4})
order; again (see \Fig{diag1} for details) the coupling of the tidal interaction (through 
the $b_{\rn}$ coefficients) and the incoming infall (expanded over 
the $c_{\rn}$ coefficients) yields the coefficient of the response inside $R_{200}$. 
The coupling is achieved via the operator $\bK_{i}$ and 
$\bQ_{i}$ 
as explained in \Fig{diag11}; the curly brace 
in front of the diagrams account for the number of such 
diagrams entering the expansion, corresponding
to the permutation of the 
 input (recalling that the order matters).
 Note also  for each branch the sum of the order of the sub branch 
 correspond to the order of the expansion.
 }
\label{f:diag2} 
\end{figure}

\subsection{Perturbative dynamical equations}

In this section, we `solve' the dynamical equation to order $\rn$, 
which will allow us in the next section to present the N-point correlation
 to order $\rn$.
\subsubsection{Perturbation theory to all orders}
\begin{figure} 
\centering
\resizebox{0.95\columnwidth}{0.45\columnwidth}{\includegraphics{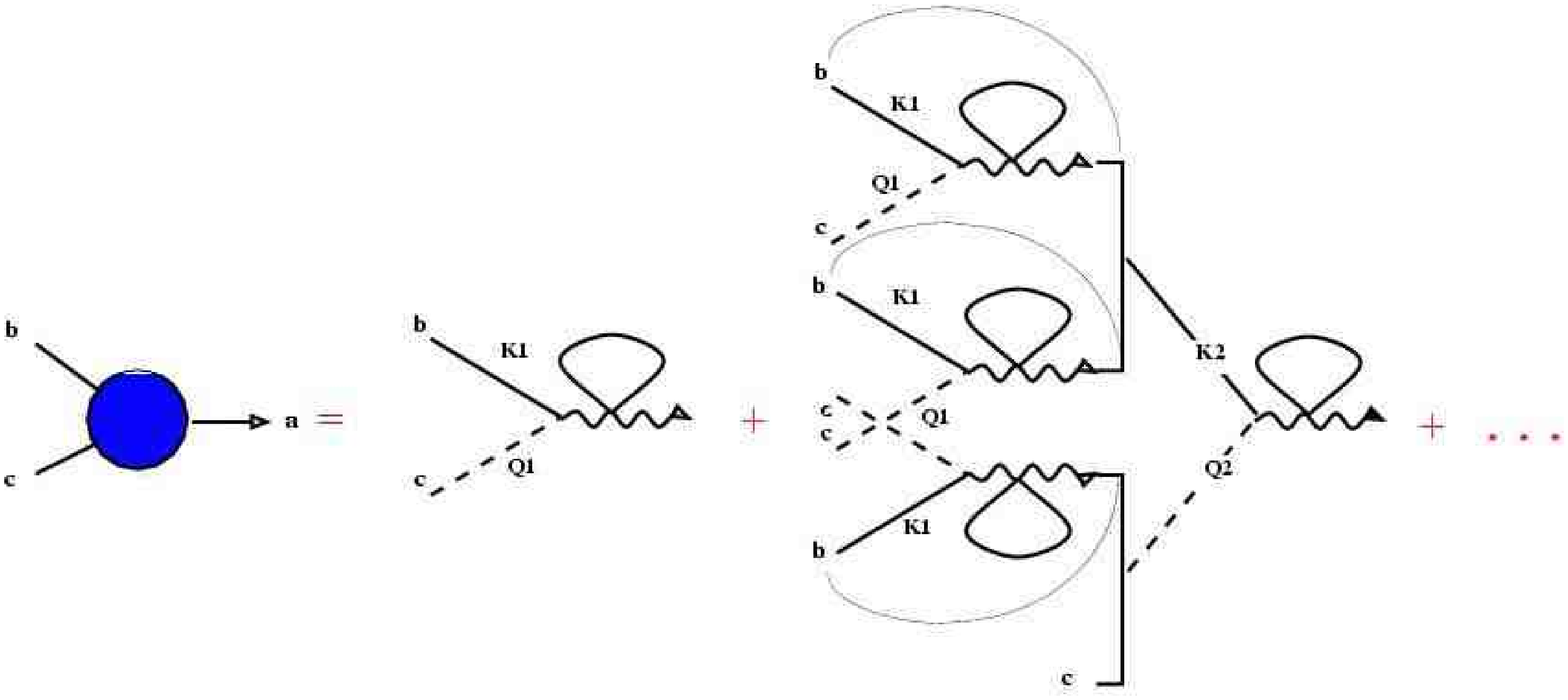}}
\caption{Reordered diagrammatic representation to second-order (first-order included)
 of the expansion given  in \Eqs{defdiag1}{defdiag2}.
 This time only $b_{\rn}$ and $c_{\rn}$ are inputs.
 The closed loop accounts for the self-gravity and represents $(1-{\hat \bK_{1}})^{-1}$.
 The thin loop traces the fact that the perturbed potential  contributes 
 also directly to the second-order term via $\ba_{1}+\bb$ (see \Eq{defdiag2} for details).
 Note that in each diagram, each oblique line represents a sum over $\bk$ and a time integral.
 The dashed line stands for infall coupling, while the thick line stands for the tidal coupling.
 }
\label{f:diag11} 
\end{figure}

Recall that for  $n \geqslant 2$,
$  f^{( \rn )}_{\bk} ( \bI, t ) 
  $obeys \Eq{eqnn}.
Given \Eq{defsum}
it follows that 
\begin{eqnarray}
  a^{( \rn )}_\bp ( t ) &=& \sum_\M{q,\bk} \int \d \tau \exp (
 \imath{\bk} \cdot \MG{\omega} [ \tau - t ] ) \hspace{0.25em} [ a^{( n
  )}_\M{q} ( \tau ) + \delta_1^n \hspace{0.25em} b_\M{q} ( \tau ) ] \left( (2\pi)^{3} \int
  \d\bI \psi^{[ \bn ]}_{\bk} (\bI) \psi^{[ \bp ]*}_{\bk}
  (\bI) \hspace{0.25em} \frac{\partial F}{\partial \bI}
  \cdot\imath{\bk} \right) - \nonumber \\ && \hskip -0.5cm
   \sum^{n - 1}_{k = 1} \sum_{\M{q},\bk} \int \d \tau \exp (
 \imath{\bk} \cdot \MG{\omega} [ \tau - t ] ) [ a^{( k )}_\M{q} ( \tau ) +
  \delta_1^k \hspace{0.25em} b_\M{q} ( \tau ) ] \left( (2\pi)^{3} \int \d\bI \left\{
  \psi^{[\M{q} ]} (\tmmathbf{w},\bI), f^{( n - k )}
  (\tmmathbf{w},\bI, t ) \right\}_{\bk} \psi^{[\bp
  ]*}_{\bk} (\bI) \right)\,,
  \label{e:an}
\end{eqnarray}
where the first term in \Eq{an} corresponds to the usual self-gravity coupling 
at order $\rn$, and the sum corresponds to the feed of lower 
order  potential coupling into the n$^{\rm th}$ order equation.
Here $  f^{( \rn )}_{\bk} (\bI, t ) $ obeys
\begin{eqnarray}
  f^{( \rn )}_{\bk} (\bI, t ) &=& \sum_\M{q} \int \d \tau \exp (\imath{\bk}
  \cdot \MG{\omega} [ \tau - t ] ) \times \left[ \frac{\partial
  F}{\partial \bI} \cdot\imath{\bk} \hspace{0.25em} \psi^{[\M{q}
  ]}_{\bk} [ a^{( \rn )}_\M{q} ( \tau ) + \delta_1^\rn \hspace{0.25em} b_\M{q} (
  \tau ) ] + \right.
   \left. \sum^{\rn - 1}_{k = 1} [ a^{( k )}_\M{q} ( \tau ) + \delta_1^k
  \hspace{0.25em} b_\M{q} ( \tau ) ] \{ f^{( n - k )}, \psi^{[\M{q} ]}_{\bk}
  \}_{\bk} \right]\,
   + \nonumber \\ &&
  \sum_\M{q} \int \d \tau \exp (\imath{\bk} \cdot \MG{\omega} [ \tau - t ]
  ) c_\M{q} ( \tau ) \delta_\bn^1 \sigma_{\bk}^{e, [\M{q} ]} .
   \label{e:defn}
\end{eqnarray}
%
%
Note that the response in \Eq{defn} is, as expected, out of phase with respect to the 
potential excitation, $a_{\bn}(\tau)$ because of inertia (hence the modulation in
$\exp(\imath \bk \cdot \MG{\omega}(\tau-t))$).
Now, to $\rn^{\op{th}}$ order \Eq{an}, \Ep{defn} may be rewritten formally as 
(using the contraction rule \Eq{defcontractn}):
\begin{eqnarray}
  && \hskip -1cm {\ba^{( \rn )} = \bK_1 \cdot \ba^{( \rn )} + \bK_2 \cdot \left( \sum_{i_1 +
  i_2 = \rn} [ \ba^{( i_1 )} + \delta_{i_1}^1 b ] \otimes [ \ba^{( i_2 )} +
  \delta_{i_2}^1 \bb ] \right) + \cdots +}  \bK_j \cdot \left( \sum_{i_1  +
  \cdots + i_j = \rn} \bigotimes_j [ \ba^{( i_1 )} + \delta_{i_1}^1 \bb ] \right)
  \cdots +
  \bK_\rn \cdot \bigotimes_\rn [ \ba^{( 1 )} + \bb ] +  \hskip -1.5cm \nonumber \\ && \M{Q}_2 \cdot \left( \ba^{( \rn - 1 )}
  \otimes \bc \right) + \cdots + \bQ_j \cdot \left( \sum_{i_1+ \cdots + i_j
  = \rn - 1} \left[ \bigotimes [ \ba^{( i_1 )} + \delta_{i_1}^1 \bb ] \right]
  \otimes \bc \right) + \cdots
  + \bQ_\rn \cdot \left[ \bigotimes_{\rn - 1} [ \ba^{( 1 )} + \bb ] \right] \otimes \bc\,,
  \EQN{soln}
\end{eqnarray}
where the kernels $\bK_{1}$ and $\bK_{2}$ are given by \Eqs{defK1}{defK2}, 
while $\bK_{\rn}, \rn\ge 2$  obey formally:
\begin{eqnarray}
&&  \left. ( \bK_\rn )_{\bp, \M{q}_1, \M{q}_2 \cdots, q_\rn} [ \tau_1 - t, \tau_2 - \tau_1,
  \cdots, \tau_\rn - \tau_{\rn - 1} ] = [2\pi]^{3} \sum_{\bk} \int \op{\d\M{I}} \exp ( \imath {\bk} \cdot\MG{\omega}
  [ \tau_1 - t ] )\sum_{{\bk}_1 +{\bk}_2 = {\bk}} \left\llbracket \exp ( \imath {\bk}_1 \cdot\MG{\omega} [
  \tau_2 - \tau_1 ] ) \cdots 
   \times \right. \right. \nonumber \\
 && \left.  \left. \sum_{{\bk}_{2 \rn - 1} +{\bk}_{2 \rn} ={\bk}_\rn}
  \left\llbracket \exp ( \imath{\bk}_{2 \rn - 3} \cdot\MG{\omega} [ \tau_\rn - \tau_{\rn - 1} ] )
  \frac{\partial F}{\partial \M{I}} \cdot \imath{\bk}_{2 \rn - 3} \psi^{[\M{q}_\rn ]}_{{\bk}_{2 \rn -
  3}}, \psi^{[\M{q}_{\rn - 1} ]}_{{\bk}_{2 \rn - 2}} \right\rrbracket \cdots,
  \psi_{{\bk}_4}^{[ \M{q}_2 ]} \right\rrbracket, \psi_{{\bk}_2}^{[ \M{q}_1 ]}
  \right\rrbracket \psi^{[ \bp ]*}_{\bk} \hspace{0.25em} . \label{e:defKn} 
\end{eqnarray}
Note that the $\rn^{\rm th}$ order Kernel involves ``only'' one integral over 
action space, but $\rn$ coupling in configuration space and $\rn+1$ time ordered
instants $(t,\tau_{1},\cdots \tau_{\rn}$). 
Note also that \Eq{an} implies that secular perturbation theory accounts
for both the rate of change in frequency of the system, via $\partial^n\MG{\omega} /
\partial \M{I}^\rn,$ the rate of change in equilibrium via $\partial^\rn F / \partial \M{I}^\rn$ but
also the rate of change in the incoming flow via $\partial^\rn \sigma^{[\bp],e} / \partial \M{I}^\rn$. Note
{finally} that the relative phases (causality) are accounted for via the ordered
time integrals.
For instance
 \Eq{soln} reads to third order:
\begin{eqnarray}
 \ba^{( 3 )} &=& \bK_1 \cdot \ba^{( 3 )} + \bK_2 \cdot \left( [ \ba^{( 1 )}_1 + \bb
   ] \otimes \ba^{( 2 )} + \ba^{( 2 )} \otimes [ \ba^{( 1 )} + \bb ] \right) +
   \nonumber \\
&&  \hskip -1cm \bK_3 \cdot \left( [ \ba^{( 1 )} + \bb ] \otimes [ \ba^{( 1 )} + \bb ] \otimes
   [ \ba^{( 1 )} + \bb ] \right) + \bQ_3 \cdot \left( [ \ba^{( 1 )}_1 + \bb ]
   \otimes [ \ba^{( 1 )}_1 + \bb ] \otimes \bc \right) + {\bQ}_2 \cdot \ba^{( 2 )}
   \otimes \bc \,,
  \EQN{defdiag3} 
  \end{eqnarray}
and is illustrated in \Fig{diag2} together with 
$\ba^{(4)}$:
\begin{eqnarray}
  \ba^{( 4 )} &&= \bK_1 \cdot \ba^{( 4 )} + \bK_2 \cdot \left( \ba^{( 3 )} \otimes [
  \ba^{( 1 )} + \bb ] + [ \ba^{( 1 )} + \bb ] \otimes \ba^{( 3 )} + \ba^{( 2 )} \otimes
  \ba^{( 2 )} \right) + \nonumber \\
 && \hskip -0.5cm \bK_3 \cdot \left( \ba^{( 2 )} \otimes [ \ba^{( 1 )} + \bb ] \otimes [ \ba^{( 1 )} +
  \bb ] + [ \ba^{( 1 )} + \bb ] \otimes \ba^{( 2 )} \otimes [ \ba^{( 1 )} + \bb ] + [ \ba^{(
  1 )} + \bb ] \otimes [ \ba^{( 1 )} + \bb ] \otimes \ba^{( 2 )} \right)
  + \nonumber \\
  && \hskip -0.5cm \bK_4 \cdot [ \ba^{( 1 )} + \bb ] \otimes [ \ba^{( 1 )} + \bb ] \otimes [ \ba^{( 1 )} +
  \bb ] \otimes [ \ba^{( 1 )} + \bb ] + \nonumber \\
&& \hskip -0.5cm  \bQ_2 \cdot \left( \ba^{( 3 )} \otimes \bc \right) + \bQ_3 \cdot \left( \ba^{( 2 )}
  \otimes [ \ba^{( 1 )} + \bb ] \otimes \bc + [ \ba^{( 1 )} + \bb ] \otimes \ba^{( 2 )}
  \otimes \bc \right) + \bQ_4 \cdot [ \ba^{( 1 )} + \bb ] \otimes [ \ba^{( 1 )} + \bb ]
  \otimes [ \ba^{( 1 )} + \bb ] \otimes \bc \,.\EQN{defdiag4}
\end{eqnarray}
Note that \Eq{defdiag4} depends recursively on \Eq{defdiag3} and both 
depends recursively on \Eq{defdiag2} and \Ep{defdiag1}. When the recursion is 
carried through, (see \Fig{diag11}) the expected relative complexity of the 
non-linear evolution appears clearly.

\subsubsection{Reordering to higher order}
\label{s:reorderannexe}

In the main text, we give in \Eq{defAbb} and above the first- and second-order reshuffling 
of the perturbation in $\bb$ and $\bc$.
Similarly, the third-order term reads
 in terms of products of $\bb$ and $\bc$ as
\begin{eqnarray}
  \ba^{( 3 )} \!\!=\!\! A_{bbb} \cdot \bb \otimes \bb \otimes \bb + A_{ccc} \cdot \bc \otimes
  \bc \otimes \bc + A_{bbc} \cdot \bb \otimes \bb \otimes \bc + A_{ccb} \cdot \bc \otimes
  \bc \otimes \bb + \nonumber \\ A_{bcb} \cdot \bb \otimes \bc \otimes \bb +
  A_{cbb} \cdot \bc \otimes \bb \otimes \bb + A_{bcc} \cdot \bb \otimes \bc \otimes \bc +
  A_{cbc} \cdot \bc \otimes \bb \otimes \bc \nonumber\,,
\end{eqnarray}
where (following the same convention as in the main text for the brackets)
\begin{eqnarray}
  A_{bbb} &=& \bK'_3 \circ \bK''_1 + \bK'_2 \circ [ \bK''_1, \bK'_2 \circ \bK''_1 ] + \bK'_2
  \circ [ \bK'_2 \circ \bK''_1, \bK''_1 ]  \,,\nonumber \\
  A_{ccc} &=& \bK'_3 \circ \bQ'_1 + \bK'_2 \circ \left\{ [ \bQ'_1, \bK'_2 \circ \bQ'_1 +
  \bQ'_2 \circ [ \bQ'_1, \b1 ] ] + [ \bK'_2 \circ \bQ'_1 + \bQ'_2 \circ [ \bQ'_1, \b1 ], \bQ'_1
  ] \right\} + \bQ'_3 \circ [ \bQ'_1, \bQ'_1, \b1 ] + \nonumber \\ && \bQ'_2 \circ [ \bK'_2 \circ \bQ'_1 +
  \bQ'_2 \circ \bQ'_1, \b1 ], \b1 ]  \,,\nonumber \\
  A_{bbc} &=& \bK'_3 \circ [ \bK''_1, \bK''_1, \bQ'_1 ] + \bQ'_3 \circ [ \bK''_1, \bK''_1, \b1 ]
  + \bK'_2 \circ [ \bK''_1, \bK'_2 \circ [ \bK''_1, \bQ'_1 ] + \bQ'_2 \circ [ \bK''_1, \b1 ] ]
   \,,\nonumber \\
  A_{bcb} &=& \bK'_3 \circ [ \bK''_1, \bQ'_1, \bK''_1 ] + \bK'_2 \circ [ \bK'_2 \circ [
  \bK''_1, \bQ'_1 ] + \bQ'_2 \circ [ \bK''_1, \b1 ], \bK''_1 ]  \,,\nonumber \\
  A_{ccb} &=& \bK'_3 \circ [ \bQ'_1, \bQ'_1, \bK''_1 ] + \bK'_2 \circ [ \bQ'_1, \bK'_2 \circ [
  \bQ'_1, \bK''_1 ] ]  \,,\nonumber \\
  A_{cbb} &=& \bK'_3 \circ [ \bQ''_1, \bK''_1, \bK''_1 ] + \bK'_2 \circ [ \bQ'_1, \bK'_2 \circ
  \bK''_1 ]  \,,\nonumber \\
  A_{bcc} &=& \bK'_3 \circ [ \bK''_1, \bQ'_1, \bQ'_1 ] + \bQ'_3 \circ [ \bK''_1, \bQ'_1, \b1 ] +
  \bK'_2 \circ [ \bK''_1, \bK'_2 \circ \bQ'_1 + \bQ'_2 \circ [ \bQ'_1, \b1 ] ] + \bK'_2 \circ
  [ \bK''_1, \bK'_2 \circ \bQ'_1 + \bQ'_2 \circ [ \bQ'_1, \b1 ] ] + \nonumber \\ && \bQ'_2 \circ [ \bK'_2
  \circ [ \bK''_1, \bQ'_1 ] + \bQ'_2 \circ [ \bK''_1, \b1 ], \b1 ]  \,,\nonumber \\
  A_{cbc} &=& \bK'_3 \circ [ \bQ'_1, \bK''_1, \bQ'_1 ] + \bQ'_3 \circ [ \bQ''_1, \bK'_1, \b1 ] +
  \bK'_2 \circ [ \bK'_2 \circ [ \bQ'_1, \bK''_1 ], \bK''_1 ] + \bK'_2 \circ [ \bQ'_1, \bK'_2
  \circ [ \bK''_1, \bQ'_1 ] + \bQ'_2 \circ [ \bK''_1, \b1 ] ] +\nonumber \\ && \bQ'_2 \circ [ \bK'_2 \circ
  [ \bQ'_1, \bK''_1 ], \b1 ]  \,.
\end{eqnarray}

Generically, after reordering, \Eq{defAbb} becomes
\begin{eqnarray}
  a^{( \rn )}_\bp ( t ) &=& \left( \sum_{\M{i}_1, \cdots \M{i}_\rn \in [ \bb, \bc ]} A_{ \M{i}_1
  \cdots \M{i}_\rn} \cdot \left( \M{i}_1 \otimes \cdots \otimes \M{i}_\rn \right) \right)_\bp (
  t )\,,  \nonumber \\
  &=&\sum_{\M{i}_1, \cdots\M{i}_\rn \in [ \bb, \bc ]} \int_{- \infty}^t \d \tau_1 \cdots
  \int_{- \infty}^{\tau_n - 1} \d \tau_\rn \sum_{\bq_{1}\cdots \bq_{\rn}}\left[ A_{\M{i}_1 \cdots\M{i}_\rn} \right]_{\bp,
  \M{q}_1 \cdots \M{q}_\rn} ( t - \tau_1, \cdots \tau_\rn - \tau_{\rn - 1} ) [\M{i}_1
  ]_{\M{q}_{_1}} ( \tau_1 ) \cdots [ \M{i}_\rn ]_{\M{q}_{_\rn}} ( \tau_\rn ) \,,\EQN{defapn}
\end{eqnarray}
which involves $2^\rn$ terms.
Here $\left[ A_{\M{i}_1 \cdots \M{i}_\rn} \right]_{\bp, \M{q}_1 \cdots \M{q}_n} ( \theta_1,
\cdots \theta_\rn )$ is some linear tensor of order $\rn + 1$ which returns the
$\rn^{\op{th}}$order response to the excitation $b_i ( \theta ), c_j ( \theta )$
at various times $\theta_1, \theta_2 \cdots \theta_p$. Note that it involve
the equilibrium distribution function, $F_{0}$ and its derivatives
with respect to the actions, $\bI$, together with
the properties of the basis function.

\subsection{The N-point correlation function}
\label{s:npointcorrel}

In the main text, we presented the calculation of the two-point correlation of the fields
within the $R_{200}$ sphere.
More generally we are interested in the N-point correlation of, say, the density (at
various times):
\begin{eqnarray}
 C_\rN &\equiv& \langle \rho ( x_1 ) \rho ( x_2 ) \cdots \rho ( x_\rN ) \rangle =
\sum^{\infty}_{\rn = \rN} \varepsilon^{\rn} \sum_{p_1 + p_2 + \cdots p_\rN = \rn}
  \langle \rho^{( p_1 )} ( x_1 ) \rho^{( p_2 )} ( x_2 ) \cdots \rho^{( p_\rN
  )} ( x_\rN ) \rangle \,,  \nonumber \\
   &=& \sum_{\rn = \rN} \varepsilon^{\rn} \sum_{p_1 + p_2 + \cdots p_\rN = \rn}
  \sum_{\,\,\,\M{q}_1, \cdots, \M{q}_\rN} \rho^{[ \M{q}_1 ]} ( \br_{1} ) \cdots \rho^{[\M{q}_\rN ]} ( \br_{\rN}
  ) \langle a_{\M{q}_1}^{( p_1 )} ( \tau_1 ) \cdots a_{\M{q}_\rN}^{( p_\rN )} ( \tau_\rN
  ) \rangle   \,. \EQN{CN1}
\end{eqnarray}
Now, solutions to the $\rn^{\op{th}}$ order perturbation theory are given by \Eq{defapn}.
It follows that
\begin{eqnarray}
  \langle  a_{\M{q}_1}^{( p_1 )} ( \tau_1 ) \cdots a_{\M{q}_\rN}^{( p_\rN )} ( \tau_\rN )
  \rangle &=& \sum_{\M{i}_{1}, \cdots \M{i}_{p_{1}} \in [ \bb, \bc ]} \!\!\!\! \cdots\!\!\!\!
   \sum_{\,\, \M{i}_{1}, \cdots \M{i}_{p_\rN} \in [ \bb, \bc ]} 
    \sum_{\,\,\, \bq_{1,1}\cdots \bq_{p_{\rN},p_{\rN}}}
   \int \d{}^{p_1} 
   \theta \left[ A_{\M{i}_1 \cdots \M{i}_{p_{1}}} \right]_{\M{q}_1, \bq_{1, 1}
  \cdots 
      \bq_{1, p_{1}}} ( \tau_1, \theta_{1, 1}, \cdots, \theta_{ 1,p_{1}} ) \cdots  
\times         \nonumber \\&& 
  \int \d{}^{p_\rN} \theta \left[ A_{\M{i}_1 \cdots \M{i}_{p_{\rN}}} \right]_{\bq_\rN, \bq_{p_{\rN},1} \cdots \bq_{p_{\rN}, p_\rN}} (
  \tau_\rN, \theta_{1,p_{1}} \cdots, \theta_{p_{\rN}, p_{\rN}} ) \times         \nonumber \\&& 
  \langle [\M{i}_1]_{\bq_{1,1}} ( \theta_{1, 1}
  ) \cdots [\M{i}_{p_1}]_{\bq_{1,p_{1}}} ( \theta_{p_{\rN}, 1} ) \cdots [\M{i}_{p_\rN}]_{\bq_{1,p_{\rN}}} ( \theta_{1, p_{\rN}} )
  \cdots [\M{i}_{p_\rN}]_{\bq_{p_{\rN}p_{\rN}}} ( \theta_{p_{\rN},p_{\rN}} ) \rangle   \,.\EQN{CN2}
\end{eqnarray}
If the perturbation is a centered Gaussian random field, Wick's theorem states that:
\begin{eqnarray}
  \langle [\M{i}_1]_{\bq_{1,1}} ( \theta_{1, 1}
  ) \cdots [\M{i}_{p_1}]_{\bq_{1,p_{1}}} ( \theta_{p_{\rN}, 1} ) \cdots [\M{i}_{p_\rN}]_{\bq_{1,p_{\rN}}} ( \theta_{1, p_{\rN}} )
  \cdots [\M{i}_{p_\rN}]_{\bq_{p_{\rN}p_{\rN}}} ( \theta_{p_{\rN},p_{\rN}} )
  \rangle = \hskip 2cm\nonumber \\ 
  \sum_{\mathrm{all \,\,\, permutations}} \prod \langle
  [\M{i}_1]_{\bq_{1,1}} ( \theta_{1, 1}
  )  [\M{i}_{p_1}]_{\bq_{1,p_{1}}} ( \theta_{p_{\rN}, 1} )  \rangle \cdots \langle 
  [\M{i}_{p_\rN}]_{\bq_{1,p_{\rN}}} ( \theta_{1, p_{\rN}} )
  [\M{i}_{p_\rN}]_{\bq_{p_{\rN}p_{\rN}}} ( \theta_{p_{\rN},p_{\rN}} )
  \rangle  \,.\EQN{CN3}
\end{eqnarray}
Putting \Eqs{CN2}{CN3} into \Eq{CN1} yields formally the N-point correlation function 
to arbitrary order.  A special case in given in the main text corresponding 
to third order expansion of the two-point correlation, \Eq{expC2}. 
The N-point correlation of other (possibly mixed)  moments of the distribution 
function may be computed following the same route.

\subsubsection{Synthetic hierarchy }

\subsection{Perturbation theory in the complex Fourier plane}
\label{s:pertComplex}
Let us close this appendix by a presentation of 
the perturbative solutions in the complex Fourier plane.
In frequency space, \Eq{defa22} reads:
\begin{eqnarray}
  \hat{a}^{( 2 )}_\M{p} ( \omega ) &=& \sum_{\M{q}_1} \hspace{0.25em} \hat{a}^{( 2
  )}_{\M{q}_1} ( \omega ) \left(  [2\pi]^{3}\sum_{\bk} \int \d\bI \psi^{[
  \M{q}_1 ]}_{\bk} (\bI) \psi^{[ \bp ]}_{\bk}
  (\bI) \hspace{0.25em} \frac{\partial F}{\partial \bI}
  \cdot{\bk} \frac{1}{\bk \cdot \MG{\omega} - \omega}
  \right) + \nonumber \\&&
  [2\pi]^{3} \sum_{\bk} \int \d\bI \sum_{\M{q}_1, \M{q}_2} \int \d \omega' [
  \hat{a}^{( 1 )}_{\M{q}_1} ( \omega' ) + \hat{b}_{\M{q}_1} ( \omega' ) ]
  \frac{\imath}{\bk \cdot \MG{\omega} -\omega'} \times \nonumber \\
&&  \sum_{\bk_1 + \bk_2 =\bk} \left\llbracket
  \frac{\imath}{ \bk_1 \cdot \MG{\omega} - ( \omega - \omega' )}
  \left[ \frac{\partial F}{\partial \bI} \cdot \imath \bk_1
  \psi^{[ \M{q}_2 ]}_{\bk_1} (\bI) \hspace{0.25em} [ \hat{a}^{(
  1 )}_{\M{q}_2} ( \omega' ) + \hspace{0.25em} \hat{b}_{\M{q}_2} ( \omega' ) ] +
  \sigma^{e, [ \M{q}_2 ]}_{\bk_1} (\bI) \hat{c}_{\M{q}_2} [ \omega'
  ] \right], \psi_{_{\bk_2}}^{[ \M{q}_1 ]} \right\rrbracket \psi^{[\bp
  ]}_{\bk} \,. \EQN{defa22Complex}
\end{eqnarray}
Following \Eq{defcontract1},
let us also define  in frequency space the contraction rule: 
\begin{equation}
  {( \hat{\bK} \cdot \hat{ \M{Z}} )_\bp ( \omega ) \equiv \sum_\M{\bq} {\hat K}_{\bp, \bq} (
  \omega ) {\hat  Z}_\bq ( \omega )} \hspace{0.25em}, \EQN{defcontract1hat}
\end{equation}
(note that \Eq{defcontract1hat} only involve a sum and no integral) and the
higher order contraction rule (\cf  \Eq{defcontractn}):
\begin{equation}
  ( \hat{\bK}_\rn \cdot {\hat  \M{Z}}^1 \otimes \cdots \otimes {\hat  \M{Z}}^\rn )_\bp ( \omega ) \equiv
  \sum_{\M{q}_1, \cdots \bq_n} \int \d \omega_1 \cdots \int \d \omega_\rn
  \hat{K}_{\bp, \M{q}_1, \cdots \bq_n} ( \omega_1, \cdots, \omega_\rn ) \delta_D (
  \omega - \sum^\rn_{i = 1} \omega_i ) {\hat  Z}^1_{\M{q}_1} ( \omega_1 ) \cdots {\hat  Z}^\rn_{\bq_\rn} (
  \omega_\rn ) \hspace{0.25em} .
\end{equation}
The operator, ${\hat \bK}_\rn [ \omega_1, \omega_2,
  \cdots, \omega_\rn ]$, obeys
\begin{eqnarray}
&&  \left. ( \hat{\bK}_\rn )_{\bp, \M{q}_1, \M{q}_2 \cdots, q_n} [ \omega_1, \omega_2,
  \cdots, \omega_\bn ] =  [2\pi]^{3}\sum_{\bk} \int \d \op{\M{I}} \frac{\imath}{\bk \cdot
  \MG{\omega} - \omega_1} \times \right.  \nonumber \\
&  &  \hskip -0.5cm \left. \sum_{\bk _1 + \bk _2 = \bk } \left\llbracket \frac{\imath}{\bk_1 \cdot
  \MG{\omega} - \omega_2} \cdots \!\!\!\! \sum_{{\bk}_{2 n - 1} + {\bk}_{2 \rn} = {\bk}_\rn}
  \left\llbracket \frac{\imath}{{\bk}_{2 \rn - 3} \cdot\MG{\omega} - \omega_\rn}
  \frac{\partial F}{\partial \M{I}} \cdot \imath{\bk}_{2 \rn - 3} \psi^{[\M{q}_\rn ]}_{{\bk}_{2 \rn -
  3}}, \psi^{[\M{q}_{\rn - 1} ]}_{{\bk}_{2 \rn - 2}} \right\rrbracket \cdots,
  \psi_{{\bk}_4}^{[ \M{q}_2 ]} \right\rrbracket, \psi_{{\bk}_2}^{[ \M{q}_1 ]}
  \right\rrbracket \psi^{[ \bp ]}_{\bk}  \,. \EQN{defKnk}
  \end{eqnarray}

\twocolumn
\section{ Other cosmological probes }
\label{s:AppendAppli}

In this appendix, we discuss  other non-linear statistical probes of the cosmic 
environment of haloes, expanding over \Sec{cosmo}.

\subsection{Dark matter disintegration }

\label{s:neutralino}

It has been claimed that dark matter could be made of neutralinos which can 
be traced indirectly via their disintegration signature, which scales like the
square of the local dark matter density (\citet{stoehr}).
The total number of $\gamma$ photons received during integration time, $t_{\gamma}$ reads
\begin{eqnarray}
N_{\rm a}(\bO,t_{\gamma})\hskip -0.3cm &=& \hskip -0.3cm D_{\rm eff} \, t_{\gamma} \, \frac{N_{\rm cont}}{2} \frac{\langle\sigma v \rangle }{m_{\chi}^{2}}\frac{\Delta\Omega}{4 \pi}\,\frac{1}{\Delta \Omega}\int \d \bO \int \d r \rho^{2}_{\rm DM}(r) \,,\nonumber \\&\equiv&  W_{\chi} \, \int \d \bO \int \d r \rho^{2}_{\rm DM}(r)
\,, \EQN{defFDM}
\end{eqnarray}
where $D_{\rm eff}$ is the effective size of the telescope, $\Delta \Omega$
the angular resolution of the telescope, and $N_{\rm cont}(E_{\gamma})$ is the number of continuum photons and $\langle \sigma v \rangle$ is the continuum cross-section of neutralinos of mass, $m_{\chi}$.
The integral  accounts for measured flux of $\gamma$ photons arising from neutralinos desintegrating in the 
direction, $\bO$.  
(See \citet{stoehr} for details about the computation of $N_{\rm cont}$, $D_{\rm eff}$ and  $\langle \sigma v \rangle$).
Since the $N_{\rm a}(\bO,T)$ scales like the line integral of the square of the density along 
the line of sight, it is straightforward to propagate the statistical properties of the density fluctuations 
to that of $N_{\rm a}$. For instance, the cosmic mean will scale like
\begin{equation}
\langle N_{\rm annih}(\bO) \rangle = W_{\chi} \int  \d r  \langle \rho_{\rm DM}(\br)\rangle^{2}+W_{\chi}
\int  \d r  \langle \delta \rho_{\rm DM}^{2}(\br) \rangle \,,
\end{equation}  where 
\begin{equation}
\langle \delta \rho_{\rm DM}^{2}(\br) \rangle =\sum_{\bn,\bn'} \langle a_{\bn} a_{\bn'} \rangle \rho^{[\bn]}(\br)
\rho^{[\bn']}(\br)\,.
\end{equation}
Hence we expect an excess of annihilation because of the polarized clumps within the halo.
Similarly, we may predict the angular correlation function, or the related variance as a function 
of smoothed angular scale as 
\begin{eqnarray}
\langle \delta N_{\rm a}(\bO) 
 \delta N_{\rm a}(\bO') \rangle
&=& \hskip -0.3cm W_{\chi}^{2} \iint   \d r   \d r'  \langle \delta \rho_{\rm DM}^{2}(\bO,r)  \delta \rho_{\rm DM}^{2}(\bO',r') \rangle  \nonumber
  \\ && \hskip -2cm = W_{\chi}^{2} \hskip -0.3cm \sum_{\bn_{1}, \bn_{2}, \bn_{3},\bn_{4}} 
  \langle a_{\bn_{1}}  a_{\bn_{2}} a_{\bn_{3}} a_{\bn_{4}} \rangle \iint   \d r  \d r'  \times \nonumber \\ 
  &&  \hskip -2cm \rho^{[\bn_{1}]}(r,\bO)    \rho^{[\bn_{2}]}(r,\bO)  \rho^{[\bn_{3}]}(r',\bO')  \rho^{[\bn_{4}]}(r',\bO') \EQN{dNan} \,,
\end{eqnarray}
where $\delta N_{\rm annih}(\bO)\equiv N_{\rm annih}(\bO)-\langle N_{\rm annih}(\bO)\rangle $.
Note that we assumed here that the resolution of the telescope was effectively infinite
(\ie $\Delta \Omega \rightarrow 0$ in \Eq{defFDM}).
 Now we may rely on Wick theorem to express the 
four-point correlation entering \Eq{dNan}  as products of two-point correlations. 
Calling $\delta {a} \equiv  a - \langle a \rangle$, we have $  \langle\delta a_{\bn_{1}} \delta a_{\bn_{2}}\delta a_{\bn_{3}}\rangle =0$, and 
\begin{eqnarray}
 \langle\delta a_{\bn_{1}} \delta a_{\bn_{2}}\delta a_{\bn_{3}}\delta a_{\bn_{4}} \rangle&\!\!\!=\!\!\!& \langle\delta a_{\bn_{1}} \delta a_{\bn_{2}}\rangle \langle\delta a_{\bn_{3}}\delta a_{\bn_{4}} \rangle
 + \hskip -1cm
 \nonumber \\
&& \hskip -2cm \langle \delta a_{\bn_{1}} \delta a_{\bn_{3}} \rangle \langle\delta a_{\bn_{2}}\delta a_{\bn_{4}} \rangle +  \langle\delta a_{\bn_{2}} \delta a_{\bn_{3}} \rangle \langle\delta a_{\bn_{1}}\delta a_{\bn_{4}} \rangle\,. \EQN{Wick}
\end{eqnarray}
If the infall  is  statistically isotropic,  
\Eq{dNan} may be averaged over the direction, $\bO$ and reads:
\begin{equation}
\langle \delta N_{\rm annih}(\bO) 
 \delta N_{\rm annih}(\bO') \rangle_{\Omega}=
 \sum_{\ell} C^{\rm annih}_{\ell}P_{\ell}[\bO\cdot \bO']\,,
 \end{equation} { where} 
 \begin{equation}
C^{\rm annih}_{\ell}= W_{\chi}^{2} \sum_{\ell} C^{\rm DM}_{\ell_{1}} C^{\rm DM}_{\ell_{2}} U^{\ell}_{\ell_{1},\ell_{2}}\,.
\end{equation}
Note that the geometric factor, $U^{\ell}_{\ell_{1},\ell_{2}}$, only depends on the basis function, $
\rho^{[\bn]}(\br)$ and possibly the resolution of the telescope if it is not assumed to be infinite:
\[
U^{\ell}_{\ell_{1},\ell_{2}}=\sum_{\bn_{1}, \bn_{2}, \bn_{3},\bn_{4}}\int \d \bO Y^{m *}_{\ell}(\bO') \int \d \bO' Y^{m *}_{\ell}(\bO')
\times 
\]
\begin{equation}
 \iint  \d r \d r'  \rho^{[\bn_{1}]}(r,\bO)    \rho^{[\bn_{2}]}(r,\bO)  \rho^{[\bn_{3}]}(\bO',r')  \rho^{[\bn_{4}]}(\bO',r') \,,
\end{equation}
given that $ \rho^{[\bn]}(\br) = u^{n}_{\ell m}(r) Y_{\ell}^{m}(\bO)$ and given the properties of 
spherical harmonics, the integral
\(\displaystyle
\int \d \bO 
Y_{\ell_{1}}^{m_{1}}(\bO)Y_{\ell_{2}}^{m_{2}}(\bO)Y_{\ell_{3}}^{m_{3}}(\bO') Y_{\ell_{4}}^{m_{4}}(\bO') \d \bO
\)
can be re expressed iteratively in terms of Clebsch-Jordan coefficients.
Ensemble average and comparison with the observation is possible at the high $\ell$ limit corresponding to the 
small-scale structure of the dark matter halo, for which we may expect independent angular regions of the  Galactic 
halo to be representative of an ensemble average.

\subsection{Bremsstrahlung X-ray emission  of stacked haloes} 

\label{s:xray}

Assuming that the gas traces the dark matter, we may reproduce the thought experiment of \Sec{neutralino},  though the ensemble average is constructed while staking projections
 of haloes on the sky rather than in a galactocentric  framework.

The emissivity per unit volume at frequency  $\nu$, $\varepsilon_{\nu}(\br)$,  for a hydrogen plasma is given by (\citet{Peacock})
\begin{eqnarray}
\varepsilon_{\nu}(\br) \d \br \d \nu&=&\frac{\epsilon_{\rm X } n_{\rm e}^{2}(\br) }{\sqrt{T_{e}(\br)}}\left(1+\log_{10}\left[\frac{k_{B}T_{e}(\br)}{h \nu}\right]\right)
\times \nonumber \\ 
&&\exp\left(-\frac{h \nu}{k_{B} T_{e}(\br)}\right)  \d \br \d \nu \,,
\end{eqnarray}
where 
 $T_{e}$ is the temperature in Kelvin, $\nu$ the frequency in Hz, $k_{B}$ the Boltzmann constant, $h$ the Planck 
constant,
and $\epsilon_{\rm X}\equiv 6.8 10^{-32}$ for an emissivity in $\rm W m^{-3} Hz^{-1}$.
Let us assume here that the cluster is isothermal, hence the variation of $T_{e}$ with $z$ are neglected compared to that of $n_{e}$ squared.\footnote{this is a better approximation than for the SZ effect}. Let us also assume that $\mathrm{L}/\mathrm{M}$ is the mass to light ratio of the cluster is constant.
Hence the  emissivity per unit surface, $\sigma_{\nu}$, is given by
\begin{equation}
\sigma_{\nu}(\M{R})  =
 \int \d z \left(\mathrm{L}/\mathrm{M}\right)^{2} \varepsilon_{\nu}(\M{R},z)
 \equiv  W_{X} \,   \int \d z \rho^{2}_{\rm DM}(\M{R},z) \,. \EQN{defXB}
\end{equation}
Hence, taking an ensemble average yields:
\begin{equation}
\langle \sigma_{\nu}(\M{R}) \rangle
=   W_{X} \int \d z \langle \rho_{\rm DM}\rangle^{2}(\M{R},z)+ W_{X} \int \d z \langle \delta \rho_{\rm DM}^{2}\rangle(\M{R},z)\,,
\end{equation}
where 

\begin{equation}
 \langle \rho_{\rm DM}^{2}\rangle(\M{R},z)= \sum_{\bn,\bn'}  \langle a_{\bn} a_{\bn'}  \rangle \rho^{[\bn]}(\br)\rho^{[\bn']}(\br)\,.
\end{equation}

The two-point correlation of the cosmic fluctuation of the emissivity is given by
\begin{displaymath}
\frac{\langle \delta \sigma_{\nu}(\M{R}) 
\delta \sigma_{\nu}(\M{R}')\rangle}{\langle \sigma_{\nu}({R}) \rangle^{2}}
= \frac{\mbox{$\displaystyle \iint \d z  \d z ' \langle
\delta \rho^{2}_{\rm DM}(\M{R},z)
\delta \rho^{2}_{\rm DM}(\M{R}',z')\rangle  $}}{\langle \sigma_{\nu}({R}) \rangle^{2}} \,,
\end{displaymath}
where
\begin{eqnarray}
 \langle
\delta \rho^{2}_{\rm DM}(\M{R},z)
\delta \rho^{2}_{\rm DM}(\M{R}',z')\rangle&=& \hskip -0.5cm \sum_{\bn_{1}, \bn_{2}, \bn_{3},\bn_{4}} 
  \langle a_{\bn_{1}}  a_{\bn_{2}} a_{\bn_{3}} a_{\bn_{4}} \rangle \times \nonumber \\ &&
\hskip -4cm   \rho^{[\bn_{1}]}(\M{R},z)\rho^{[\bn_{2}]}(\M{R},z)  \rho^{[\bn_{3}]}(\M{R}',z')\rho^{[\bn_{4}]}(\M{R}',z')\,.
   \EQN{emissfluc}
\end{eqnarray}
Note the cancellation of the dependence on $W_{\rm X}$ (hence $T$ or $\mathrm{ M}/\mathrm{L}$) in \Eq{emissfluc}.
Relying again on Wick's theorem, \Eq{Wick}, we may express the four-point correlations as a products of known
(\cf \Eq{Wick}) two-point correlations.

\subsection{Galactic halo's ellipticity}
\label{s:ellip}


\label{s:ellip}
More generally, let us consider a problem which depends non trivially on the 
perturbed distribution function, \eg the ellipticity, $e_{\rm H}$, of the departure from sphericity of the substructures induced
by the environment around a given halo. The ellipticity is defined as
\begin{equation}
e_{\rm H}= \frac{3 \lambda_{1}}{\sum_{i} \lambda_{i}}-1\equiv{\cal G}(\delta \rho(\br) )\,,
\,\,\,\, {\rm with}\,
\,\, \{\lambda_{i}\}={\rm Eigenval}(\bI_{\rm H} )\,,
\end{equation}
and 
\begin{equation}
 \left.
\bI_{{\rm H},i,j}= \int_{\le R_{\rm 200}} \d \br \delta \rho(\br) x_{i} x_{j}\right/ 
 \int_{\le R_{\rm 200}} \d \br  \rho_{\rm NFW}(\br) \,,
 \end{equation}
(so that $\lambda_{1}$ is the largest eigenvalue of $\bI_{\rm H} $ and   $ e_{\rm H}=0$  if the halo' s perturbation is spherical).
Since we know the statistical properties of $\delta \rho(\br)$, we may  
predict the statistical properties  of $e_{\rm H}$.
In practice, 
assuming $\cal G$ is a well-behaved function of its arguments, we may Taylor-expand 
$e_{\rm H}$ with respect to $\delta \rho$ as :
\begin{equation}
e_{\rm H}= \sum_{n} \left(\pdrvn{\cal G}{\delta \rho}{n} \right) \cdot \left[\delta \rho(\br_{1}) - \langle \delta \rho(\br_{1})\rangle  \right]\cdots \left[\delta \rho(\br_{n}) - \langle \delta \rho(\br_{n})\rangle  \right] \EQN{derivE}
\end{equation}
Note that the derivative in \Eq{derivE} is a Frechet functional derivative, so that the dot
involves an integration over $\br$.
Hence the ensemble average, $\langle e_{\rm H} \rangle$ will involve N-point correlations,
and reads
\begin{equation}
\langle e_{\rm H} \rangle\!=\! \sum_{n} \!\!
\sum_{\M{i}_{1} \cdots \M{i}_{n}} \langle a_{\M{i}_{1}} \cdots a_{\M{i}_{n}} \rangle
\left(\pdrvn{\cal G}{\delta \rho}{n} \right) \cdot
  \rho^{[\M{i}_{1}]}(\br)   \cdots  \rho^{[\M{i}_{n}]}(\br_{n})\,, \EQN{defmneanE}
\end{equation}
where, once again, we may rely on Wick's theorem to reexpress 
$\langle a_{i_{1}} \cdots a_{i_{n}} \rangle$ as products of two-point correlations.
Since the relationship between the density perturbation and the ellipticity is not 
linear, we expect a non zero ellipticity on average. 

Note that in principle, we may reconstruct the  full 
PDF of $e$. Formally, calling $z\equiv (e,a_{2},\cdots a_{\bn})$ (so that 
$z=({\cal G}(a_{1},\cdots, a_{\bn}), a_{2}\cdots, a_{\bn})=g(a_{1},\cdots,a_{\bn})$),
inverting for $z$ as a function of $\{a_{i}\}$ (provided the ellipticity is not degenerate in 
$a_{1}$), and marginalizing over 
the other coefficients yields:
\[
\PDF(e)=\int \d a_{2}\cdots \d a_{\bn} \PDF\left(g^{-1}(z)\right)/\left|\pdrv{z}{a_{\bn}}
 \right| \,.
\]
Now in practice, \Eq{defmneanE} might not be the simplest procedure to
compute  $\langle e_{\rm H} \rangle$, and monte carlo resimulation may
turn out to be more practical.

\subsection{Metal lines in QSO  DLA  systems}
%
%
Let us finally consider a more convolved observable, which will depend on both
the clump distribution within the haloes, but also on their velocities. 
\parn
In the red part of a high resolution spectrum of  quasars, 
groups of absorption features are found, corresponding to the physical situation 
where the light emitted by the quasar is partially absorbed by the  metal-rich\footnote{Since we make predictions at lower redshift we need to concentrate on 
metals such as $\rm Mg_{{}_{\rm II}}$, or  $\rm Fe{{}_{\rm II}}$ which are found 
typically at redshift $z\le 1.5$ in the visible} clumps
which the line of sight happens to intercept. 
Formally, the normalized flux in a QSO is proportional to minus the log of the optical 
depth along the line of sight. 
 The optical 
depth in the metal transition  is (\citet{pichly}):
\begin{eqnarray}
\tau(w,\M{R})&=& \frac{c \, \sigma_0 }{H(\overline{z}) \sqrt{\pi}}
\int_{-\infty}^{+\infty} \frac{n_{{\rm Z}}(v,\M{R})}{b(v,\M{R})}\times \nonumber\\ && \exp\left(- \frac{(w-
v-v_{z}(v,\M{R}))^2}{b(v,\M{R})^2} \right) \d v \,, \EQN{DefTau}
\end{eqnarray} 
where 
$c$ is the velocity of light, $\sigma_0$ is the metal absorption
cross-section, $H(\bar z)$ is the Hubble constant at redshift $\overline{z}$,
$n_{{\rm Z}}(v,\M{R})$ the ionized
metal number  density field, $b(v,\M{R})$ the Doppler parameter (accounting for the thermal broadening of the line),
and $v_{p}(v,\M{R})$ is the peculiar velocity, at impact parameter, $\M{R}$ from the centre of the cluster. The observed normalized flux, $F$, is simply $F=\exp(-\tau)$.
If we assume here again constant biasing, so that $n_{\rm Z}\propto \rho_{\rm DM}$. This assumption may be lifted once the identification of 
virialized substructure described in \Sec{satcount} is carried through.
The two-point correlation of the optical depth fluctuation will involve statistical properties of both the  density and the velocity field in a non trivial manner.
\begin{equation}
\frac{1}{\langle \tau \rangle^{2}(w,{R})} \langle \delta \tau(w,\M{R}) \delta \tau(w',\M{R}) \rangle
\,,\end{equation}
with
\begin{equation}
\quad \delta \tau(w,\M{R}) =  \tau(w,\M{R})-\langle \tau \rangle(w,{R})\,.
\EQN{corelmetal}
\end{equation}
Note that the distance to the halo center, $\M{R}\equiv b (\cos[\vartheta_{b}],\sin[\vartheta_{b}])$ still occurs in \Eq{corelmetal}.
Since we do not know in general the impact parameter of the line of sight with respect to the 
halo center, let us marginalize over its {\it a priori} probability distribution, which we may infer from \eg the PT model (which at these scales corresponds
essentially to the autocorrelation of the unperturbed 
universal halo profile).
Given that we consider systems at the redshift of a damped Lyman-$\alpha$,
we may assume that we fall close to a galactic structure.
Calling $p_{b}(b,\bar z,M) \d b \d \bar z $ the probability of  a given point in space to be 
at a distance, $b$ within $\d b$ of an object of mass larger than $M$, which is at redshift $\bar z$ within $\d \bar z$, 
we may construct the weighted sum :
\begin{eqnarray}
C_{\tau}(\Delta w)&=&
\int_{0}^{\infty} \d b \int_{0}^{\infty}  \d \bar z \int_{0}^{2 \pi} \d \vartheta_{b}  p_{b	}(b,z,M) \times
\EQN{defCt} \\
&& \hskip -2.cm
\left\langle \delta \tau(w, b \cos[\vartheta_{b}] ,b \sin[\vartheta_{b}]) \delta \tau(w+ \Delta w, b \cos[\vartheta_{b}] ,b \sin[\vartheta_{b}]) \right\rangle_{w}\,. \nonumber
\end{eqnarray}
This quantity may now be compared to the observable.
Let us assume some equation of state for the metal phase, 
so that $b(\M{R},z)=b_{0}\left({ \rho}(\M{R},z)\!\!\left.\right/{\bar \rho}\right)^{\gamma}$.
\Eq{DefTau} may then be written formally as 
\(
\delta \tau(w,\M{R}) = {\cal T}[\delta \rho(v,\M{R}),  v_{z}(v,\M{R})] 
\,. 
\)
Let us Taylor expand this expression in the neighborhood of the mean 
density fluctuation as:
\begin{eqnarray}
\delta \tau(w,\M{R})&=& 
\sum_{n} \left(\pdrv{{}^{n}\cal T}{\delta \rho\,\cdots \partial \delta v_{z} } \right) \cdot \nonumber \\ && \hskip -1cm
\left[\delta \rho(\br_{1}) - \langle \delta \rho(\br_{1})\rangle  \right]\cdots \left[\delta v_{z}(\br_{n}) - \langle \delta v_{z}(\br_{n})\rangle  \right]\,.
\EQN{derivT}
\end{eqnarray}
Again the derivative in \Eq{derivT} is a functional derivative (\cf \Sec{ellip}).
\Eqs{defCt}{derivT} together with \Eq{correlA}
yield the expected correlation as a function of the statistical environment.


\end{document}